\documentclass[12pt,preprint]{aastex}

\begin{document}
\title{Deep MIPS Observations of the IC 348 Nebula: Constraints on the 
Evolutionary State of \textit{Anemic} Circumstellar Disks and the Primordial-to-Debris 
Disk Transition}
\author{Thayne Currie\altaffilmark{1}, Scott J. Kenyon\altaffilmark{1}}
\altaffiltext{1}{Harvard-Smithsonian Center for Astrophysics, 60 Garden St. Cambridge, MA 02140}
\email{tcurrie@cfa.harvard.edu, skenyon@cfa.harvard.edu}
\begin{abstract}
We describe new, deep MIPS photometry and new high signal-to-noise optical spectroscopy of 
the 2.5 Myr-old IC 348 Nebula.  To probe the properties of the IC 348 
disk population, we combine these data 
with previous optical/infrared photometry and spectroscopy 
to identify stars with gas accretion, to examine their mid-IR colors, 
and to model their spectral energy distributions.  IC 348 contains 
many sources in different evolutionary states, including 
protostars and stars surrounded by primordial disks, two kinds of 
transitional disks, and debris disks.   
Most disks surrounding eary/intermediate spectral-type stars ($>$ 1.4 M$_{\odot}$ at 
2.5 Myr) are debris disks; most disks surrounding solar and subsolar-mass 
stars are primordial disks.  
At the 1--2 $\sigma$ level, more massive stars also have a smaller frequency of 
gas accretion and smaller mid-IR luminosities than lower-mass stars.
These trends are suggestive of a stellar mass-dependent evolution of disks, where
 most disks around high/intermediate-mass stars 
shed their primordial disks on rapid, 2.5 Myr timescales.  The frequency of MIPS-detected transitional 
disks is $\approx$ 15--35\% for stars plausibly more massive than 0.5 M$_{\odot}$.  
The relative frequency of transitional disks in IC 348 compared to that for 1 Myr-old Taurus and 5 Myr-old NGC 2362 
is consistent with a transition timescale that is a significant fraction of the 
total primordial disk lifetime.
\end{abstract}
\keywords{stars: pre-main-sequence--planetary systems: formation --- planetary systems: protoplanetary disks
}
\section{Introduction}
Most $\le$ 1 Myr-old stars are surrounded by optically-thick, accreting {\it primordial} disks, 
which produce strong near-to-mid infrared (IR) emission \citep[L$_{d}$/L$_{\star}$ $\ge$ 0.1;][]{Kh95}.  
The gas and dust in these disks comprise the building blocks of planets.  As stars age, 
the dust grains grow, settle toward the midplane, and become incorporated into planetesimals. 
  Circumstellar gas depletes by accretion onto the star \citep{Ha98}.  
All of these processes reduce the amount and 
 frequency of disk emission on timescales of $\sim$ 3--7 Myr for most stars \citep{He07a, Hi08b,Clp09}.

  By $\sim$ 10 Myr, most stars do not have optically-thick primordial 
disks \citep{Cu07a,Hi08b}.  Disks around these older stars are 
typically debris disks \citep{Cu08a}, which have weaker, optically-thin dust emission 
(L$_{d}$/L$_{\star}$ $\lesssim$ 10$^{-3}$) and lack evidence for circumstellar gas accretion.
Because stellar radiation -- radiation pressure and Poynting-Robertson 
drag -- or stellar wind drag can remove dust in debris disks on timescales much 
less than the age of the star, their dust requires a replenishment 
source, which is supplied by active icy or rocky planet formation \citep[e.g.][]{Bp93, 
Kb04, Kb08}.  Because debris disks lack copious amounts of 
circumstellar gas needed to form gas giant planets, identifying 
when most primordial disks turn into debris disks pinpoints 
an empirical upper limit for the formation timescale for gas giant planets.

Recent \textit{Spitzer Space Telescope} studies of 1--5 Myr-old clusters 
indicate that primordial disks disappear faster around 
early type, high/intermediate-mass stars than they do around late type, low-mass stars 
\citep{Ca06, La06, He07a, Clp09}.  By 5 Myr, debris disks completely dominate 
the disk population around high/intermediate-mass stars.  At 5 Myr, most disks around lower-mass stars
appear to have inner holes and/or extremely low warm dust masses \citep{Dahm09,Clp09}, which indicates that 
they may be disappearing and actively making the primordial-to-debris 
disk transition.  Thus, the epoch at which most high/intermediate-mass stars make the 
primordial-to-debris disk transition must occur at an age prior to 5 Myr and may be identified 
by mid-IR observations of younger clusters.

With a median age of $\sim$ 2--3 Myr, stars in the IC 348 Nebula \citep{He98} 
may help to constrain the epoch 
when most primordial disks around high/intermediate-mass 
stars turn into debris disks.  Cluster membership 
is well determined \citep[e.g.][]{He98,Lu98,Lu03,Mu07}.
At only 320 pc distant, the cluster provides a sensitive probe of 
low-levels of disk emission in spite of its high infrared 
background.  

Recently, \citet{La06} and \citet{Muzerolle2006} analyzed Spitzer Cycle 1 IRAC and MIPS IC 348 data 
  and found evidence for substantial disk evolution.  
Based on the IRAC 3.6 $\mu m$ to 8 $\mu m$ flux slope, 
\citet{La06} show that only $\approx$ 30\% of cluster stars have strong ('thick')
IRAC excess emission comparable to stars with primordial disks (e.g. most disk-bearing stars in Taurus).  
Another $\approx$ 23\% have weaker ('anemic') IRAC excess emission, and $\approx$ 44\% lack 
IRAC excess emission ('diskless').  However, inferring the physical state of 
disks from the observed IRAC slope is not straightforward.
Both debris disks and 'evolved' primordial disks ('transitional' disks: those with lower disk masses and/or inner holes)
 can have weak IRAC-excess emission \citep{Cu07b, Clp09}.  Furthermore, while SED analysis 
including longer wavelength data (e.g. 24 $\mu m$) helps to break this 
degeneracy \citep{Clp09}, only $\approx$ 30\% of IC 348 stars have Cycle 1 MIPS detections.  
Therefore, deeper MIPS data coupled with other diagnostics of the disk 
evolutionary state (e.g. gas accretion signatures) 
would help to make IC 348 a better laboratory for studying disk evolution.

In this paper, we combine new, high signal-to-noise 
optical spectra and new, deeper MIPS 24 $\mu m$ and 70 $\mu m$ 
photometry of IC 348 with archival data to study disk evolution 
in more detail.  We first use the optical spectra to search for evidence of 
circumstellar gas accretion, which is absent in debris disks.  
Deeper MIPS data is then used to establish a much larger sample 
of disk-bearing stars.  From these data, we compare 
the disks SEDs to models to constrain their evolutionary states, 
to compare disk properties around high-mass stars to low-mass stars, and to investigate the 
timescale for and the morphology of the primordial-to-debris 
disk transition.

\section{New Observations and Archival Data}
\subsection{New Optical Spectroscopy of High/Intermediate-Mass IC 348 Stars}
To investigate accretion signatures for IC 348 stars,
we took high signal-to-noise spectra of cluster stars with published 
spectral types earlier than K0.  Our primary motivation for 
selecting this sample is that these stars lack recent high signal-to-noise 
observations.  High signal-to-noise observations are effective 
in identifying gas accretion signatures around solar and subsolar-mass 
cluster stars not previously known as accretors \citep{Dahm08}.
New observations of early/intermediate type stars may identify 
sources with emission line reversal in the cores of their H$_{\alpha}$ 
absorption lines, a feature which is consistent with gas accretion 
\citep[][]{Dahm08}.

Selected IC 348 stars were observed on March 1, 2008 with the 
FAST spectrograph on the 1.5m Tillinghast telescope at Fred Lawrence 
Whipple Observatory \citep{Fa98}.  The spectra were taken using a 300 g mm$^{-1}$ 
grating blazed at 4750 \AA\ and a 3" slit.  These spectra cover 3700-7500 \AA\ 
at 6 \AA\ resolution.  The data were processed using the standard FAST reduction 
pipeline \citep{Fa98}. The continuum signal to noise at $\approx$ 6550 \AA\ ranged 
from 80 to 170.  To derive line fluxes in H$_{\alpha}$, we fit the 
line to a gaussian profile with standard IRAF routines.  The properties
 of each source and their spectral features are listed in Table \ref{spectra}.

\subsection{Archival Optical/Near-IR Ground-based Data and Archival Spitzer 3.6--24 $\mu m$ Data}
We collected published optical and infrared photometric data and spectroscopic data
for IC 348 stars from the \citet{La06} catalog.  This catalog yields photometric data from 
3.6 $\mu m$ through 24 $\mu m$, optical extinction (A$_{V}$), spectral types, and IRAC SED slopes (thick, anemic or 
diskless).  We added BVI photometry of IC 348 sources from \citet{Lu03}, \citet{He98}, and \citet{Lu05}, 
near-IR JHK$_{s}$ photometry from 2MASS, and bolometric luminosities from \citet{Mu07}.  

Nearly all of the 307 sources in the \citet{La06} sample have spectral types (304) and A$_{V}$ measurements (295).  
Most have 2MASS data (280), I-band data (270), and IRAC 8 $\mu m$ data (265).  
Completeness in the MIPS 24 $\mu m$ band is poorer (88 sources). 
Many sources without MIPS detections are either not detected in the 8 $\mu m$ IRAC channel (35) and/or have
photospheric colors through 8 $\mu m$ (128).  
\subsection{New MIPS 24 $\mu m$ and 70 $\mu m$ Photometry}
IC 348 was reobserved in five separate epochs with the Multiband Imaging Photometer for Spitzer (MIPS) on
September 23--27, 2007 (Observing Keys 22961920, 22962176, 22962432, 22962688, and 22962944; 
PI J. Muzerolle).  The area of coverage is $\approx$ 0.5 square degrees.  
A series of papers focused on analyzing the IC 348 sources with variable 
mid-IR emission will be published later using these MIPS data and 
IRS spectra (J. Muzerolle, K. Flaherty, et al. 2009a,b, in prep.).  Our motivation for using 
these data is to identify new 24 $\mu m$ detections of cluster stars from stacking together 
pointings from the five epochs.  Preliminary findings indicate that 
some low-mass IC 348 stars have 24 $\mu m$ emission varying by up to several tenths of magnitudes 
from epoch to epoch (J. Muzerolle, pvt. comm.).  Therefore, by combining observations from these epochs together, we sacrifice 
any information about disk variability.  The resulting photometry is an epoch-averaged value: 
photometry for low-mass stars from individual epochs may differ by several tenths of magnitudes. 

\subsubsection{MIPS 24 $\mu m$ Image Processing, Photometry, and Source Matching}
Using \textit{MOPEX} \citep{Mak05b}, we first mosaic the Basic Calibration 
Data (BCD data) together using a bicubic spline interpolation.  
In constructing the final 
mosaic, we rejected and reinterpolated over pixels with values 
more than 5 $\sigma$ away from the (clipped) mean value.  
Standard deviation mosaics were also constructed.
The Basic Calibration Data (BCD) frames have an integration time of 3.67s; most pixels 
had $\gtrsim$ 110 separate pointings.  
The typical cumulative integration time/pixel is then $\approx$ 410s, 
about a factor of five greater than the Cycle 1 data analyzed in \citet{La06}.

Source detection and photometry was performed 
 using \textit{APEX} \citep{Mak05} in single-frame mode.  In a first pass, we identified sources as 
clusters of 4--50 pixels in the signal-to-noise mosaic, which is produced by dividing the background-subtracted 
science mosaic by a map of the background fluctuations (produced from the \textit{gaussnoise} module from APEX).  
We used the 'peak' algorithm with a 45x45 pixel box to define the background and a 2.5$\sigma$ detection threshold.
While these settings yielded robust detections of many faint stars, they sometimes 
'deblended' single stars, especially those with bright Airy rings.  Furthermore, these settings 
missed detecting some sources located in regions with smaller-scale fluctuations in the 
background (e.g. near the center of the nebula).  Therefore, in a second pass, we chose a 
25x25 pixel box, using a 5$\sigma$ detection threshold and the 'combo' algorithm, which  
alleviated these problems. 

Pixel Response Function (PRF) photometry was performed on detected sources, using 
an empirical PRF derived from bright (S/N $>$ 50-100) stars in low, uniform background regions 
resampled from the native pixel scale ($\sim$ 2.45"/pixels) by a factor of four.  
The empirical PRF and model PRF supplied by the \textit{Spitzer Science Center} show 
strong agreement.  PRF fitting was performed over a 5x5 pixel area with fitting tolerance of 10$^{-3}$. 
Photometric errors were estimated from both the signal-to-noise from background fluctuations (due to both 
photon noise and nebulosity) and errors in the PRF fitting.  We detect $\approx$ 5,900 candidate 
point sources; the source counts peak at [24] $\sim$ 11.

\begin{figure}
\centering
\epsscale{0.99}
\plottwo{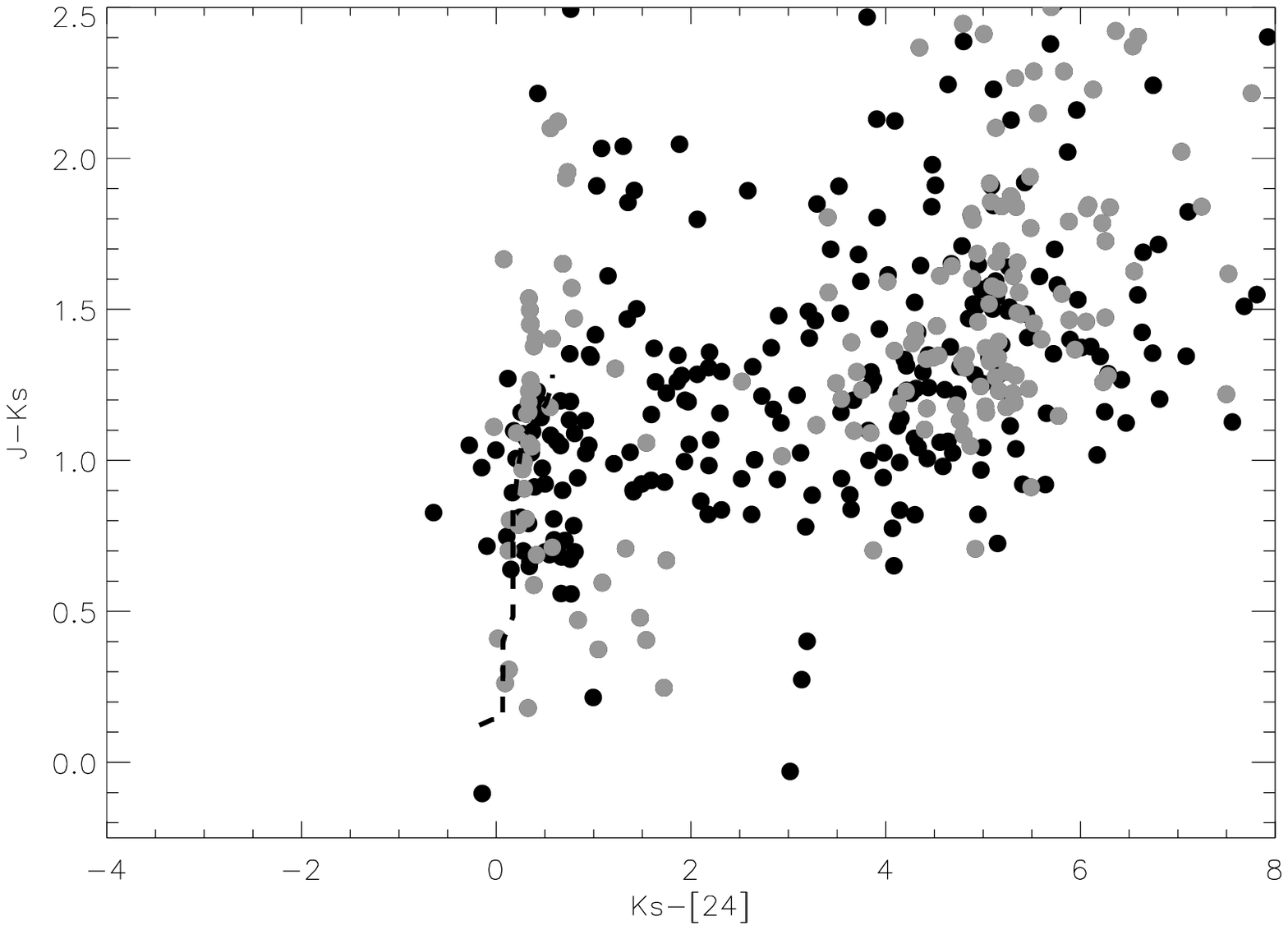}{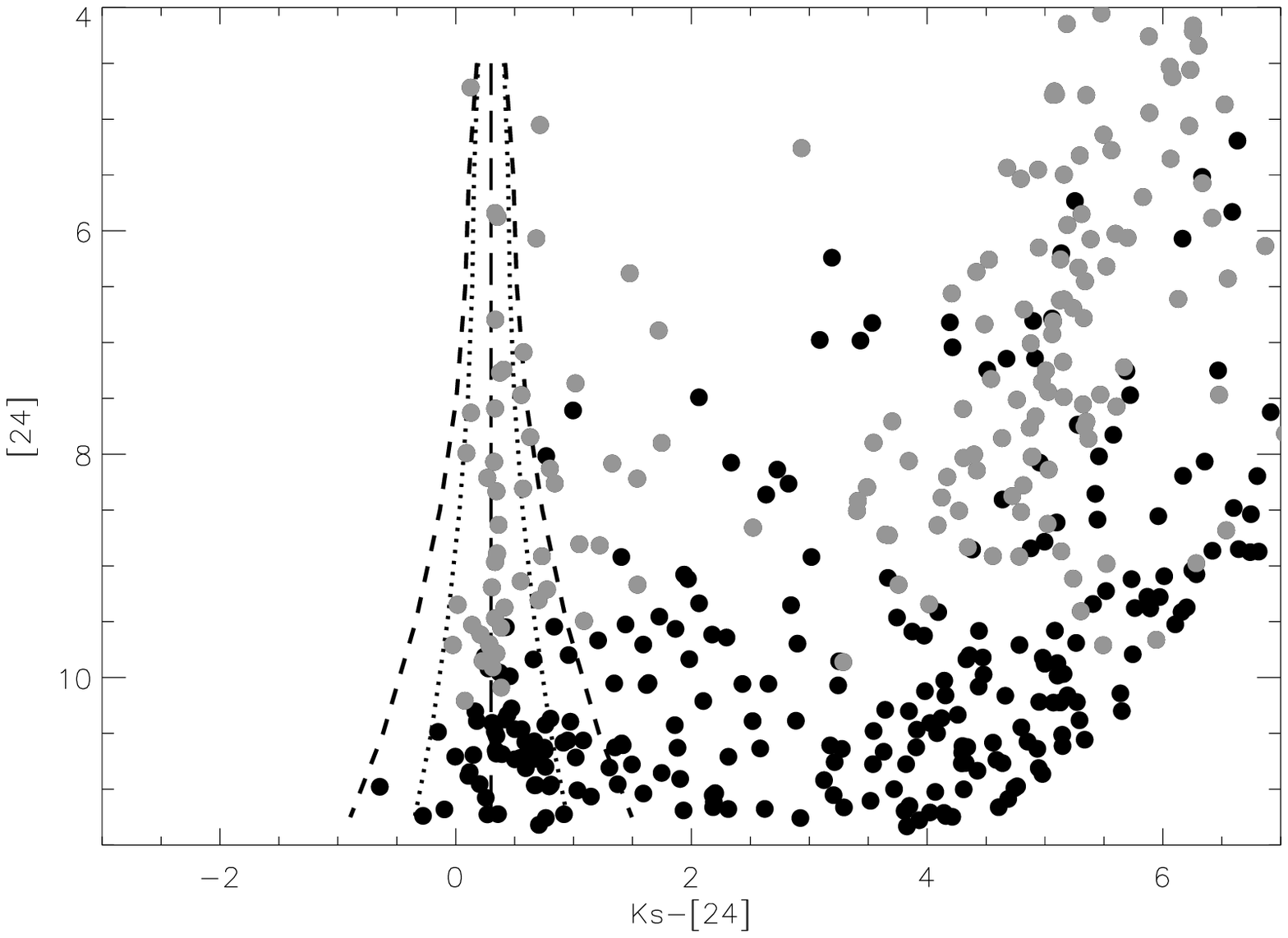}
\caption{(left) J-K$_{s}$ vs. K$_{s}$-[24] for all 451 MIPS-24 $\mu m$ detections with 2MASS counterparts.  The dashed 
line identifies the locus of photospheric colors for a star reddened by A$_{V}$=2.  (right)
The [24] vs. K$_{s}$-[24] color-magnitude diagram for the same sources.  The long dashed line identifies the 
photospheric locus for a star with K$_{s}$-[24]= 0.25: a mid-K star reddened by A$_{V}$=2 or an unreddened 
M2--M3 star.  The dotted line, short-dashed line, and dash-three dots line correspond to the 1 and 2 $\sigma$ 
limits from the combined K$_{s}$ and [24] photometric uncertainties assuming a $\sim$ 1 magnitude spread in 
extinction.}
\label{colmagall}
\end{figure}

To understand the empirical photometric scatter in our MIPS photometry and to identify candidate 
pre-main sequence stars, we cross-correlated our MIPS-24 $\mu m$ catalog with the 2MASS 
point source catalog.  Using a 4" matching radius, we found 451 matches.  The left panel of 
Figure \ref{colmagall} shows the J--K$_{s}$ vs. K$_{s}$-[24] colors for these sources.  
Overplotted as a dotted line is the locus of photospheric colors from the Kurucz-Lejeune stellar atmosphere 
models \citep{Kurucz1993,Lejeune1997} as listed by the STAR-PET tool available on the 
\textit{Spitzer Science Center} website.  Sources with 
photometric uncertainties in their K$_{s}$-[24] colors $<$ 0.2 magnitudes (grey dots) either lie well redward of the 
locus (IR excess sources) or are tightly clustered along the locus.  Sources with 
larger photometric uncertainties (black dots) appear to have a wider dispersion in color.  

Figure \ref{colmagall} (right panel) shows the [24] vs. K$_{s}$-[24] 
color magnitude diagram of these sources.  
The vertical curves (dotted, dashed, and three-dash dots lines) 
show the median 1 and 2 $\sigma$ uncertainties in K$_{s}$-[24] colors computed in 1-magnitude bins, assuming an 
intrinsic dispersion of 0.1 magnitudes centered on K$_{s}$-[24] = 0.25.  There is little blueward deflection 
of colors outside of the 1 $\sigma$ limit; the 2 $\sigma$ limit completely contains all sources with colors 
bluer than K$_{s}$-[24] = 0.25.  We do not find a significant trend between photometric scatter and 
background level (not shown).  Thus, if there is an intrinsic $\gtrsim$ 0.1 magnitude dispersion in colors 
due to differences in spectral type/reddening, the observed photometric scatter is explainable 
in terms of the computed photometric uncertainties for MIPS and 2MASS.

Using a matching radius of 4", we identify 27 IC 348 cluster stars with new 24 $\mu m$ detections.
  These detections were confirmed by inspection of the final
mosaic image, the background-subtracted mosaic, and the signal-to-noise mosaic.  
The coordinates were then checked against the positions of other sources to identify 
possible blends.  One source, ID 9099, is $\sim$ 3" from ID 99.  Its catalog position is 
offset from the MIPS detection by 2.83".  Another source, ID 1937, is within $\sim$ 4.3" of ID 1928 and has 
an offset of 3" between its catalog position and position on the MIPS-24 $\mu m$ mosaic.
We remove these sources from the list, leaving 25 non-blended sources detected using APEX.

To find missed sources, we \textit{manually} searched the positions of the remaining 192 stars in the \citet{La06} catalog 
without APEX detections.  This search yielded two additional detections (IDs 124 and 373).
PSF-fitting photometry was performed on these two sources using the IDL version of 
the \textrm{DAOPHOT} package with empirical PSFs constructed from 
bright stars uncontaminated by nebulosity (S/N $>$ 50--100).  With these new data 
added, the \citet{La06} catalog includes 27 sources with new detections, bringing 
the total number of MIPS-detected stars to 115.
This new sample is a $\sim$ 30\% larger than the sample analyzed in \citet{La06}.  
Table \ref{m24det} lists the properties of the IC 348 stars with new MIPS 24 $\mu m$ 
detections.

\subsubsection{MIPS 70 $\mu m$ Image Processing, Photometry, and Source Matching}
Using \textit{MOPEX}, we mosaiced both the MIPS-70 $\mu m$ BCD data  
and the filtered BCD data.  As with the 24 $\mu m$ data, we combined 
the individual frames together, rejecting and reinterpolating over 
pixels more than 5$\sigma$ from the (clipped) mean pixel value.
We constructed the final mosaic images and standard deviation 
mosaics.  The MIPS 70 $\mu m$ data have an intergration time/pixel of 
 4.19s for the BCD frames and a cumulative integration 
time of $\approx$ 210s, a factor of five improvement 
over previous data presented in \citet{Mu07}.  

Source identification and photometry were performed using \textit{APEX} in 
single-frame mode.  For source detection on both the unfiltered and filtered mosaics, 
we chose a 5$\sigma$ detection threshold for initial cluster detection, 
limiting image segmentation to candidate sources between 5 and 250 pixels in area.
Because there were too few bright point sources detected at 70 $\mu m$, we used the model 
PRF for photometry instead of constructing an empirical one.
  Each source was fit by the model PRF over a 7x7 pixel area with a resampling factor of four and a 
fitting tolerance of 10$^{-3}$.  Photometric errors were calculated in 
the same manner as the 24 $\mu m$ data.  Because sources brighter than $\approx$ 1 Jy can have 
their fluxes attenuated on the filtered mosaic, we chose the unfiltered mosaic for 
 70 $\mu m$ photometry.

Because MIPS-70 $\mu m$ is not sensitive to stellar photospheric fluxes of stars 
at the distance of IC 348, we cannot investigate the photometric reliability of 
the 70 $\mu m$ as we did with the 24 $\mu m$ data.
We matched the APEX 70 $\mu m$ source list to the \citet{La06} catalog using a 
5" matching radius.  
We find 6 sources in the MIPS-70 catalog with matches in the \citet{La06} catalog, four 
of which (IDs 31, 67, 10343, and 10352) are new detections.  
We visually confirmed these detections on both the unfiltered and filtered mosaics.
Table \ref{m70det} lists the properties of stars detected at MIPS 70 $\mu m$.

\subsubsection{MIPS Upper Limits for Unmatched Sources}
\begin{figure}
\centering
\epsscale{0.85}
\plotone{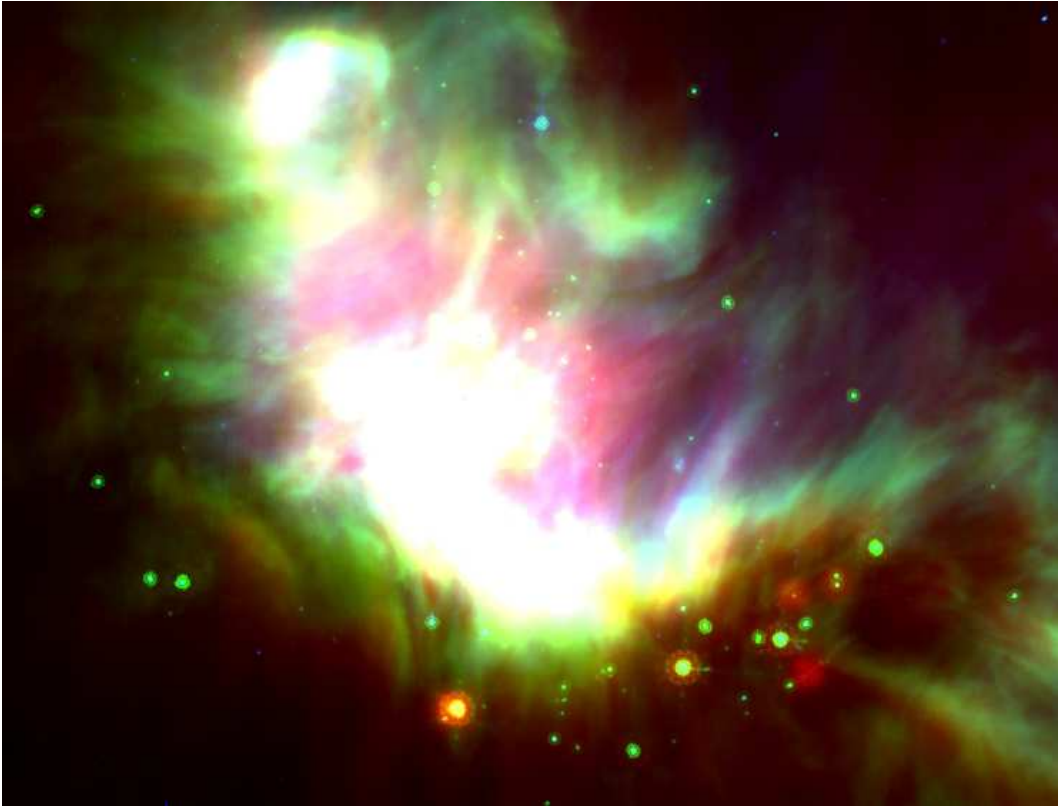}
\caption{Three color mosaic of the IC 348 Nebula.  Blue represents IRAC 8 $\mu m$ Cycle 1 data produced by the post-bcd pipeline 
mosaic; green and red colors represent the MIPS-24 $\mu m$ and 70 $\mu m$ emission from the processed data described in 
this paper.}
\label{threecolor}
\end{figure}

Figure \ref{threecolor} shows a RGB mosaic image produced from our 24 $\mu m$ 
and 70 $\mu m$ mosaics and the post-BCD mosaic of the IRAC 8 $\mu m$ data 
used by \citet{La06}.  Most of the IC 348 sources in the 
\citet{La06} catalog are located within regions of enhanced 
nebulosity.  While nebulosity does 
not serious impede detections of most of the \citet{La06} sources at 8 $\mu m$, 
the decreased sensitivity of MIPS and enhanced nebular emission at 24 $\mu m$ 
and 70 $\mu m$ greatly increase the number of nondetections at these wavelengths.

To compute the 5$\sigma$ upper limits for sources lacking detections, we 
use the 24 $\mu m$ and 70 $\mu m$ uncertainty images as maps of the 
noise.  The median 5$\sigma$ flux density is computed over a 3x3 region centered 
on the coordinates of each \citet{La06} source.  To calculate the 
5$\sigma$ flux upper limit for MIPS 24 $\mu m$, we multiply the flux 
density by a pixel area of $\pi$(2.45 pixels)$^{2}$ and use an aperture correction of 
1.697.  For the 70 $\mu m$ data, we multiply the flux density by a pixel area 
of $\pi$(1.62 pixels)$^{2}$ and use an aperture correction of 1.735.
Upper limits at 24 $\mu m$ varied from $\approx$ 0.3 mJy ([24] $\sim$ 11.0) in 
low (black) background regions to $\approx$ 180 mJy ([24] $\sim$ 4) in the 
high background regions of the nebula center (reddish-white/white).
  Typical upper limits at 70 $\mu m$ varied from $\approx$ 30 mJy 
in low background regions to $\approx$ 420 mJy in the center of the nebula.

\subsubsection{Comparisons with Previous IC 348 MIPS Photometry}
\begin{figure}
\centering
\plottwo{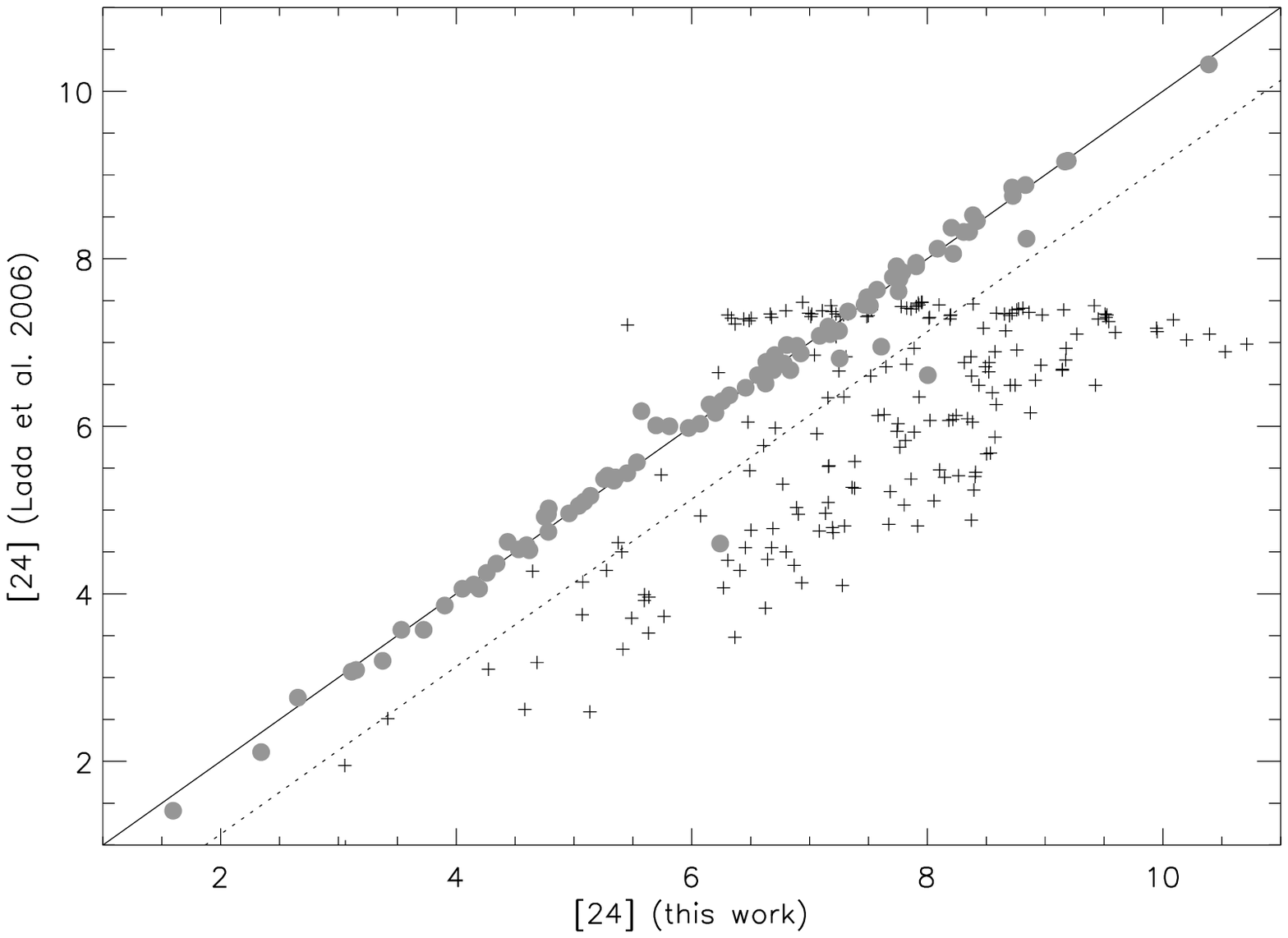}{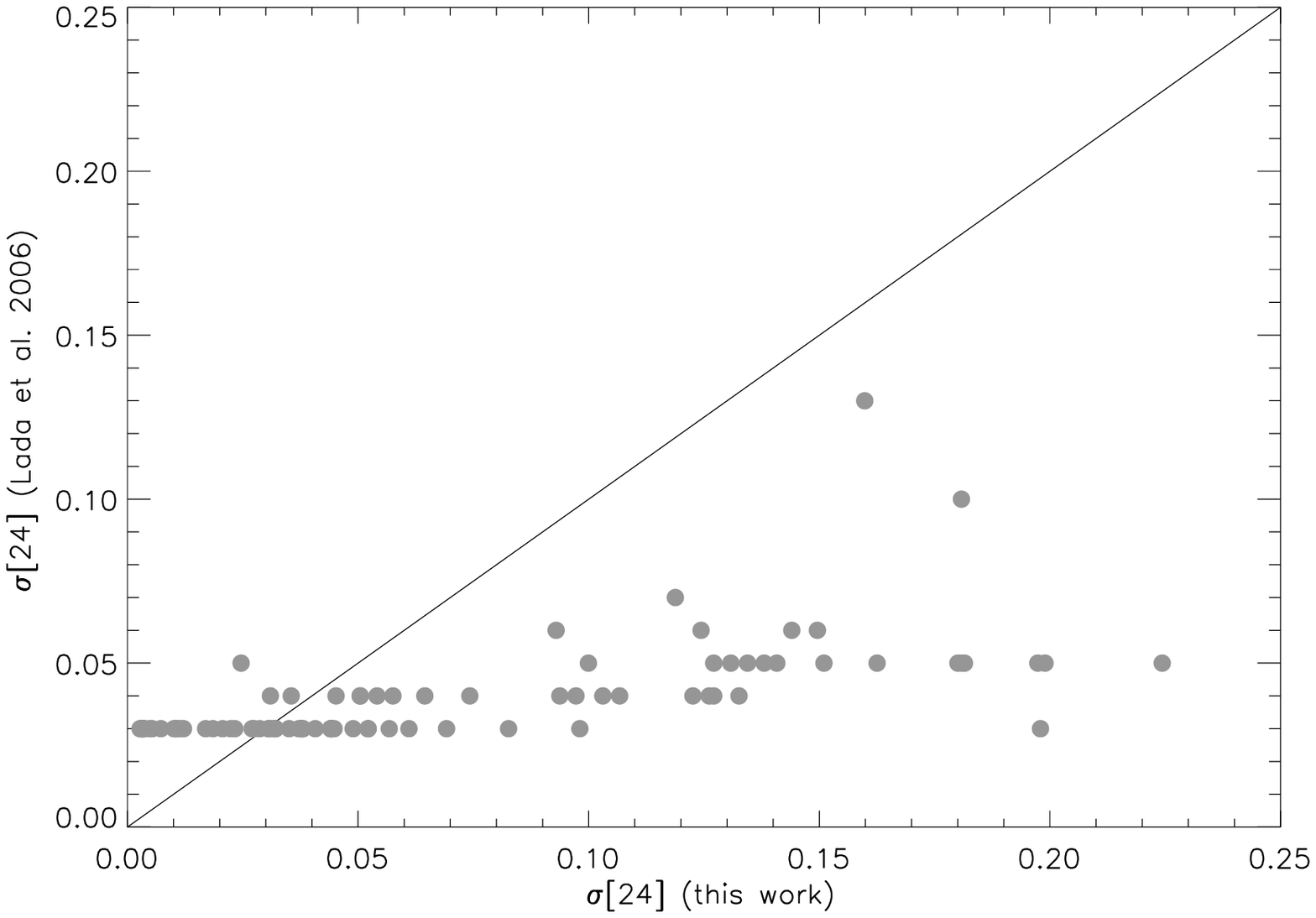}
\caption{Comparisons of our MIPS 24 $\mu m$ photometry/upper limits (left) and 
photometric errors (right) with those from \citet{La06}.  In the left panel, 
grey dots identify magnitudes for sources detected from both datasets and crosses identify 
the 2$\sigma$ upper limits for sources that lack detections in either dataset.  
The solid line identifies where our [24] photometry and that from \citet{La06}
would be in perfect agreement in the photon-noise limit.  The dotted line represents the expected locus of 
upper limits, given our exposure time, based on the \citet{La06} upper limits.  
In the right panel, grey dots represent photometric errors and the solid line identifies 
where our photometric errors are identical.} 
\label{ladacompare}
\end{figure}

In Figure \ref{ladacompare}, we compare our MIPS photometry
 with results from \citet{La06} and \citet{Mu07} derived from 
 Cycle 1 data.  The left panel contrasts
our [24] magnitudes and upper limits, where grey circles represent 
magnitudes for IC 348 members detected from both data sets and crosses represent 
2$\sigma$ upper limits for IC 348 members lacking detections from both data sets.   
Compared to the \citet{La06} photometry, our [24] photometry exhibits
 a systematic offset of -0.0075 magnitudes and a median difference of 0.05 
magnitudes.  While our 70 $\mu m$ flux for ID-13 is just over half that quoted by 
\citet{Mu07}, our measurement for ID-51 is nearly identical to that from \citet{Mu07}.

In the cases where the [24] photometry differs by more than 0.2--0.3 magnitudes, 
there is evidence that our photometry is more accurate.  For example,  
  ID-7 has [24] = 7.61 from this work and [24] = 6.95 from \citet{La06}. 
Comparing its flux densities on our mosaic and the Cycle 1 data 
post-BCD mosaic indicates that variability is unlikely to explain the strong 
disagreement in photometry.  \citet{La06} find that ID-7 is about equal in 
brightness to nearby ID-8.  However, it is clear that ID-7 is much
 fainter because its peak flux density above the local background is just over half 
that of ID-8's ($\approx$ 3.7 MJy sr$^{-1}$ vs. 7.1 MJy sr$^{-1}$).  
Thus, ID-7 should be $\approx$ 0.7 magnitudes 
fainter than ID-8, as confirmed by our measurements ([24]$_{ID-7}$=7.61 and [24]$_{ID-8}$=6.89).

ID-19 is another source whose discrepant photometry probably cannot be due to variability 
 ([24]=6.24 here, [24]=4.6 from \citealt{La06}).  This source is located near
the center of the nebula, where the background flux density increases from $\approx$ 75 MJy sr$^{-1}$ 
to $\approx$ 98 MJy sr$^{-1}$ over $\approx$ 20".  Unless we assume that the average background is 
$\approx$ 75 MJy sr$^{-1}$,  we fail to reproduce the \citet{La06} result from either PRF fitting or PSF fitting. 
However, there is a significant slope 
in the local background at constant declination near ID-19; thus, the average or median background 
is much higher than 75 Myr sr$^{-1}$.  Independent MIPS 24 $\mu m$
photometry yields values for ID-7 and ID-19 that are similar to ours
(J. Hernandez 2009, private communication).  

The left panel of Figure \ref{ladacompare} also compares our upper limits for nondetections (crosses).  
Assuming the \citet{La06} upper limits are accurate,  
The dotted line corresponds to the predicted locus of upper limits given our integration time in 
the photon noise limit (2.5log$_{10}$($\sqrt{5}$) $\sim$ 0.87). 
Because nebular emission dominates the background in most regions, our upper limits should 
lie to the left of the line.  However, most of our upper limits are clearly fainter than predictions based 
on Lada et al.'s upper limits.  Independent analyses also find significantly fainter upper limits
 (J. Hernandez 2009, private communication).  The disagreement likely arises because \citet{La06} add together the background 
flux and its uncertainty, which would determine the upper limit of the source \textit{plus} background, not the upper limit of 
the source flux, since the background is an astronomically distinct object.

Even though the new MIPS data are deeper, the right panel of Figure \ref{ladacompare} shows that our photometric uncertainties are larger than 
those listed by \citet{La06}.  Our results agree with those of Lada et al. much better 
 if we assume that the background uncertainty (from nebulousity) does not contribute to photometric uncertainty.
In spite of some disagreements with \citet{La06}'s photometry, results of the analysis presented in the proceeding sections 
are not qualitatively affected by which photometric catalog is used for previously detected stars.
\section{Results from New Data}
We now use new data to expand on Lada et al's investigations of 
the IC 348 disk population.
First, we investigate accretion signatures to probe the presence 
or absence of warm circumstellar gas in disks surrounding 
cluster stars.  We then analyze the mid-IR colors of 
stars with new MIPS detections to identify IR excess sources.
We then follow the analysis methods of \citet{Clp09} to model 
the SEDs of new sources to constrain their evolutionary 
states and to refine the \citet{La06} 
analysis of previously detected sources.  

\subsection{Gas Accretion Signatures and Upper Limits}
\begin{figure}
\centering
\plottwo{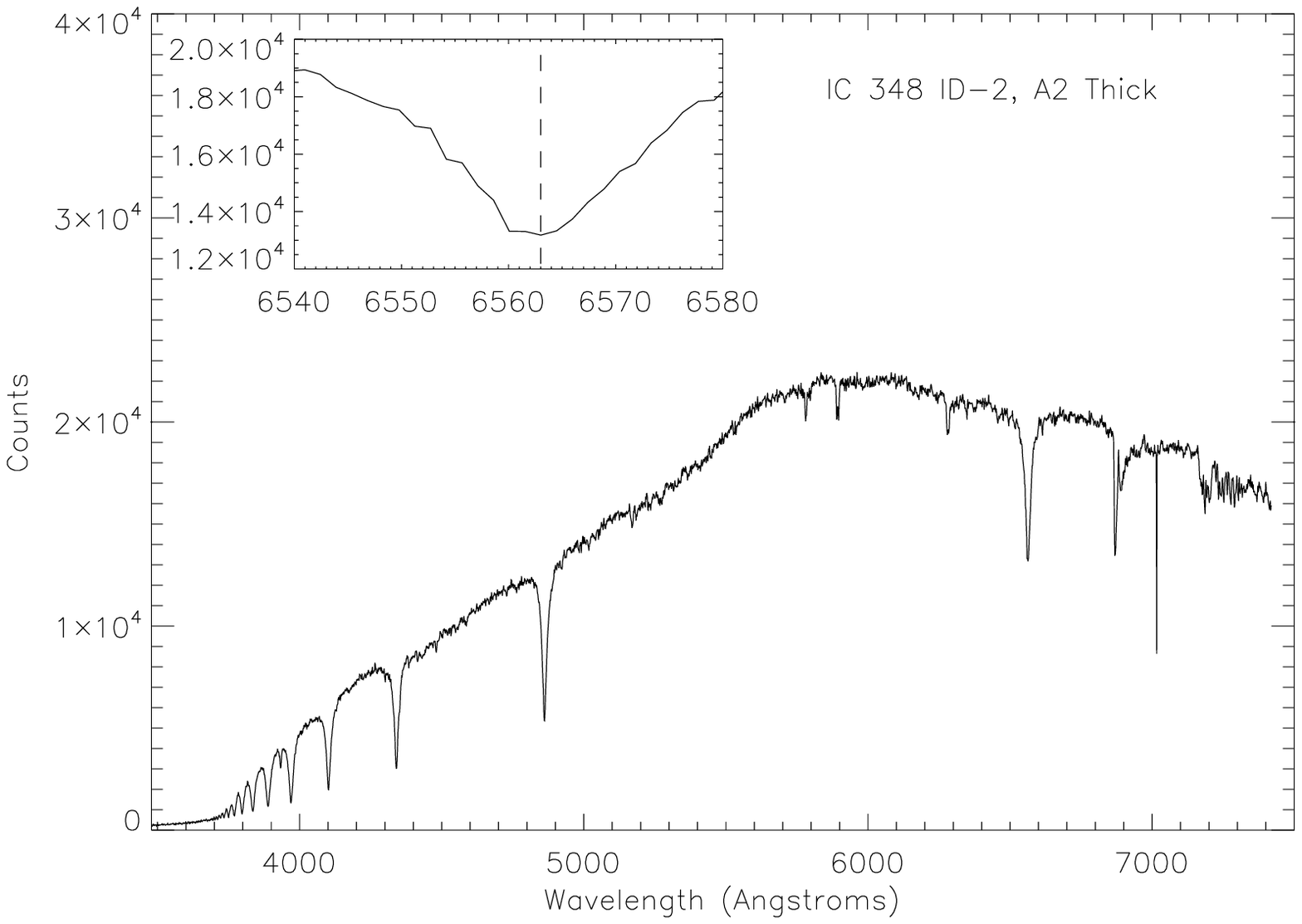}{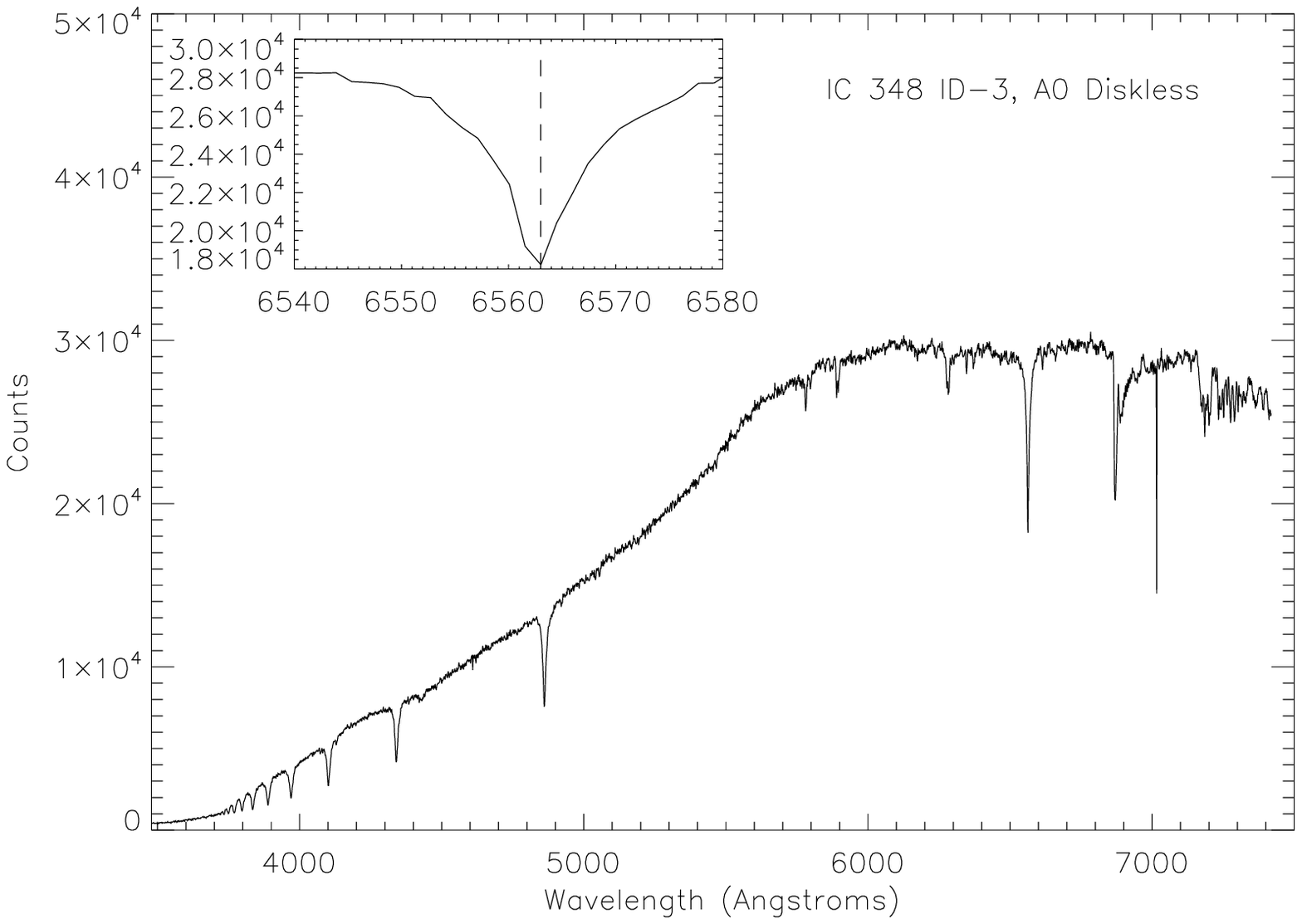}
\plottwo{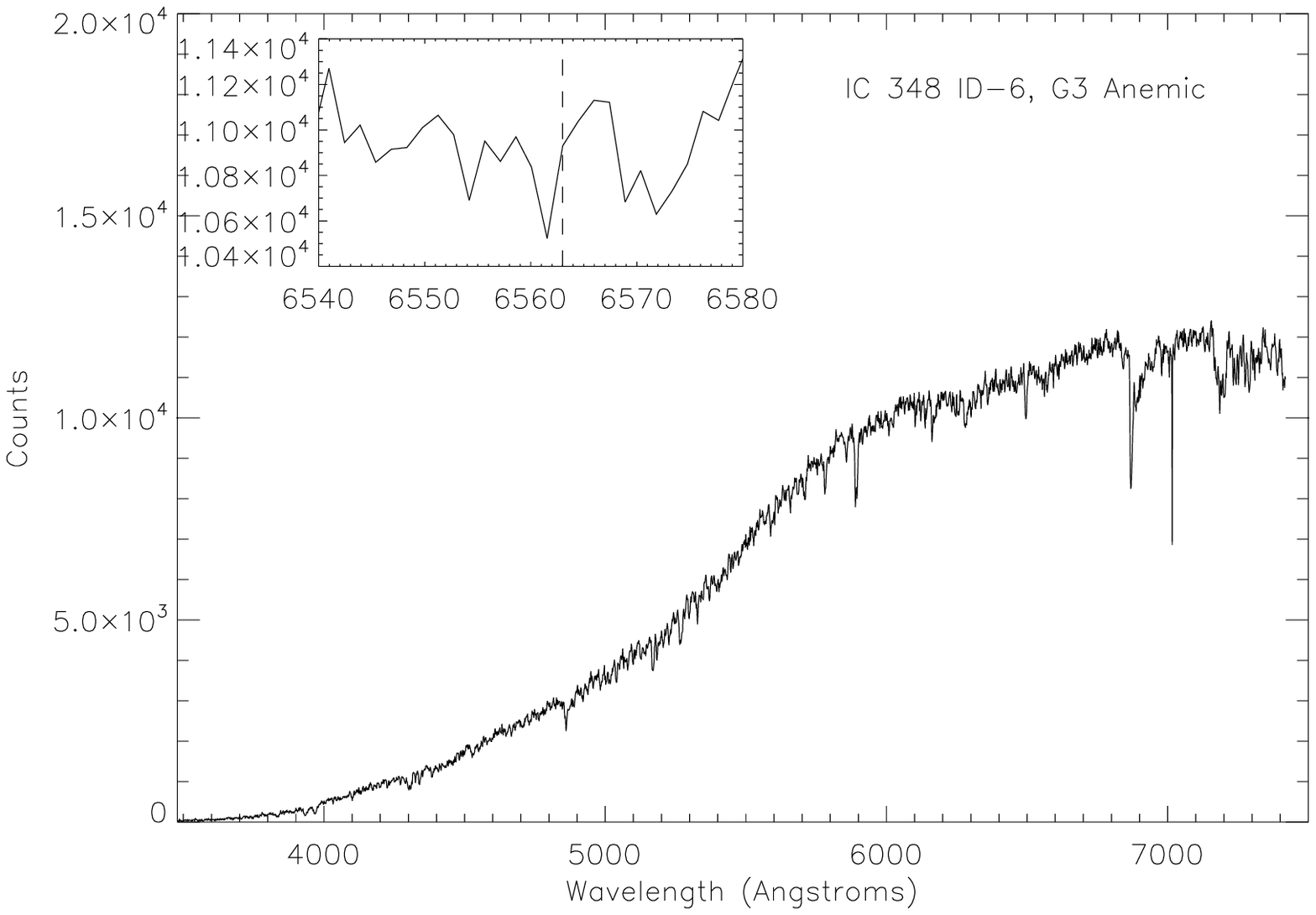}{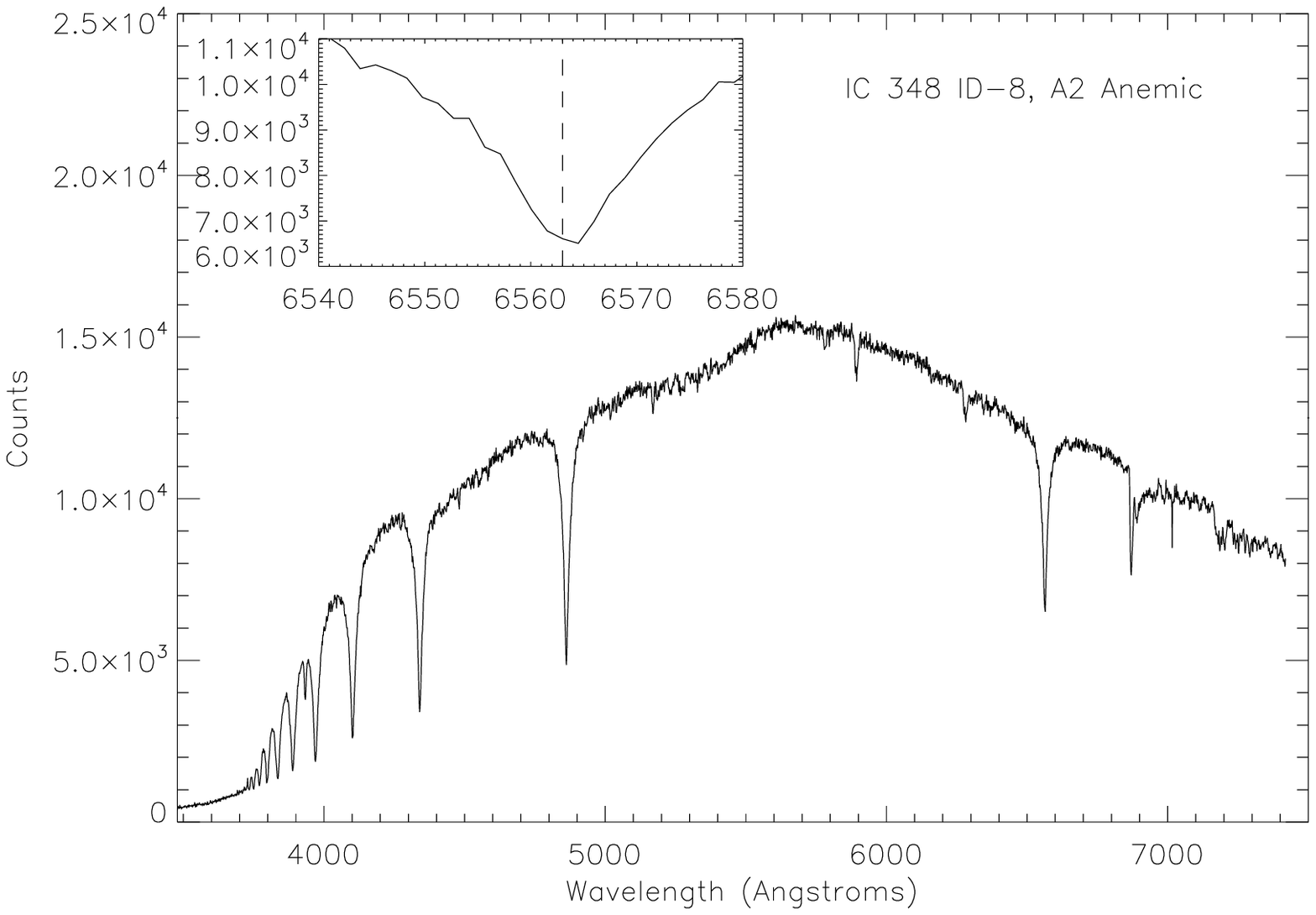}
\plottwo{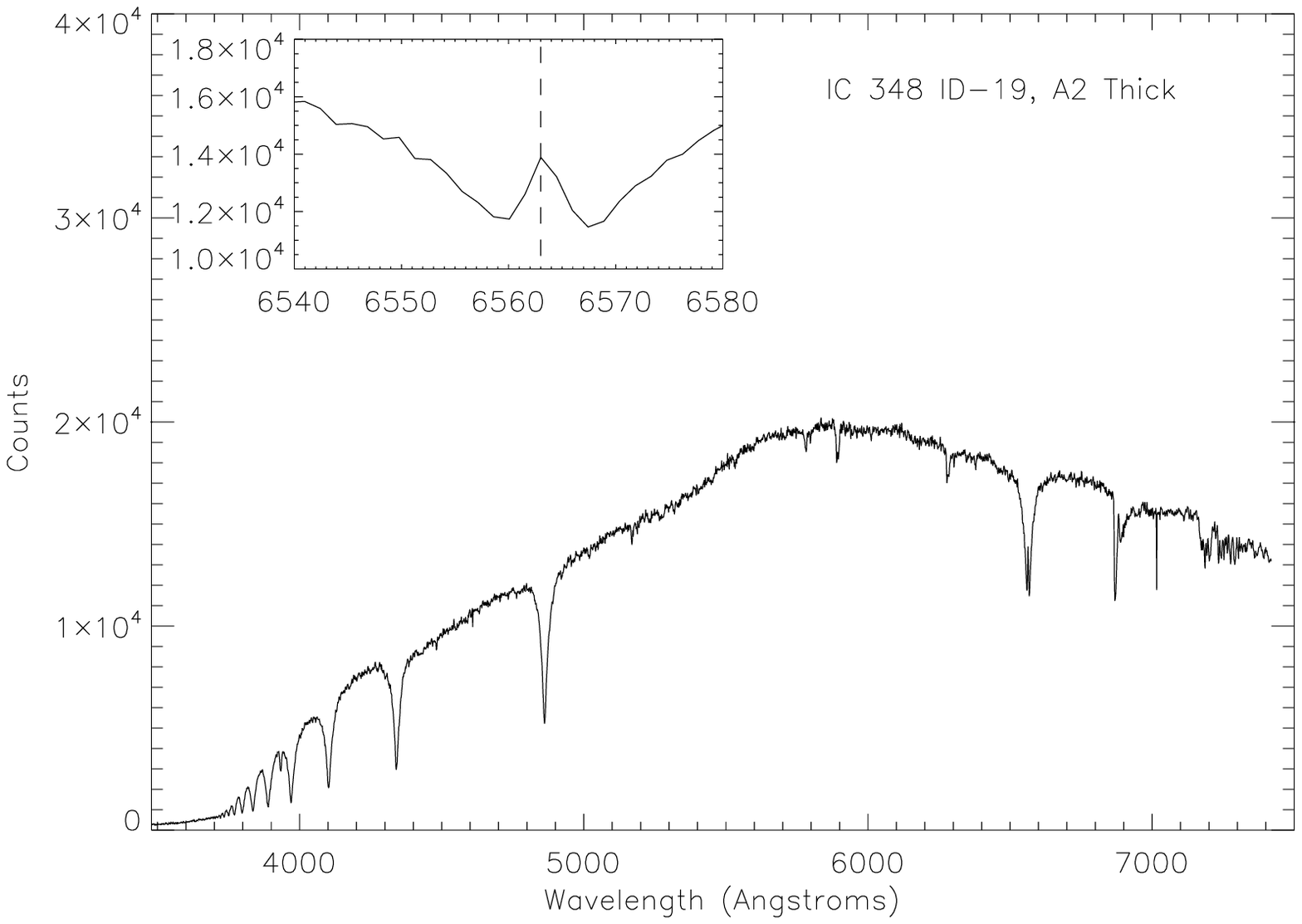}{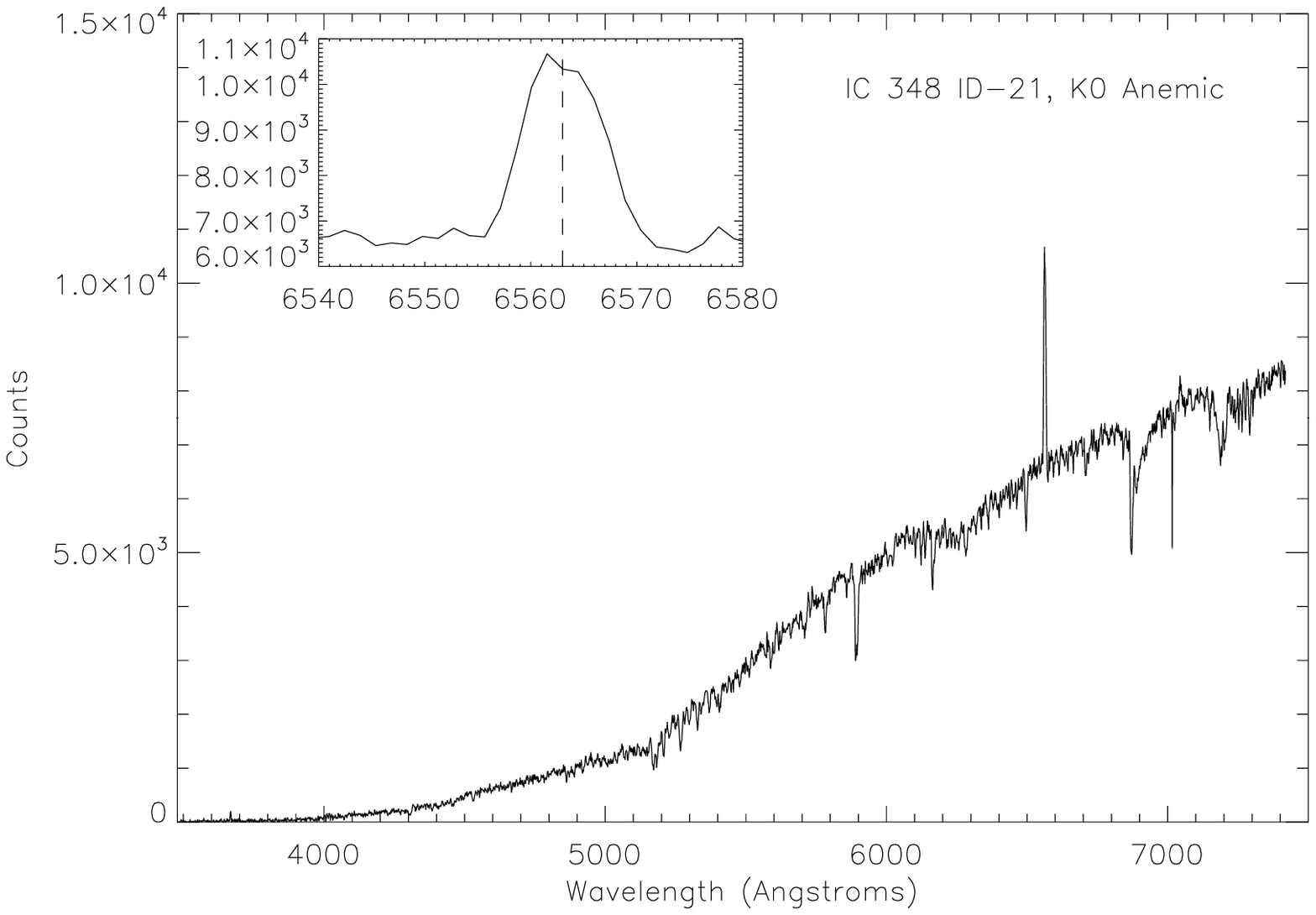}
\epsscale{0.425}
\plotone{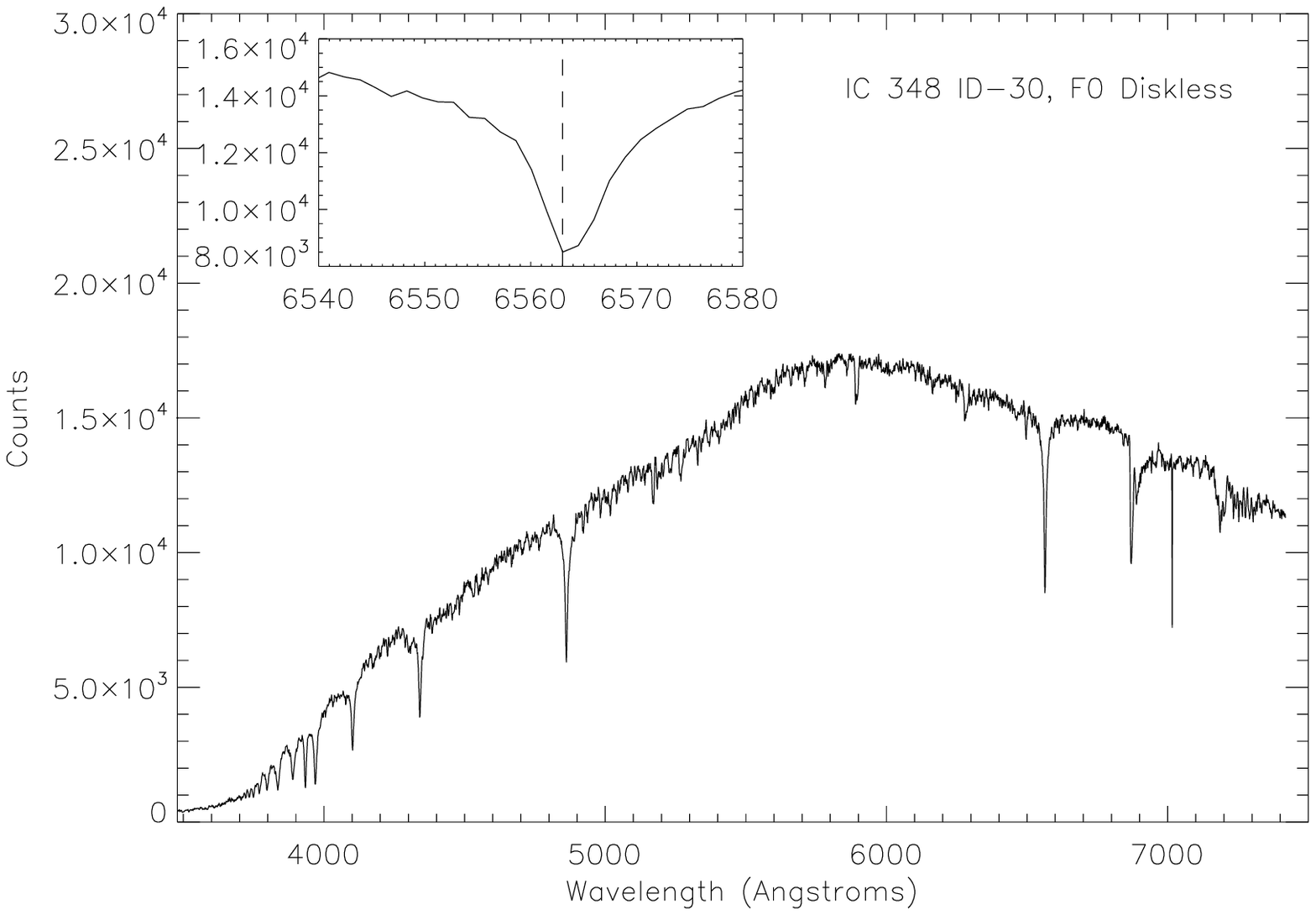}
\caption{FAST spectra of IC 348 sources.  
The spectra have not been smoothed.  Each source is labled by 
its catalog number from \citet{La06}, its spectral type, and 
its IRAC slope.  ID-21 (K0 anemic) 
has a strong, wide H$_{\alpha}$ emission line; ID-19 (A2 thick) also has 
a H$_{\alpha}$ emission core in its absorption line.  ID-6 (G3 anemic)
 has H$_{\alpha}$ nearly filled in, though this is likely due to 
strong chromospheric activity.  The other stars clearly have H$_{\alpha}$ deep 
in emission.}
\label{FASTspec}
\end{figure}

To investigate the gas content of high/intermediate-mass stars,
we consider line fluxes derived from new, higher signal-to-noise
optical spectra.  Our spectra show a variety of H$_{\alpha}$ 
morphologies (Figure \ref{FASTspec}).
IDs 3, 8, and 30 lack 2$\sigma$ fluctuations that could be interpreted 
as H$_{\alpha}$ emission in the central absorption core.
The spectra of ID-2 shows a flattening of the H$_{\alpha}$ absorption 
core, though it lacks unambiguous evidence for emission line reversal.  
Higher-resolution spectroscopy suggests that this source may have 
emission line reversal (S. Dahm, unpublished) as noted in 
\citet{Dahm08}.

ID-6 has H$_{\alpha}$ in absorption but 
a smaller measured equivalent width than the typical width 
for early/mid G stars ($\sim$ 3--4 \AA).  Thus, the H$_{\alpha}$ line may be filled in by emission.  
However, the spectrum shows no clear evidence for an emission line core.
Furthermore, ID-6 is very chromospherically active: 
 it has the largest X-ray luminosity of any Chandra-detected source in IC 348, 
L$_{x}$ $\sim$ 1.96$\times$10$^{31}$ ergs s$^{-1}$ 
\citep{Pz01, Pz02}.  Such activity should produce weak H$_{\alpha}$ emission and result in 
a weakened absorption line.  

ID-21 shows H$_{\alpha}$ in emission with EW(H$_{\alpha}$) $\sim$ 5\AA\, consistent with previous results.
Surprisingly, ID-19, an A2 star not previously known to be accreting,
 also shows evidence for H$_{\alpha}$ emission consistent with accretion.  Even though 
the line is in absorption, a central emission core is clearly visible.  

For the sources lacking clear evidence for accretion, we can 
rule out H$_{\alpha}$ line emission hiding in the noise at a level observed in late-type stars.
  We estimate the upper limit of the H$_{\alpha}$ flux following  
\citet{Al95}.  First, from the spectral type and extinction, we derive the expected dereddened continuum flux in the Cousins 
R band.  We then measure the equivalent width of the continuum noise near the H$_{\alpha}$ line.  
An upper limit to the line flux then results: Flux$_{H_{\alpha}, ul}$ $\approx$ EW(noise)$\times$R$_{flux}$.  

Finally, we derive an upper limit to the accretion rate scaled to that of ID-21 from 
L$_{H_{\alpha}}$ $\propto$ G$\dot{M}$M$_{\star}$/R$_{\star}$.  We determine R$_{\star}$ from the derived values for 
T$_{e}$ and L$_{\star}$ from \citet{Mu07} and use the \citet{Dm97} isochrones to estimate 
stellar masses.  Although the theoretical ages of young stars are uncertain 
\citep[e.g.,][]{Dm97,Si00,Ba98} the \textit{masses} derived from 
L$_{\star}$ and T$_{e}$ using different isochrones are generally consistent.   As shown in Table 1, 
the H$_{\alpha}$ line flux (accretion rate) for sources lacking evidence for gas accretion 
is $\lesssim$ 10\% ($\lesssim$ 0.5--10 \%) of the H$_{\alpha}$ emission (accretion rate) in ID-21.  
These upper limits also rule out accretion at a rate equal to that for ID-19.

As with any pre-main sequence star, it is still \textit{possible} that the stars 
lacking clear accretion signatures are accreting at undetectable 
rates.  Furthermore, it is possible that disks surrounding these stars 
lack warm circumstellar gas accreting onto the star but contain some gas in outer disk regions.  
Their optical spectra and resulting upper limits on accretion 
are more consistent with a lack of accretion and thus a heavily depleted 
or even nonexistent gas reservoir.

\subsection{Mid-to-Far IR Colors of IC 348 Stars with new MIPS Detections}
\subsubsection{MIPS 24 $\mu m$ Colors}
\begin{figure}
\centering
\plottwo{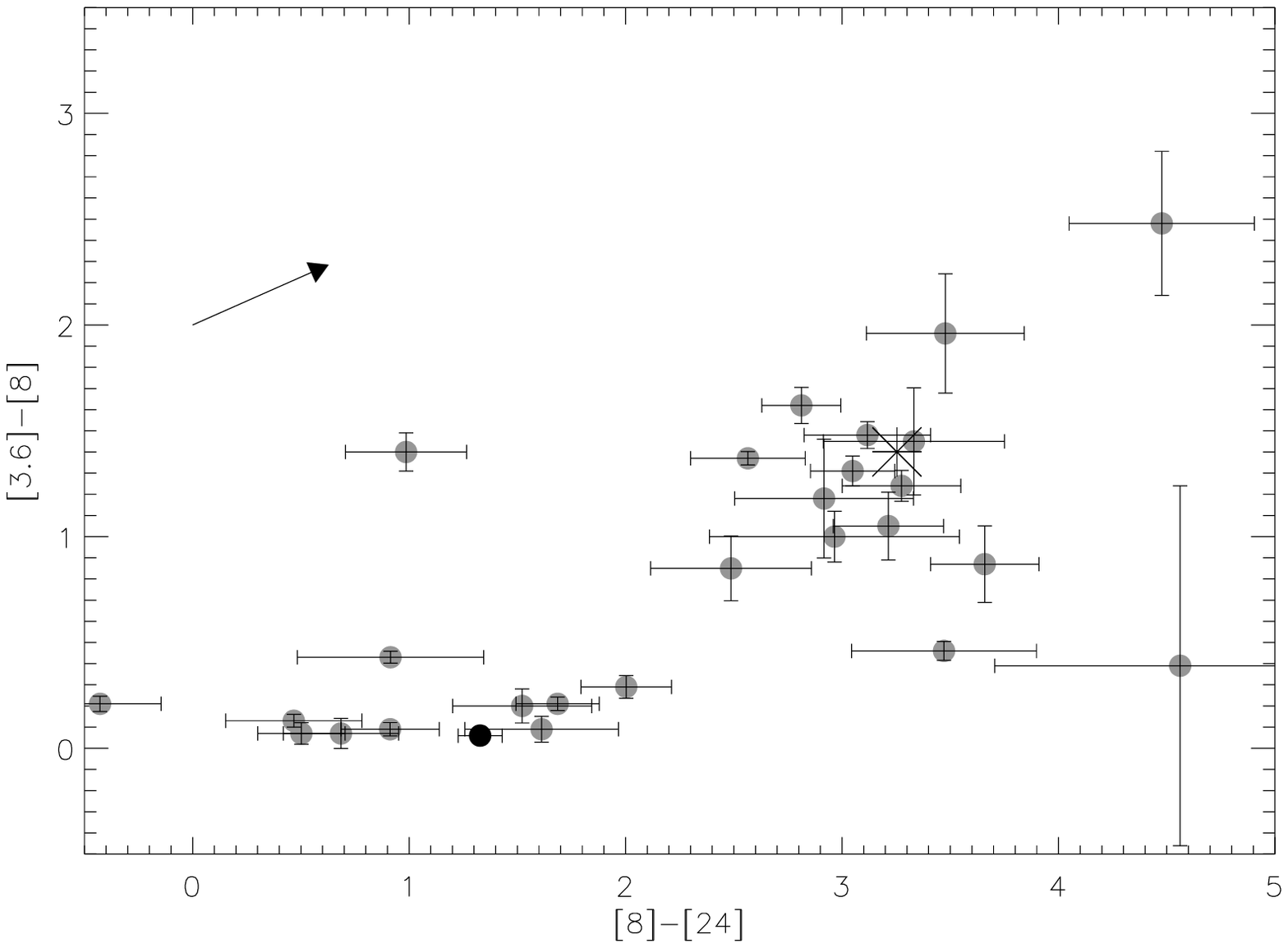}{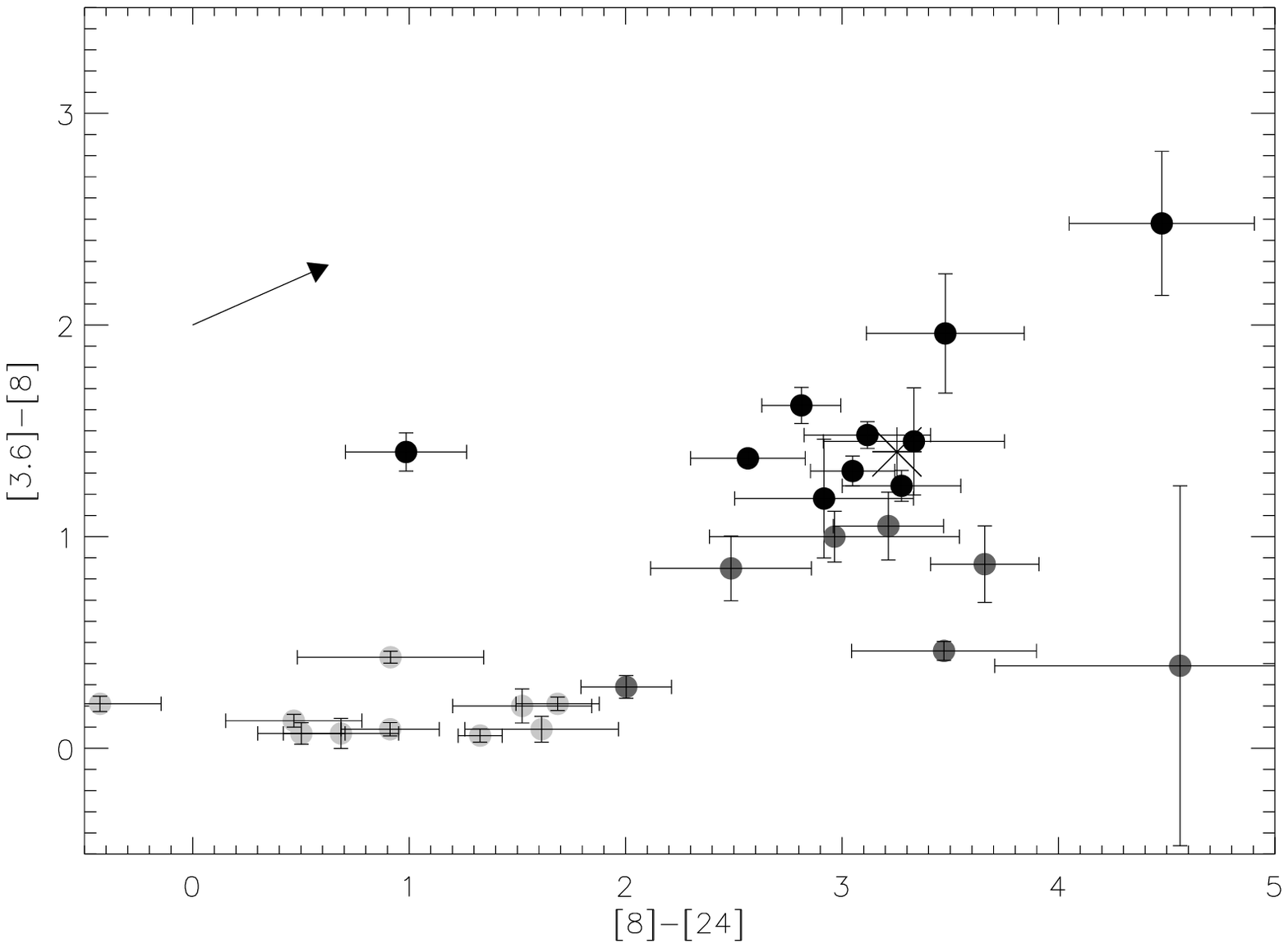}
\caption{[3.6]-[8] vs. [8]-[24] color-color diagram
for IC 348 stars with new MIPS 24 $\mu m$ detections.  Overplotted are the 
1$\sigma$ error bars.  For comparison, we show the colors for the median Taurus 
SED (asterisks).  (Left) Stars with spectral types earlier than K0 are shown 
as dark circles; stars with spectral types later than K0 are shown as 
grey circles.  (Right) Stars identified as having strong ('thick') IRAC excess emission are 
dark circles, stars with weak ('anemic') IRAC excess emission are grey, and stars lacking 
IRAC excess emission ('diskless') are light grey.  In both plots, 
a reddening vector of A$_{v}$=20 is overplotted, where 
we assume extinction laws from \citet{In05} for IRAC and approximate 
the 24 $\mu m$ extinction from \citet{Ma90}.  We note that \citet{Fl07} 
estimate that the extinction law is much flatter from 8 $\mu m$ to 24 $\mu m$ 
than assumed here.}
\label{newmips1}
\end{figure}

Figure \ref{newmips1} shows the [3.6]-[8] vs. [8]-[24] color-color diagram for new MIPS detections.   
Overplotted are the 1$\sigma$ error bars in each filter.  We include a reddening 
vector of length A$_{V}$=20 (the maximum extinction), using the extinction laws 
from \citet{In05} and \citet{Ma90}.
Most newly-detected stars have spectral types later than K0 (left panel; grey dots); only 
one new detection has an earlier spectral type (left panel; black dot).

Cluster stars with new 24 $\mu m$ detections generally have 
excess emission consistent with reprocessed radiation from circumstellar disks.
Most new detections have an observed [3.6]-[24] color redder 
than $\sim$ 1.5--2 magnitudes.  In contrast, [3.6]--[24] $\lesssim$ 0.5 for
 bare stellar photospheres for stars of most spectral types \citep[e.g.][]{Cpk08}.  
Because the maximum extinction for cluster members of 
A$_{V}$ $\sim$ 20 (shown as an arrow) reddens the stars by less than $\sim$ 0.5 magnitudes,
high extinction is unlikely to explain these colors.  Therefore, the
excess emission plausibly arises from circumstellar material.

The distribution of mid-IR colors is consistent with results for previously 
detected cluster stars analyzed in \citet{La06}.  
Most stars identified as 'diskless' (right panel; light grey dots)
 by \citet{La06} either have photospheric MIPS emission or excess emission less
than 2 magnitudes.  
Their MIPS excess emission is less than that from
 stars with strong ('thick') IRAC excess (right panel; black dots) and stars with weak 
('anemic') excess (right panel; dark grey dots).  Both 'anemic' and 'thick' disks have a range of 
[8]-[24] colors from $\sim$ 2 to 4.5 and differ primarily by their levels of 
IRAC emission.

To compare the mid-IR colors of IC 348 sources with fiducial 
values, we overplot colors of the median Taurus SED (asterisk) 
reddened to E(B-V) $\sim$ 1.56, the mean value for IC 348 stars.  To produce the Taurus colors, 
we use fluxes from \citet{Fu06} for the 8 $\mu m$ and 24 $\mu m$ filters and 
approximate the 3.6 $\mu m$ flux from \citet{Dal99}.  The Taurus colors 
lie within the distribution of colors for 'thick' sources in IC 348. 

While many newly-detected stars have levels of excess emission 
typical of primordial disks, the full range of
IRAC vs. MIPS colors hints at a diversity in disk properties.  
Two stars have nearly zero [3.6]-[8] color and [8]-[24] $\gtrsim$ 3.  
These colors are consistent with stars whose disks have inner holes 
cleared of dust and are optically-thick 
 at larger stellocentric distances.  A second group of stars also 
 have zero IRAC color but may have weak 24 $\mu m$ excess emission ([8]-[24] 
$\sim$ 0.5--2).  These colors are consistent with the colors of either 
 debris disks or remnant protoplanetary disks that are actively disappearing.
\subsubsection{MIPS 70 $\mu m$ Colors}
\begin{figure}
\centering
\plotone{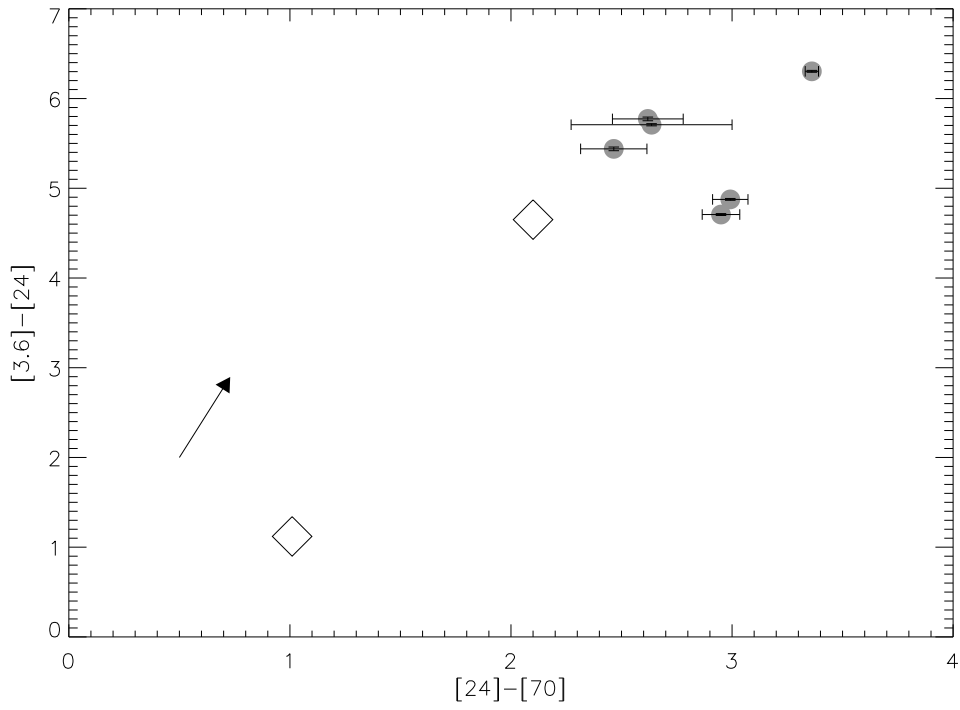}
\caption{[3.6]-[24]  vs. [24]-[70] color-color diagram for IC 348 stars 
detected at 70 $\mu m$ (dots).  Overplotted are the positions of two 
$\eta$ Cha stars (diamonds), ECH-2 (lower left) and ECH-15 (upper right).}
\label{newmips2}
\end{figure}

Figure \ref{newmips2} shows the distribution of  
[24]--[70] colors for stars detected at 70 $\mu m$.  
All of these stars have spectral types later than K0 and 
have substantial emission at both 24 $\mu m$ and 70 $\mu m$.
  The right panel of Figure \ref{newmips2} 
shows that the stars with 70 $\mu m$ detections include both
'anemic' and 'thick' sources, indicating that at least some stars with dust-poor 
inner regions may have very luminous outer disks.  While the distribution of 
MIPS colors appear correlated (stronger 24 $\mu m$ emission $\rightarrow$ stronger 70 $\mu m$ emission), 
the small number of stars with 70 $\mu m$ detections precludes any robust 
statistical analysis to determine if this trend is real.

To provide a context for our 70 $\mu m$ detected stars, we compare their colors 
to those for two stars in the $\eta$ Cha Association using data from 
\citet{Gau08}.  Even though $\eta$ Cha is substantially older than IC 348 ($\sim$ 8 Myr), 
its proximity to the Sun and lack of high IR background allow for many 70 $\mu m$ detections; 
the median Taurus SED currently does not include a 70 $\mu m$ data point.
Overplotted as diamonds are colors for ECH-15 (top center), an accreting late-type 
star with a primordial disk, and ECH-2 (lower left), which has weak, optically-thin 
 emission consistent with a debris disk.  The MIPS colors of IC 348 stars are slightly redder 
than ECH-15 at both 24 $\mu m$ and 70 $\mu m$ and far redder than ECH-2.  

\subsection{SEDs of new MIPS Detections}
\subsubsection{Constraints on Disk Properties of Sources with MIPS-70 $\mu m$ Detections}
The addition of the MIPS-70 $\mu m$ data better determines plausible circumstellar 
disk properties from models with nearly identical fluxes through 24 $\mu m$ but 
very different far-IR fluxes.  To better constrain the disk morphology of 
IC 348 sources, we compare the SEDs of MIPS-70 $\mu m$-detected 
sources with the grid of 3D radiative transfer models from \citet{Rob06,Rob07}. 
In fitting the SEDs, we assume 
a 10\% uncertainty in the IRAC and MIPS-24 $\mu m$ fluxes and a 20\% uncertainty 
in the MIPS-70 $\mu m$ fluxes to account for photometric error, zero-point uncertainty, and 
variability.  We allow for a range of distances ($\sim$ 300--350 pc) and a range of 
optical extinction, typically comparable to the standard deviation in the extinction as 
determined by \citet{Mu07}.  The online fitting tool then 
returns the SEDs of the best-matching models and lists the disk characteristics (inner and outer 
disk radius, accretion rate, disk mass, disk flaring index, etc.) of the models.

\begin{figure}
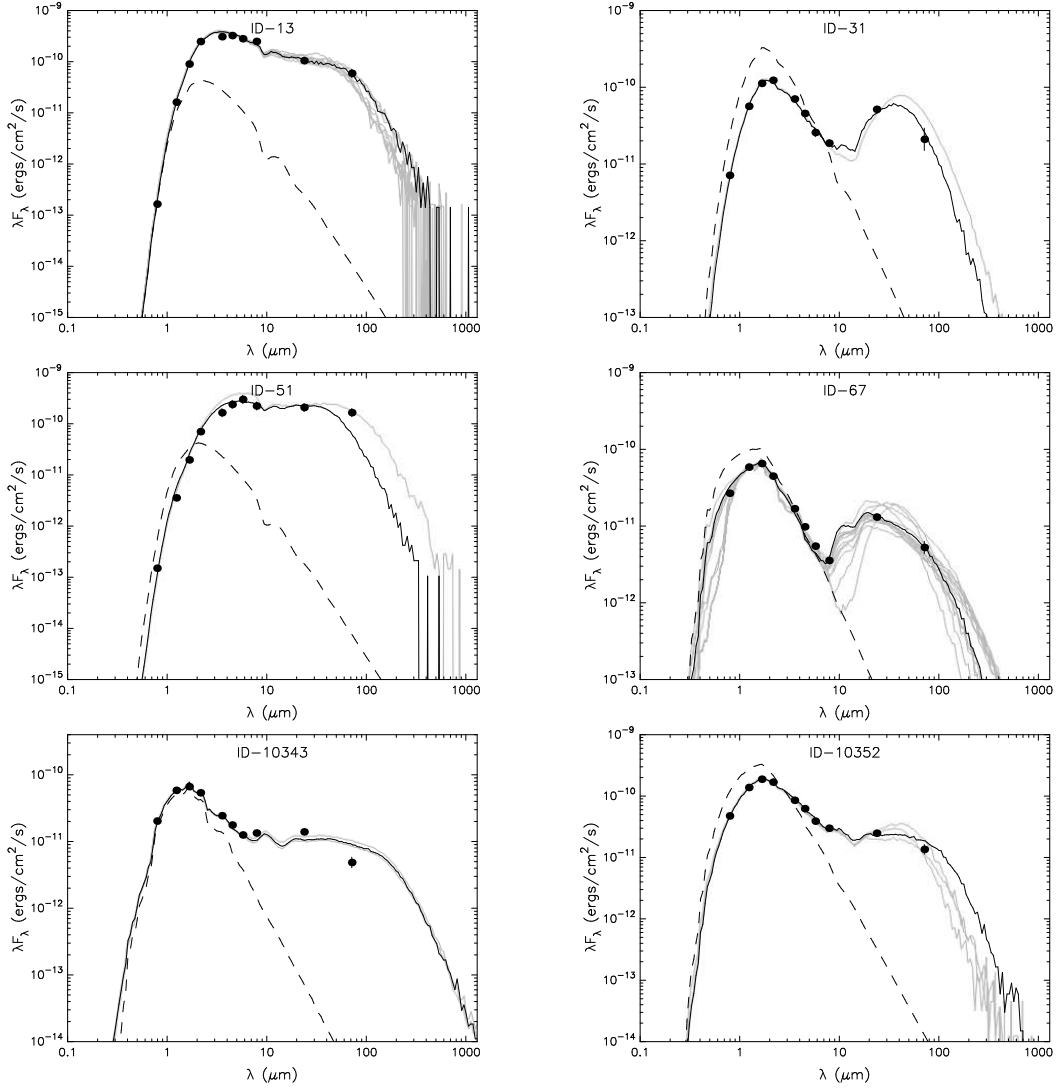

\centering
\plottwo{id13.eps}{id31.eps}
\plottwo{id51.eps}{id67.eps}
\plottwo{id10343.eps}{id10352.eps}
\caption{SED fits to IC 348 sources detected at 70 $\mu m$ using the \citet{Rob06} models.  The dotted line corresponds to the unreddened SED of 
the best-fit stellar photosphere.  The solid black line corresponds to the best-fit SED \citet{Rob06} model.  The light grey lines 
correspond to other \citet{Rob06} models with a $\chi^{2}$-$\chi^{2}_{best}$ $<$ 10.}
\label{mips70sed}
\end{figure}

Figure \ref{mips70sed} shows the fitted spectral energy distributions.  
In each panel, we overplot the SED of the best-fit bare stellar photosphere (dotted line), 
the best-fit SED model (solid black line), and other SED models with $\chi^{2}$-$\chi^{2}_{best}$ $<$ 10.  
The $\chi^{2}_{best}$ for the sources range from $\chi^{2}$=9.4 (ID 103542) to $\chi^{2}$ = 136 for 
ID 67.  While the $\chi^{2}$ value for the latter source is large, much of the power 
in the $\chi^{2}$ values comes from measurements at shorter 
wavelengths (e.g. J band) where the photometric uncertainty is very small.

One source, ID-51, has a nearly flat SED ($\lambda$F$\lambda$) from 2 $\mu m$ through 
70 $\mu m$ and is thus consistent with "flat-spectrum" protostars like those found in Taurus-Auriga \citep{Kh95}.  
Such sources may be in the process of dispersing their protostellar envelopes \citep{Calvet1994}.  
Parameters for the best-fit model include a flaring index of $\beta$ =1.1 ($\beta$ = H/r, where H is the 
disk scale height) and a disk mass of $\sim$ 3 $\times$ 10$^{-2}$ M$_{\odot}$.  
The SED of ID-51 rises from I band through 5.8 $\mu m$, which is consistent with a 360 K protostar 
as suggested by \citet{Mu07}.

Half of the 70 $\mu m$ sample -- IDs 13, 10343, and 10352 -- have SEDs consistent with pre-main sequence stars 
surrounded by primordial disks.  These sources have clear IR excesses at near-IR wavelengths and 
SEDs whose flux declines with wavelength from peak values at $\sim$ 1--2 $\mu m$.  Assuming 
a standard grain size distribution (n(a) $\propto$ a$^{-3.5}$) and a solar gas-to-dust ratio, 
these sources have disks with masses $\sim$ 10$^{-3}$--10$^{-2}$ M$_{\odot}$, comparable to 
the Minimum Mass Solar Nebula.  The disk flaring indices 
range from $\beta$ = 1.01 (for ID 13) to 1.17 (ID 10343).  While \citet{Mu07} identify ID-13 as a 
protostar, its IRAC slope is slightly steeper than their criteria for either protostars or flat spectrum sources.  
Our analysis coupled with its published spectral type (M0.5; T$_{e}$ $\sim$ 3800 K) indicates that it may be 
at the boundary between a flat spectrum source and a pre-main sequence star + primordial disk.

Two sources, IDs 31 and 67, show evidence for a strong flux deficit from 3.6 $\mu m$ through 8 $\mu m$ 
but substantial emission at both 24 $\mu m$ and 70 $\mu m$.  Disk parameters from the best-fitting 
models have a wide range of disk masses (10$^{-7}$ -- 10$^{-2}$ M$_{\odot}$) but typically have 
large inner disk boundaries ($\sim$ 1--50 AU).  While a depleted mass in warm dust and inner holes are 
consistent with disks in an intermediate stage between primordial and debris --'evolved' primordial disks or, 
as typically described, 'transitional' disks-- there are alternate possibilities.  Specifically, given that 
IC 348 contains many protostars, the strong 24 and 70 $\mu m$ emission could be due to 
a protostar lying in close proximity to IDs 31 and 67.  Because such sources would be unresolved with 
MIPS and too faint to detect from optical/near-to-mid IR imaging, their presence cannot be ruled out.
However, these two sources are located in regions well away from most IC 348 protostars identified by 
\citet{Mu07}. 
\subsubsection{Evolutionary States of Disks with New MIPS-24 $\mu m$ Detections}
In this section, we investigate the evolutionary states of disks surrounding sources 
with new MIPS-24 $\mu m$ detections.
To provide a simple but effective probe of the disks' evolutionary states, we 
compare their SEDs to the median Taurus SED and upper/lower quartiles of Taurus sources 
and debris disk SEDs.  We use the Taurus SED and upper/lower quartiles from \citet{Fu06};  
for the debris disk SED, we adopt the synthetic SED of a warm, terrestrial planet-forming debris 
disk around a solar-mass star from \citet{Kb04}.
Evolutionary states derived from these simple comparisons show good agreement with 
those obtained from more detailed SED modeling \citep[see][]{Clp09}.

Following \citet{Clp09}, we classify sources as having primordial disks if their 
3.6 $\mu m$ to 24 $\mu m$ emission is comparable to or greater than the lower quartile of the median Taurus 
SED.  Sources lacking clear excess emission in the IRAC bands but emission comparable to 
Taurus at 24 $\mu m$ are identified as disks with inner holes.  Sources with IRAC and MIPS 
excess emission significantly weaker than the lower quartile Taurus SED but stronger than the debris disk SED 
are labeled 'homologously depleted disks' \citep{Clp09}.  
Here, we define "significantly weaker" as less than half of the lower-quartile Taurus SED flux at
24 microns.  This stricter criterion is justified because, unlike in \citet{Clp09},
most (68\%) of the new detections are stars later than M3, which are intrinsically redder and fainter than
the K7--M2 stars used to derive the median Taurus SED, and thus should have systematically weaker primordial disk emission.
Finally, sources with IR excess emission comparable to the debris disk SED are identified as debris disk candidates.  

\begin{figure}
\centering
\epsscale{0.43}
\plotone{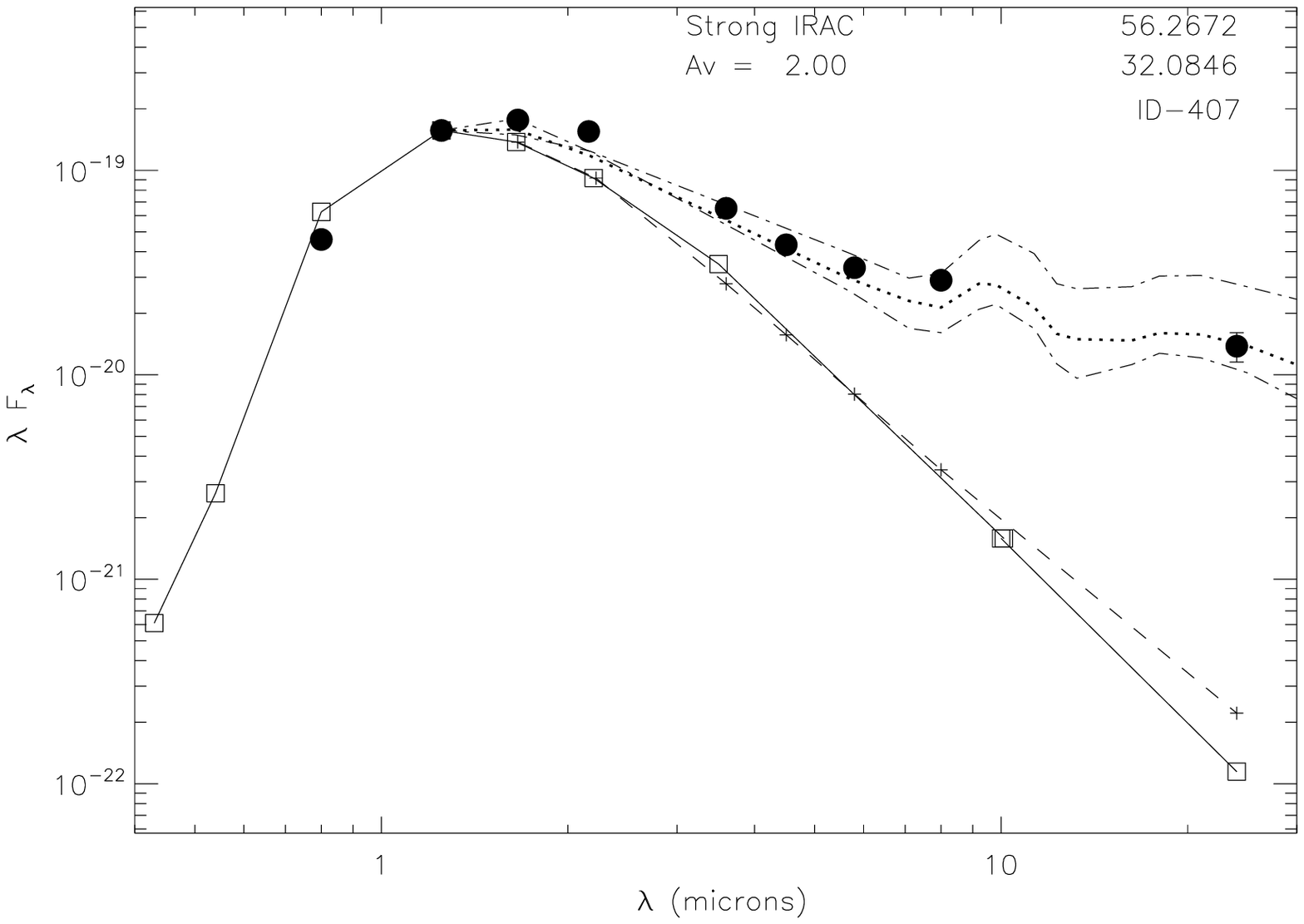}\\
\epsscale{0.85}
\plottwo{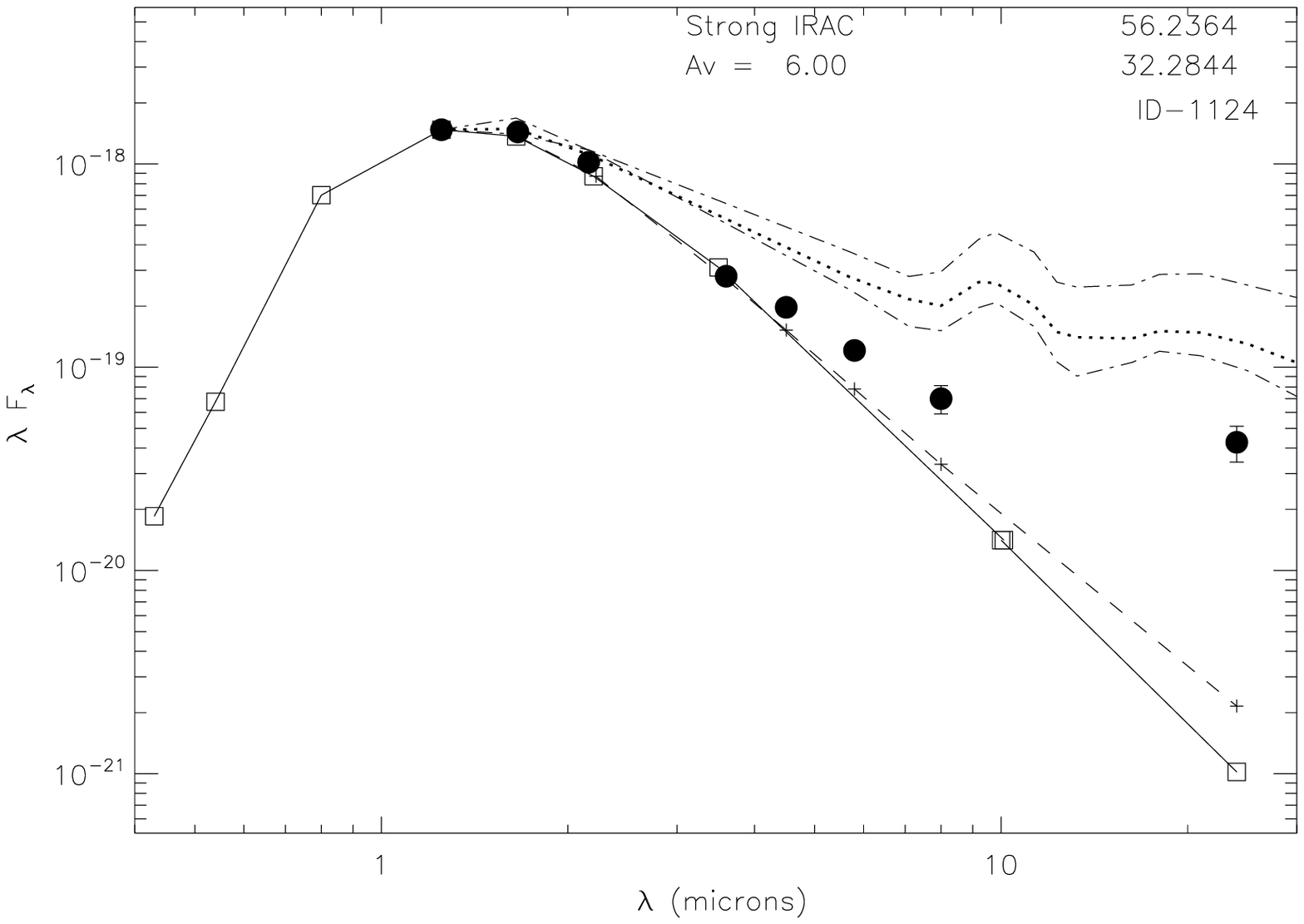}{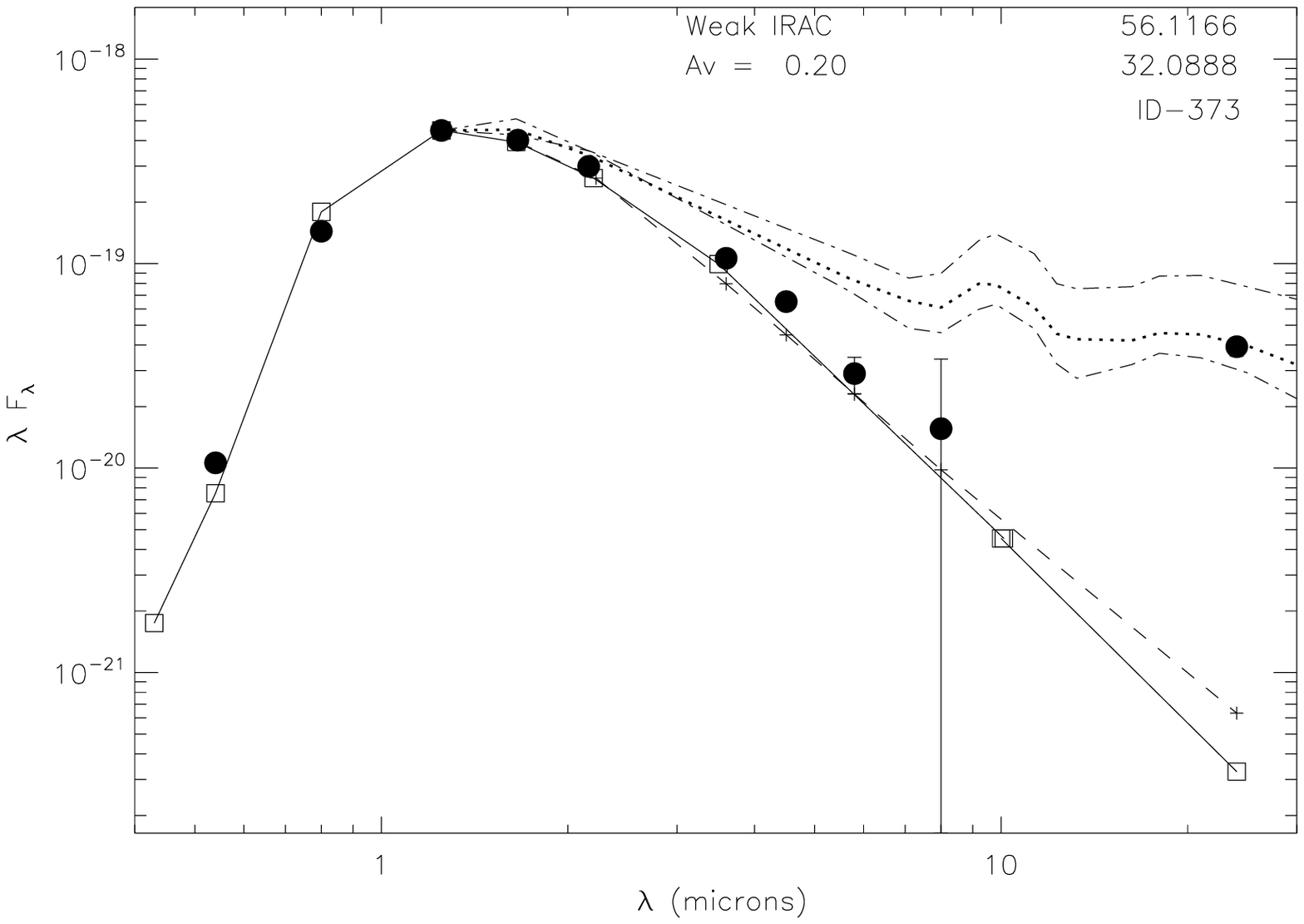}\\
\plottwo{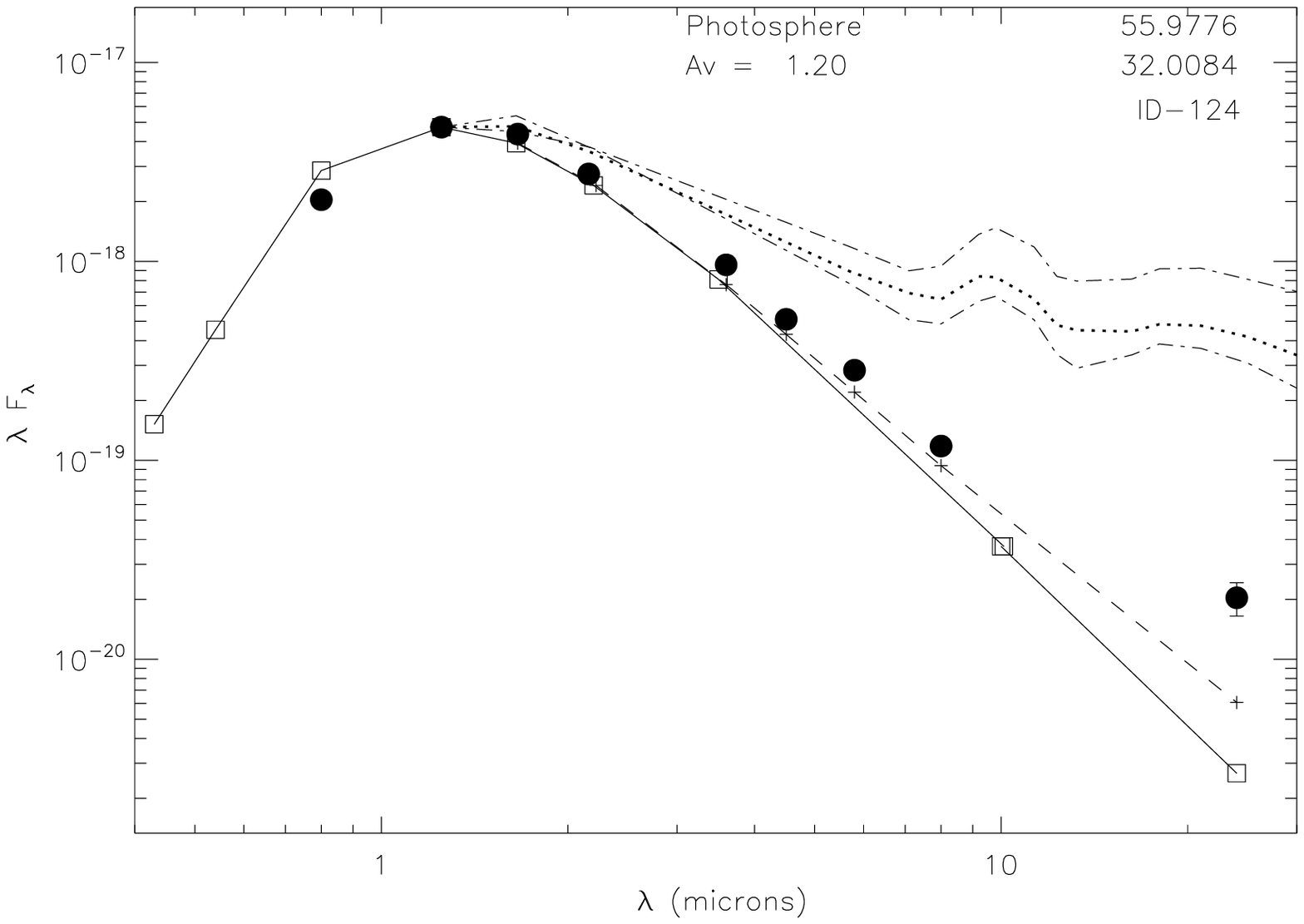}{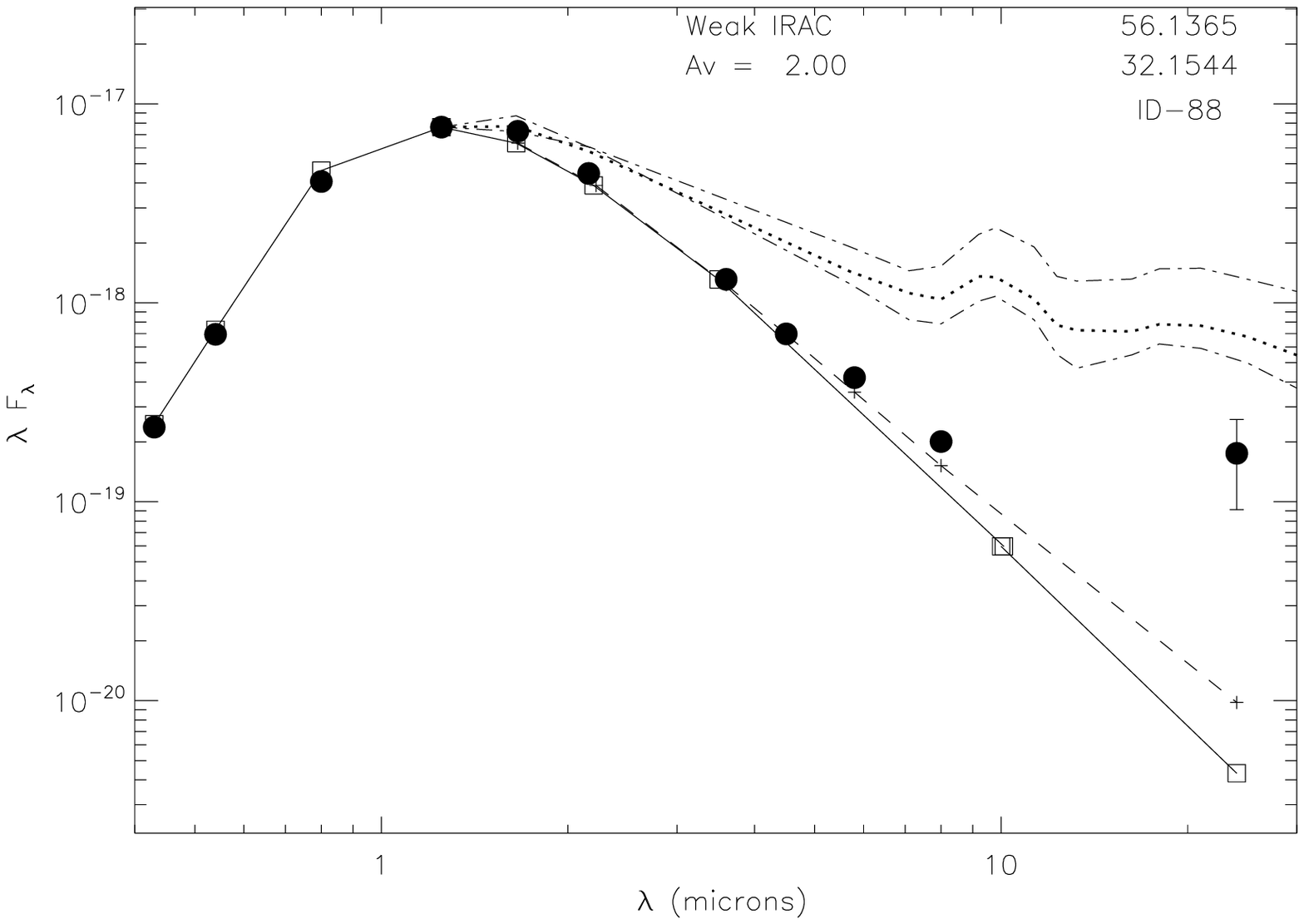}\\
\centering
\epsscale{0.43}
\plotone{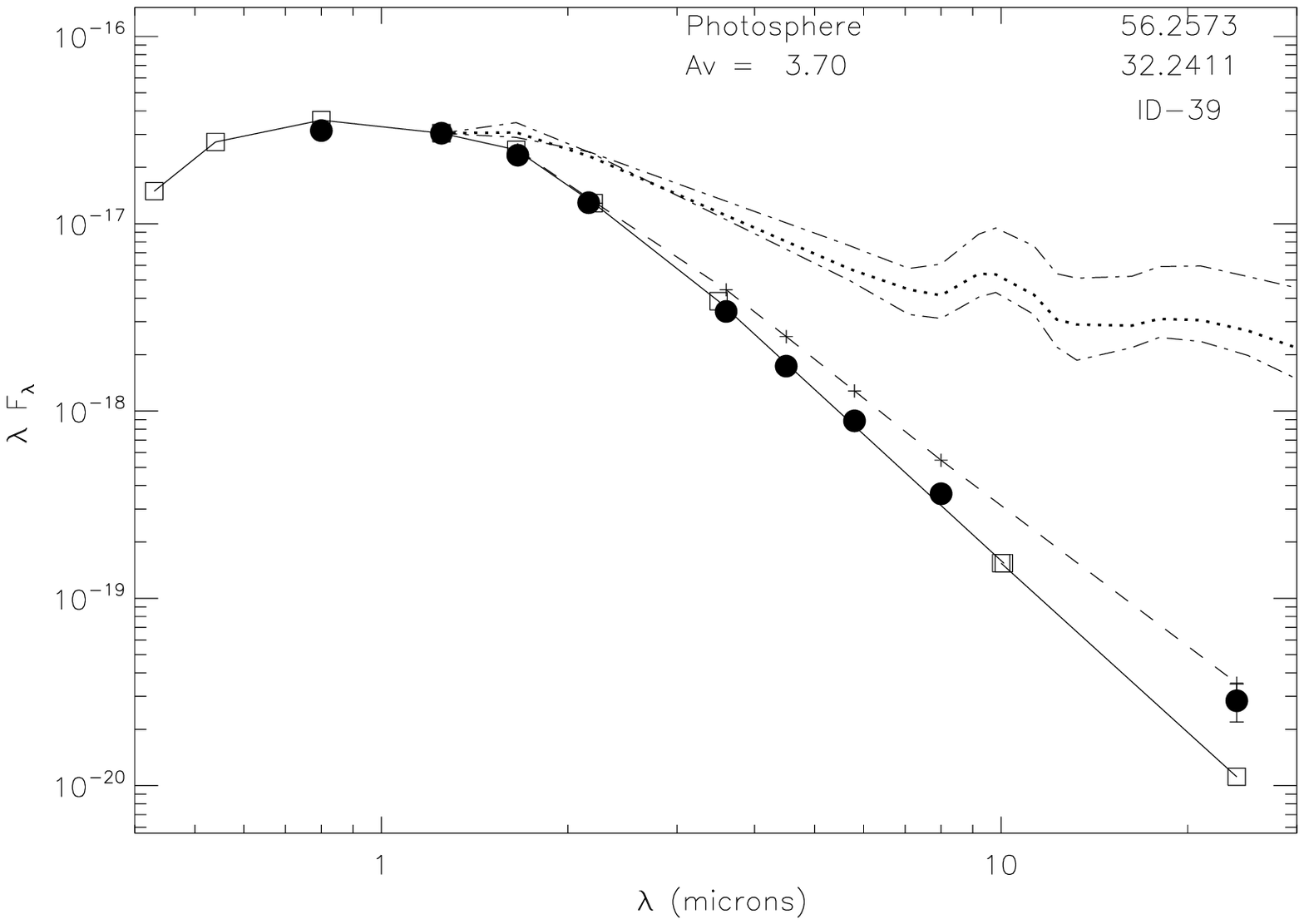}\\
\epsscale{0.85}
\caption{Representative SEDs of IC 348 stars with new MIPS 24 $\mu m$ detections.  Data points are black dots with 
photometric errors overplotted.  The solid line corresponds to a stellar photosphere appropriate for the star's spectral type, 
the dashed line corresponds to the debris disk model from \citet{Kb04}, the dotted line corresponds to the SED of the 
median Taurus SED, and the dot-dashed lines correspond to the upper and lower quartiles for the median Taurus SED.   Each source 
is identified by its IRAC slope (Strong IRAC= 'thick', Weak IRAC= 'anemic', Photosphere='diskless), its optical extinction, 
its J2000 position in decimal degrees, and its catalog ID number.}
\label{SEDex}
\end{figure}

Figure \ref{SEDex} shows SEDs of sources representative of each disk class.  
All source SEDs are labeled by their ID number, J2000 coordinates, and extinction.  
We also include the IRAC slope (stron/weak IRAC, photosphere) following the 
formalism of \citet{La06} as modified by \citet{Clp09}.
The figures are arranged to illustrate the evolution of primordial disks (top panel) 
into debris disks (bottom panel) along two separate pathways:
 homologously depleted disks (second and third row panels, lefthand side) and disks 
with inner holes (second and third row panels, righthand side) \citep[see][]{La06, Clp09}.
The Appendix contains the atlas of all SEDs; Table \ref{tabevo24} lists results for all stars.
Below we describe the frequency and characteristics of each disk class.
Because of a lack of short wavelength data, we did not determine the evolutionary
 states for IDs 746 and 2096.

\begin{enumerate}
\item \textbf{Primordial Disks}-- Nine sources have SEDs indicative of 
primordial disks.  The levels of disk excess from most of these sources 
are more comparable to the lower quartile of the median Taurus SED than the 
 median Taurus SED.  However, most sources in IC 348 have spectral types later than M4; it is 
not clear whether their slightly weaker emission is an evolutionary effect or simply a temperature (contrast) 
effect \citep{Er09} because the median Taurus SED was derived from hotter (K7--M2) stars.  
Some of these sources have mid-IR emission comparable to
 the median Taurus SED, a characteristic consistent with 
a relatively unevolved primordial disk.  

\item \textbf{Homologously Depleted Disks}-- Seven sources have disks that produce 
a clear IR excess from $\sim$ 4.5 $\mu m$ to 24 $\mu m$ but at a level significantly 
weaker than Taurus: a 'homologously depleted' disk \citep{Clp09}.  Because both the IRAC and MIPS emission is 
depleted by up to a factor of 10 relative to the Taurus SEDs, it is unlikely that this weaker 
emission is simply due to flatter disk geometries.  The weaker emission is likely due in part to 
significantly lower warm dust masses which can emerge from a combination of grain growth and accretion onto the 
star \citep[see][]{Clp09}.  The SEDs exhibit a power-law decline 
with $\lambda$F$_{\lambda}$ $\propto$ $\lambda^{-q}$, where q $\sim$ 1.5--2.5.

\item \textbf{Disks with Inner Holes} -- Two sources show evidence for a disk with little 
or no emission in the IRAC bands but substantial MIPS emission, up to a level comparable
 to the median Taurus SED.  These properties are characteristic of disks with inner regions 
mostly cleared of material.  Because these disks represent only one of the two pathways 
from primordial disks to debris disks, and thus comprise one of two types of transitional disks,
 \citep{Clp09, La06}, we do not equate them with transitional disks but call them 'disks with inner holes' instead.   

\item \textbf{Debris Disks Candidates} -- Five sources have MIPS emission  
comparable to predicted fluxes from debris disks.  This emission is a factor of $\approx$ 100
  weaker than primordial disks and is also weaker than emission from homologously 
depleted disks, and disk with inner holes.  We label these sources 'debris disk candidates'.  

Debris disks are usually defined as disks with second-generation 
dust, a property that correlates with broadband SEDs but cannot be directly inferred from SEDs.
Disks around late type, low-mass stars have a much lower contrast with the photosphere 
than do disks around high-mass stars; thus, there is less difference in broadband fluxes between debris 
disks and disks in earlier evolutionary stages.  Some of these disks could be very evolved (homologously depleted, inner hole) 
disks with first generation dust, not second generation dust.  Others have larger photometric errors at 24 $\mu m$ 
and could be bare photospheres instead.  Thus, our labeling of these sources 
as 'debris disk candidates' is tentative and requires more analysis for confirmation.
We consider these sources further in \S 3.4.
\item \textbf{Stars Without Disks} --- Two sources have SEDs that are consistent with 
bare stellar photospheres.  ID-47, a K4 star, is the source that most obviously lacks 
IR excess emission.  It lacks any IRAC-excess emission, and its
 24 $\mu m$ flux is slightly \textit{bluer} than a stellar photosphere.
  However, this source has a marginal, 5 $\sigma$ detection, making its flux 
consistent with a stellar photosphere to within $\sim$ 2 $\sigma$.   ID-49 also has a very marginal 
IR excess at 24 $\mu m$ (K$_{s}$-[24] $\sim$ 0.5) that is within 2 $\sigma$ of photospheric colors.  
Thus, there is no clear evidence for a disk surrounding this star.
\end{enumerate}

\subsubsection{Mid-IR Colors of Sources with Disks in Various Evolutionary States}
\begin{figure}
\centering
\epsscale{0.75}
\plotone{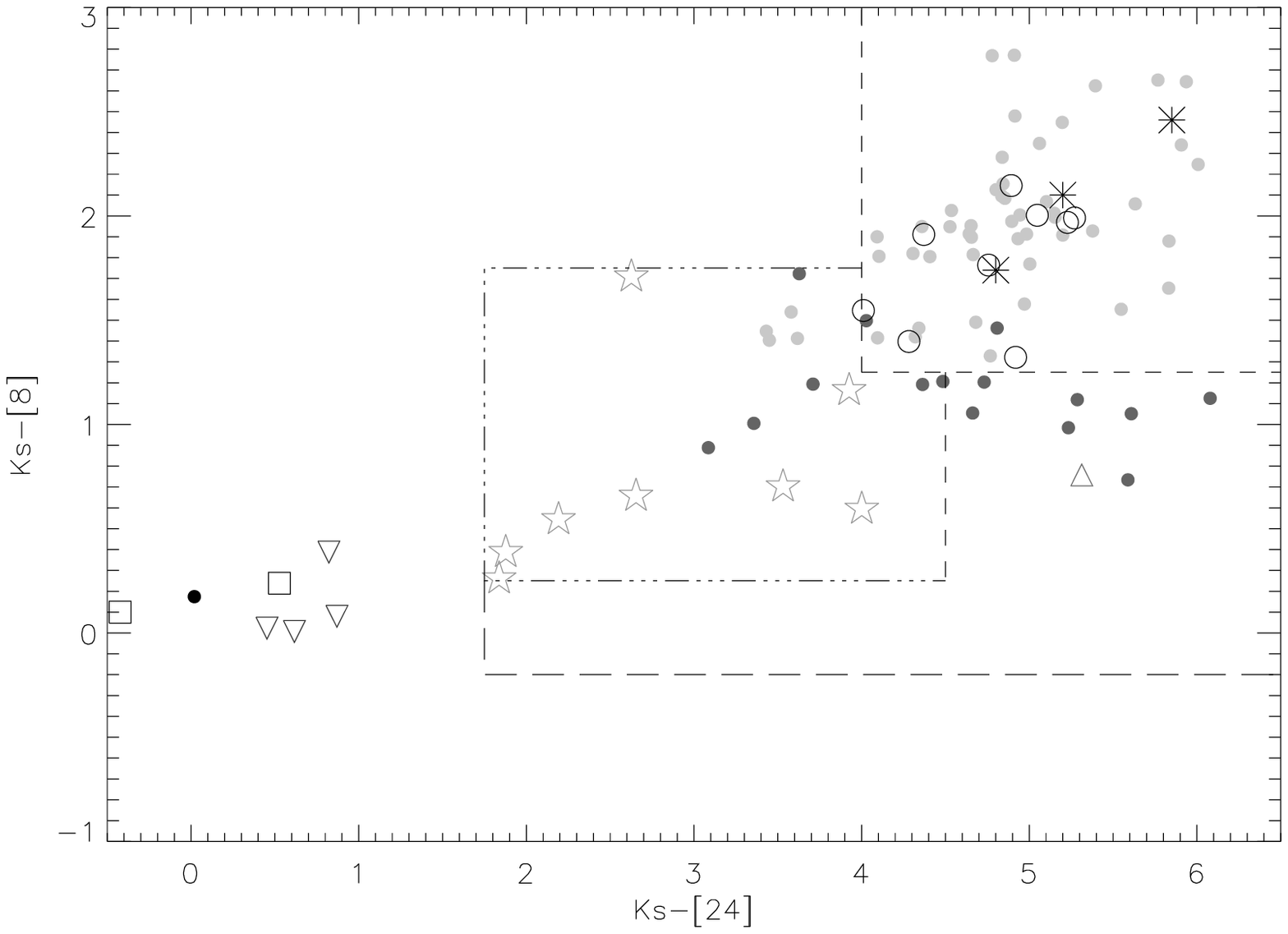}
\plotone{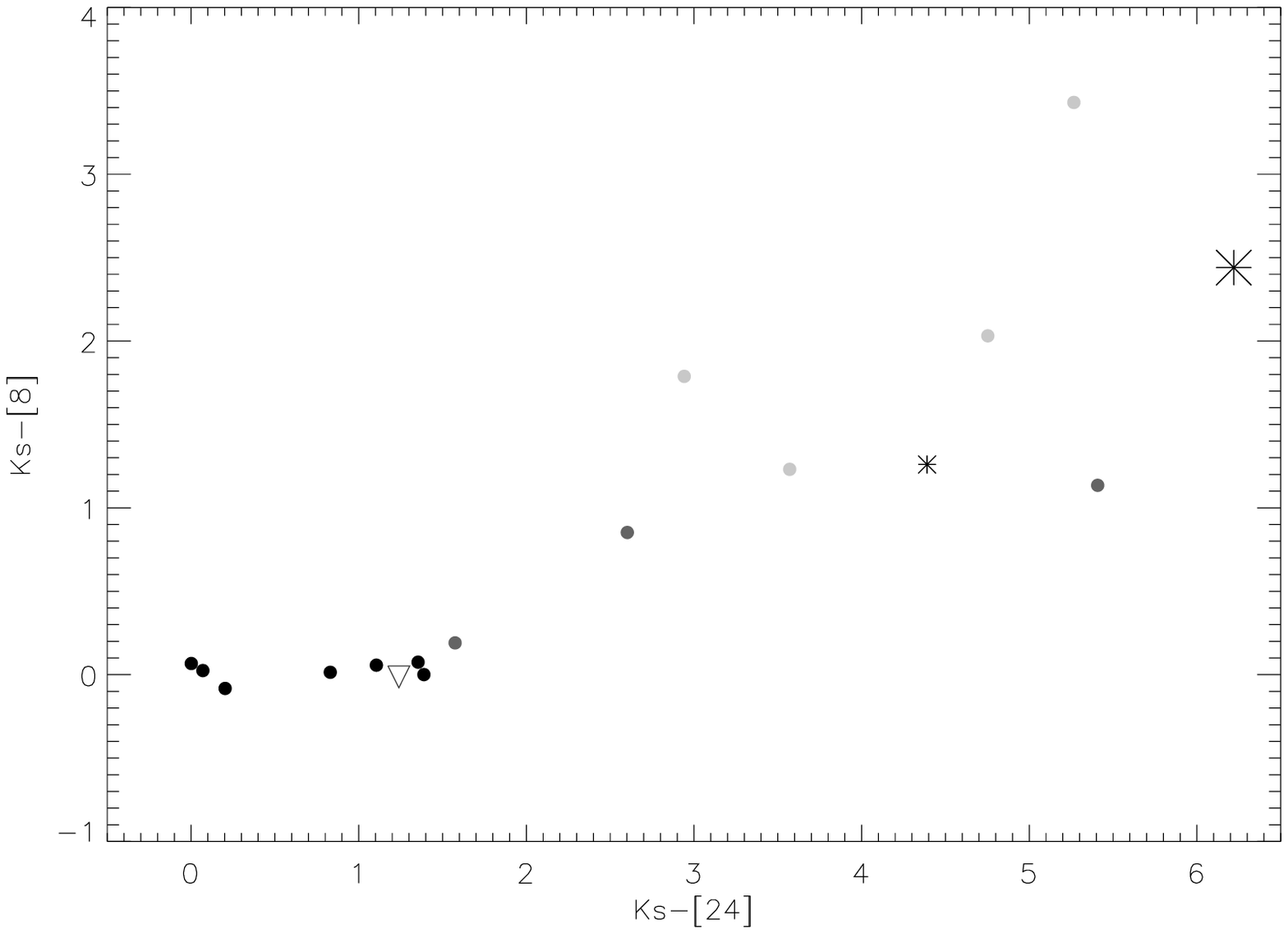}
\caption{(top) K$_{s}$-[8] vs. K$_{s}$-[24] color-color diagram for sources having spectral types 
K0 and later with new MIPS detections (open symbols) and previously-detected sources (filled dots)  
 later than K0.  The colors are dereddened using A$_{V}$ values from \citet{Mu07}.  
For sources with new detections, squares correspond to bare photospheres, 
inverted triangles are debris disk candidates, triangles are disks with inner holes, 
stars are homologously depleted disks, and open circles are primordial disks.  
The disk population is well separated in color-color space (dashed and dash-dotted lines).  
Previously detected sources labeled as 'thick' tend to reside in the primordial disk 
region, sources identified as anemic are located in the homologously depleted and inner-holed 
regions, and 'diskless' sources are located in the debris disk/photospheric region.
(bottom) The same diagram for sources with spectral types earlier than K0.  The symbols for 
IC 348 sources are the same.  The small and large asterisks represent the colors of 
a massive debris disk (h and $\chi$ Per-S5) and AB Aur, respectively.
}
\label{colcolevo}
\end{figure}

In Figure \ref{colcolevo} (top panel), we show the dereddened K$_{s}$-[8] vs. 
K$_{s}$-[24] color-color diagram for sources with new MIPS-24 $\mu m$ 
detections\footnote{Two sources -- IDs 746 and 2096 -- lack 
2MASS K$_{s}$ detections and are thus not shown.}.  
Symbols for sources with new MIPS detections correspond to their derived 
disk evolutionary state: primordial disks (circles), disks with inner holes (triangles), 
homologously depleted disks (inverted triangles), and debris disk candidates/photospheres (squares).
Sources with primordial disks lie redward of a region defined by a vertical and horizontal short-dashed line.
Polygonal regions enclose the range of colors for the two types of transitional disks: homologously depleted 
disks (dash-three dots) and disks with inner holes (dashes).  
Debris disk candidates and photospheres comprise all the newly-detected sources bluer in K$_{s}$-[24] 
than the homologously depleted/inner hole boundaries.  
The positions of sources later than K0 in IRAC/MIPS color-color diagrams 
thus show a good separation by disk evolutionary state as determined in the previous section.  

For reference, we overplot the colors of median Taurus SED with upper and lower quartiles (asterisks).  
The median Taurus SED (K$_{s}$-[8]=2.1, K$_{s}$-[24]=5.1), the upper quartile (K$_{s}$-[8]=2.5, K$_{s}$-[24]=6.3), 
and lower quartile (K$_{s}$-[8]=1.8, K$_{s}$-[24]=4.8) all lie within the primordial disk region.  
All of the sources identified as having primordial disks have 8 (24) $\mu m$ fluxes (measured relative to K-band)
 a factor of $\gtrsim$ 1.6 (2.1) smaller than the lower quartile Taurus SED or 
24 $\mu m$ fluxes reduced by a factor of $\approx$ 2.1.  The population from which the median Taurus SED was constructed 
includes sources with inner holes (e.g. DM Tau) or a significantly low dust mass (V836 Tau), so 
a median SED constructed only from sources with full disks has redder colors.  
Therefore, the fact that the boundaries enclosing primordial and non-primordial disks in Figure \ref{colcolevo} (top panel)
are \textit{bluer} than positions of the Taurus SED colors indicates that our identification of non-primordial disks is 
conservative.

We also overplot IC 348 sources with previous MIPS detections from \citet{La06} as filled dots 
with a brightness corresponding to their IRAC slope: 'thick' 
sources (light-grey dots), 'anemic' sources (grey dots), and 'diskless' 
sources (black dots).  There is generally a good correspondance between the IRAC slope and 
the inferred evolutionary states as defined by the colors 
(e.g. diskless=photospheres/debris, anemic=homologously depleted/inner holes, 
thick=primordial).  
However, some 'thick' sources are located in the region corresponding to 
homologously depleted disks and some 'anemic' sources are located in the region corresponding to 
primordial disks.  

Figure \ref{colcolevo} (bottom panel) shows the dereddened K$_{s}$-[8] vs. K$_{s}$-[24] diagram for 
sources earlier than K0, which for a 2--3 Myr-old cluster corresponds to 
stars with masses $\gtrsim$ 1--2 M$_{\odot}$ \citep{Ba98, Dm97}.  
  There is no fiducial SED (and thus no empirical colors) 
characterizing primordial disks surrounding stars of this mass range.
To provide a context for the colors of these IC 348 sources, we overplot the colors of 
a F9 star with a massive debris disk, h and $\chi$ Per-S5 \citep[small asterisk,][]{Cu07b}, and AB Aurigae (large 
asterisk), an A0 star in Taurus surrounded by a primordial disk.  The colors for AB Aur are 
estimated from the 2MASS and Spitzer-IRS fluxes at 2.2 $\mu m$, 8 $\mu m$, and 24 $\mu m$ \citep{Fu06}.
ID-4 has colors much bluer than h and $\chi$ Per-S5 and is
thus consistent with being a debris disk candidate.  Among previously-detected 
sources, one (ID-8012) clearly has primordial disk emission (K$_{s}$-[8] $\sim$ 3.4, K$_{s}$-[24] $\sim$ 5.3). 
Five others with K$_{s}$-[24]=2.5--5.5 require further analysis because their mid-IR colors 
could be consistent with either a debris disk or a remnant protoplanetary disk (i.e. homologously 
depleted, inner hole).

In summary, the dereddened K$_{s}$-[8] vs. K$_{s}$-[24] color-color diagram 
separates late-type IC 348 sources by disk evolutionary state.  This analysis is 
 more ambiguous for early/intermediate spectral type stars.  
Sources of any spectral type identified as debris disk candidates on the basis of their mid-IR colors require further 
analysis because debris disks, by definition, lack gas accretion and have second generation dust.  
These properties cannot be determined from mid-IR colors alone.  In the next section, we will 
investigate the properties of sources with ambiguous states, including all debris disk candidates.  
After determining their likely evolutionary states, we will use the results from color-color 
diagrams presented in this section to analyze the global properties of the IC 348 disk population.

\subsection{Evolutionary State of Disks Surrounding Early/Intermediate-Type Stars and 
All Debris Disk Candidates}
Many early/intermediate spectral type sources with mid-IR colors 
bluer than those harboring primordial disks (e.g. AB Aurigae) have ambiguous 
disk evolutionary states.  Many sources surrounding late type, low-mass 
stars with new MIPS-24 detections have extremely weak mid-IR disk emission that could either come from 
a debris disk or a remnant protoplanetary disk.  In this section, we constrain the evolutionary 
state of disks surrounding stars from these two populations.  We first use new and 
archival optical spectra to identify accretors 
and to remove them from the population of debris disk candidates.  For remaining debris disk candidates, we 
estimate the mass of emitting dust (through 24 $\mu m$) and calculate dust removal timescales.
The early/intermediate type stars investigated in this section are IDs 2--8, 11, 19--20, 25, and 30--31.
  The late-type stars investigated here, all of which have new MIPS-24 $\mu m$ detections,
 are IDs 39, 53, 56, and 1939.

\subsubsection{Accretors}
Using new and archival optical spectra, we can remove several accreting early-type
stars from the list of debris disk candidates.  These include ID 19 -- identified as 
an accretor from this work -- and IDs 5 and 31, which were previously identified 
as accretors.  None of the late-type stars 
show unambiguous evidence for accretion based on their H$_{\alpha}$ equivalent widths.

One source --ID 2 -- is a borderline case.  Our spectra for ID-2 may exhibit 
slightly weakened H$_{\alpha}$ absorption line compared to non-accreting 
stars of the same spectral type (ID-8).  Based on unpublished high-resolution 
echelle spectroscopy, \citet{Dahm08} identifies this star as 
having an H$_{\alpha}$ emission line reversal plausibly due to accretion.
  Thus, we remove it from our sample of debris disk candidates.  
We treat this star as lacking evidence for circumstellar gas, though we caution that 
accretion may persist at low rates.  Given that the presence or 
absence of accretion is a primary marker of disk evolutionary state, the inability to 
 rule out or identify accretion for this source renders its disk 
status uncertain.

\subsubsection{SED Modeling of Debris Disk Candidates: Estimating Dust Masses and Temperatures}
 Next, we compare the SEDs of early-to-late type debris disk candidates
 with 3D radiative transfer disk models for a range of disk masses.  
 To model the near-to-mid IR emission of debris disk candidates,
we compare their SEDs to the grid of 200,000 radiative transfer models from \citet{Rob06},
which include a range of spectral types and dust masses.
The infrared fluxes predicted for disks are sensitive to the dust masses down to
$\sim$  10$^{-10}$ M$_{\odot}$.
By selecting the best-fitting SEDs, we estimate the minimum mass of small ($\lesssim$ 100 $\mu m$ -- 1mm) 
dust grains required to reproduce the observed disk emission.  We compare this 
estimate to typical dust masses in debris disks and primordial disks.
Typical dust masses for debris disks range from 10$^{-11}$ M$_{\odot}$ to
5$\times$10$^{-9}$ M$_{\odot}$ \citep[cf.][]{Ch05, Lo05}.  Dust masses\footnote{This mass is
not the same as the \textit{disk} masses, which is more typically cited.
To derive the disk mass from the dust mass, \citet{Aw05} and other other
 authors assume a gas to dust ratio of 100.  The \citet{Rob06} models assume 
the same gas to dust ratio. } typically needed to reproduce the IRAC and MIPS excess emission
for transitional disks range from 10$^{-7}$--10$^{-5}$ M$_{\odot}$ \citep{Clp09}.

We choose a representative sample of sources to model: IDs 6, 8, 20, and 30 for 
the early/intermediate-type population and IDs 39 and 1939 for the late-type population.  
Because most of the other sources have weaker disk emission or comparable emission, the 
required mass of emitting dust in their disks is most likely comparable or lower.
Therefore, modeling the SEDs of these sources to estimate the mass of small, emitting 
dust grains is sufficient to yield approximate dust masses for all 
debris disk candidates.

\begin{figure}
\epsscale{0.99}
\centering
\plottwo{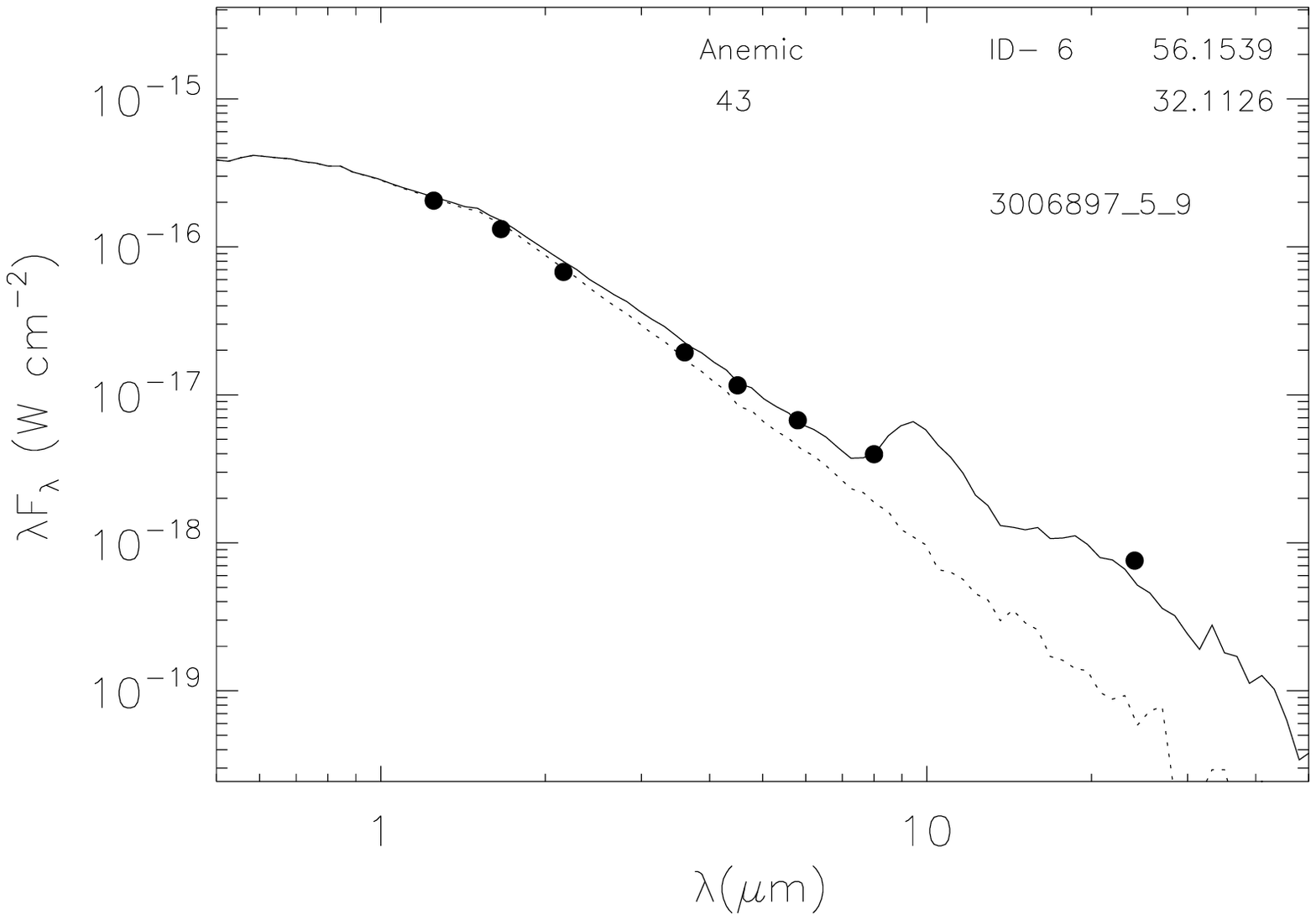}{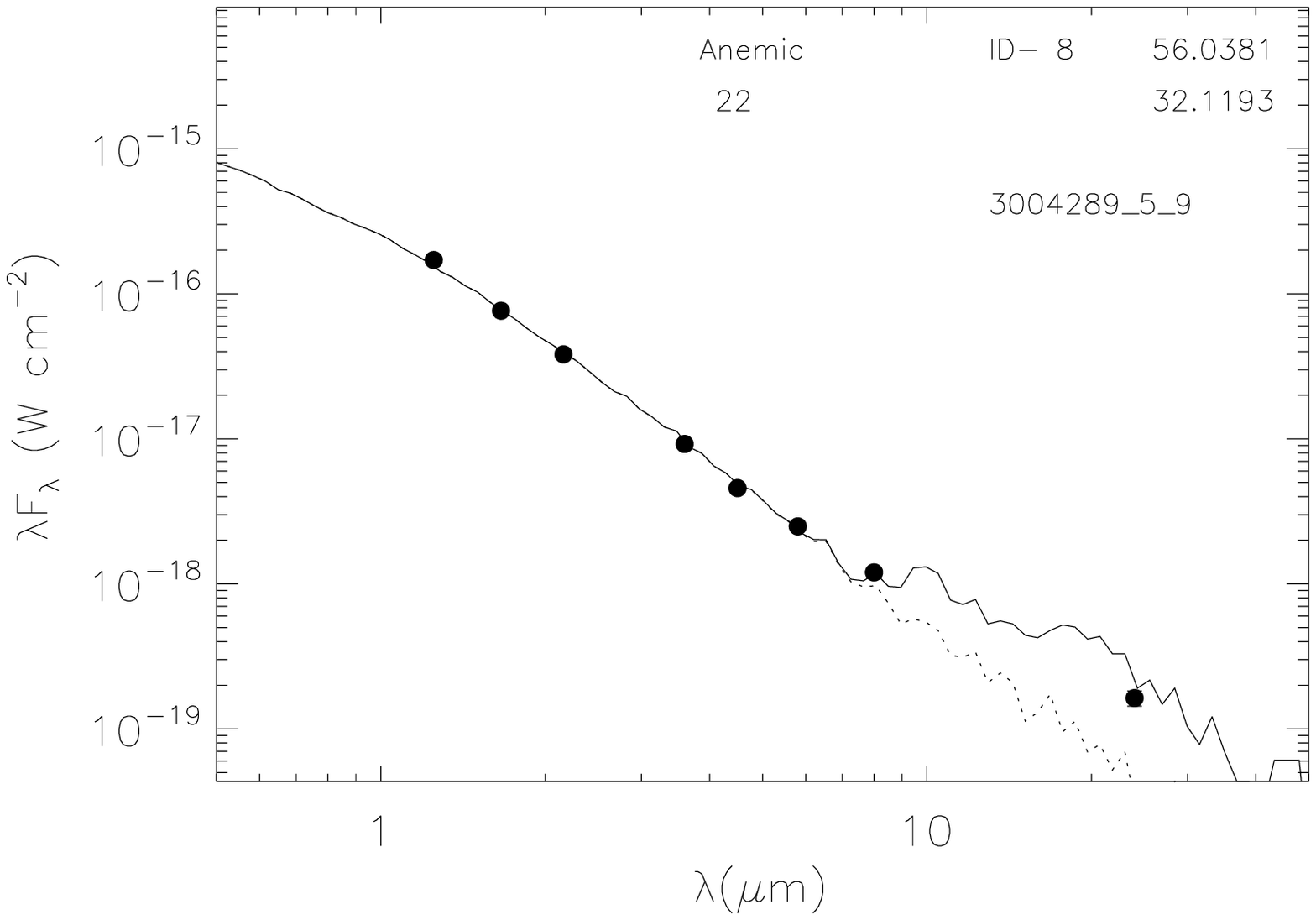}
\plottwo{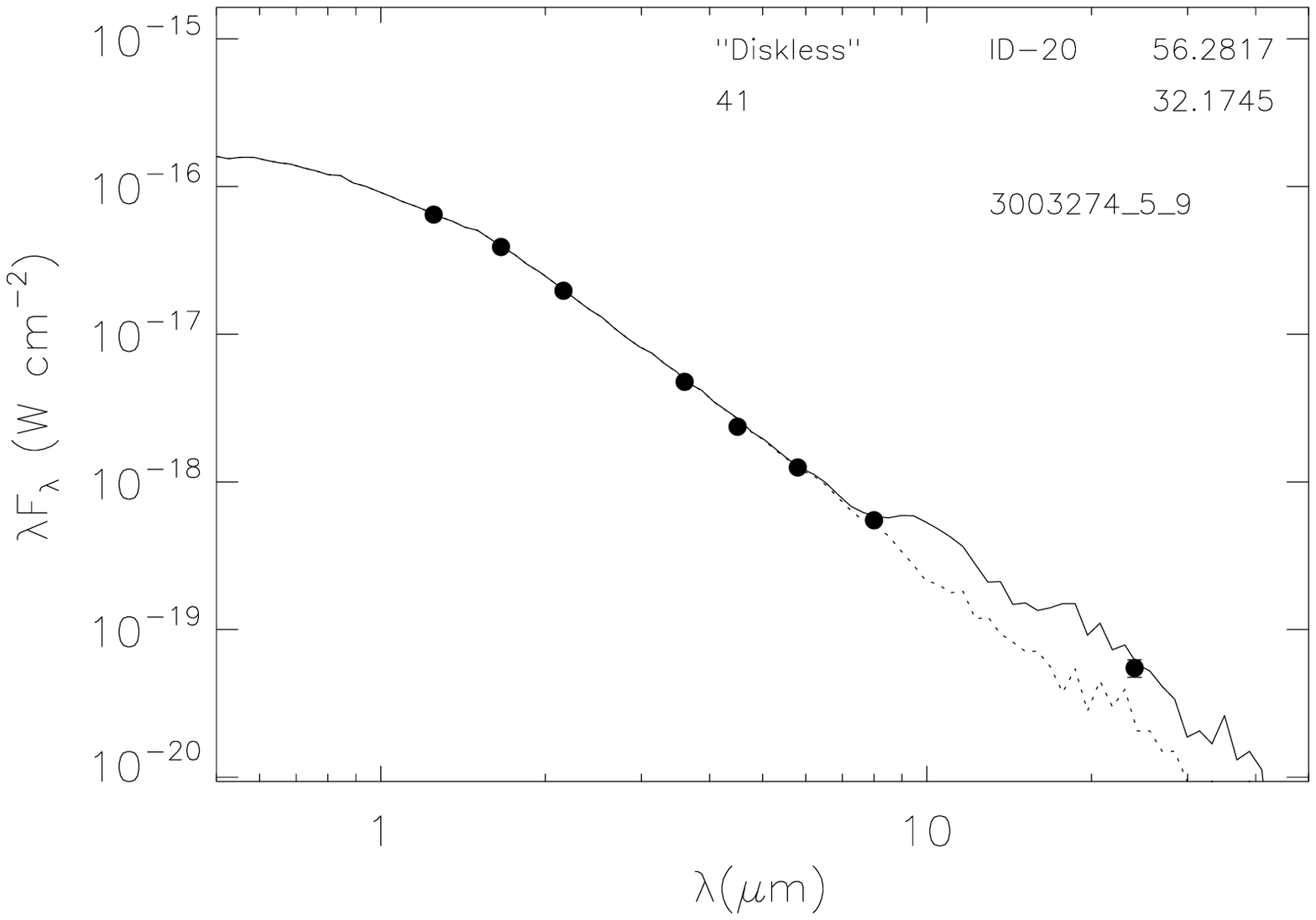}{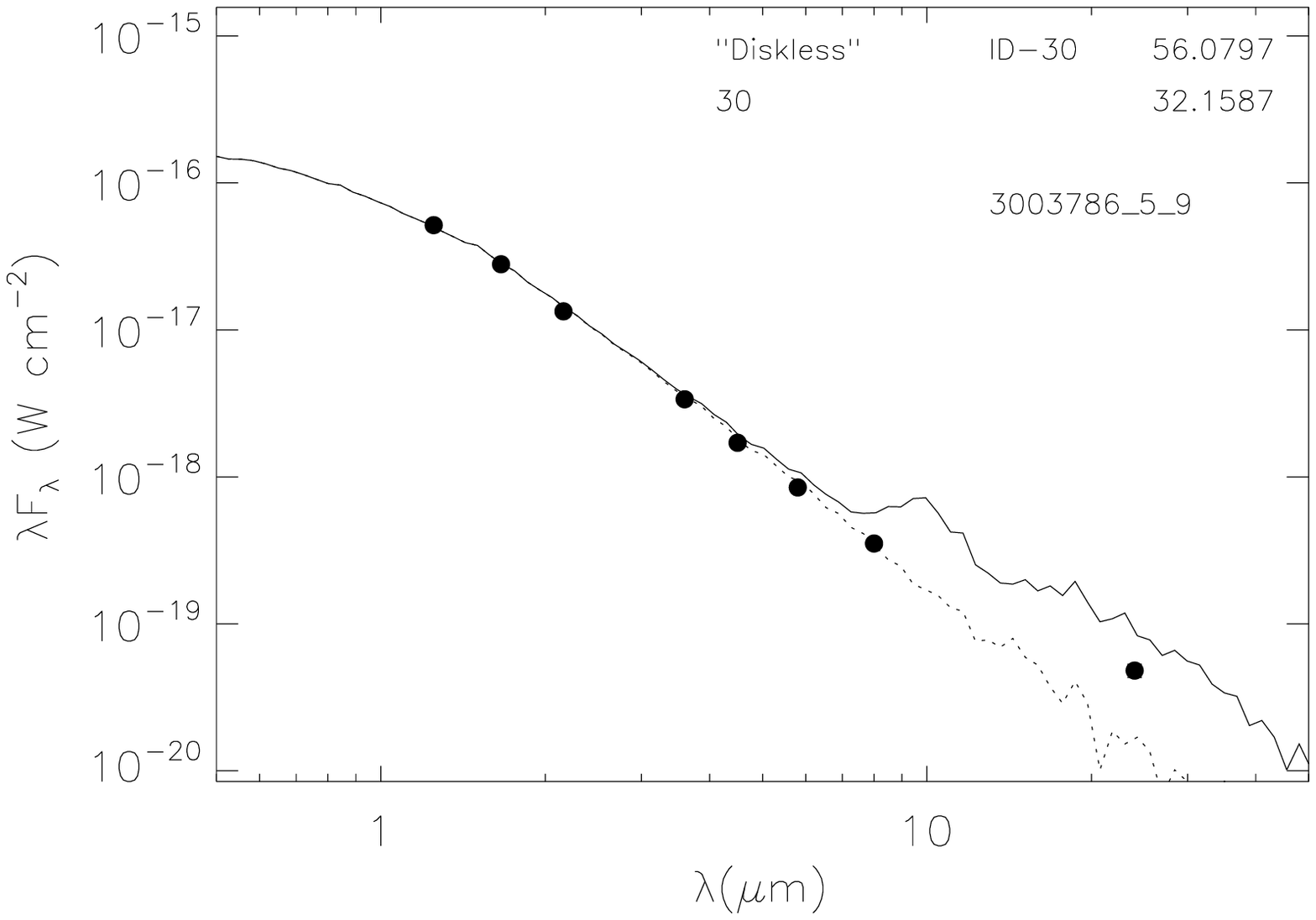}
\plottwo{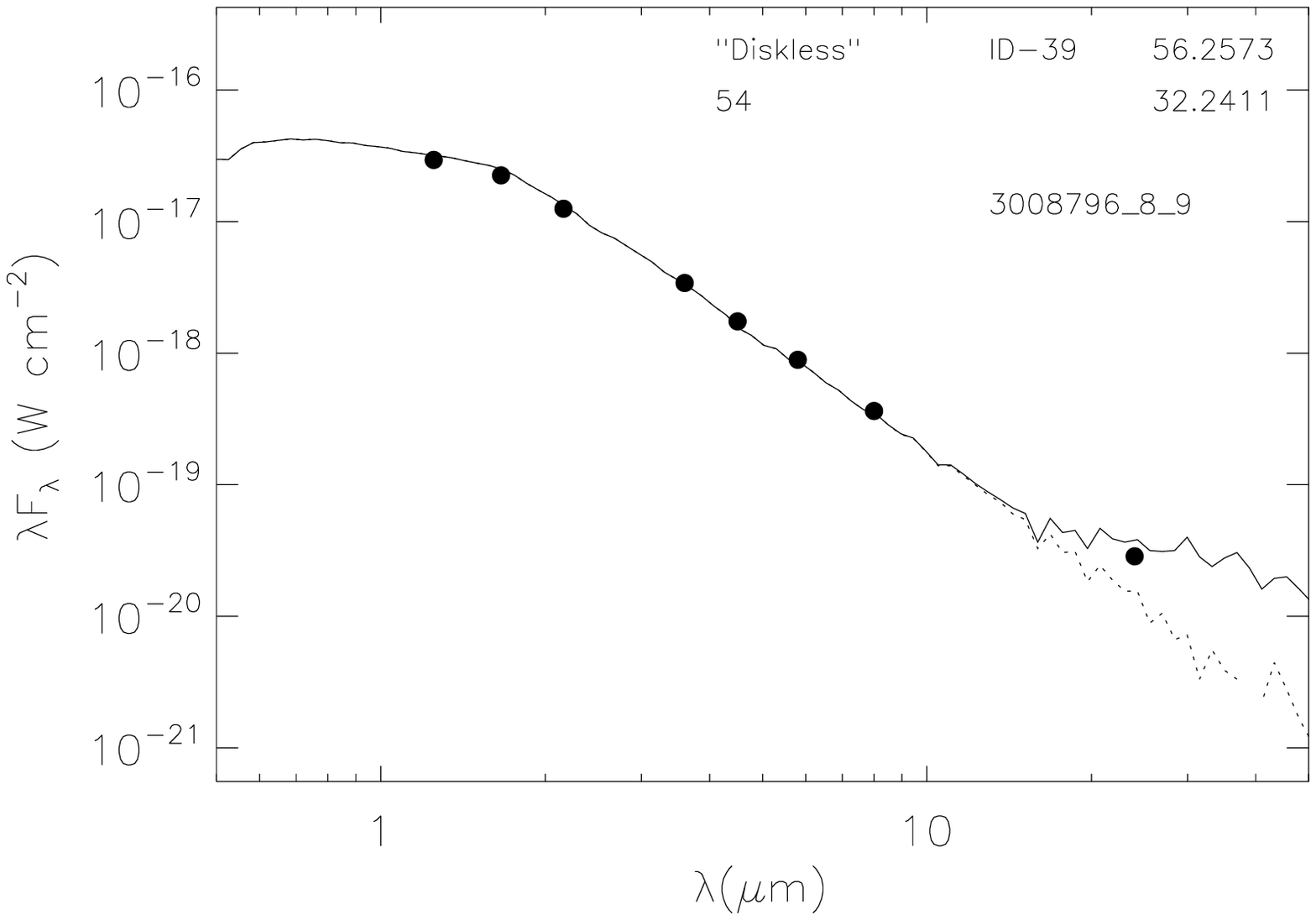}{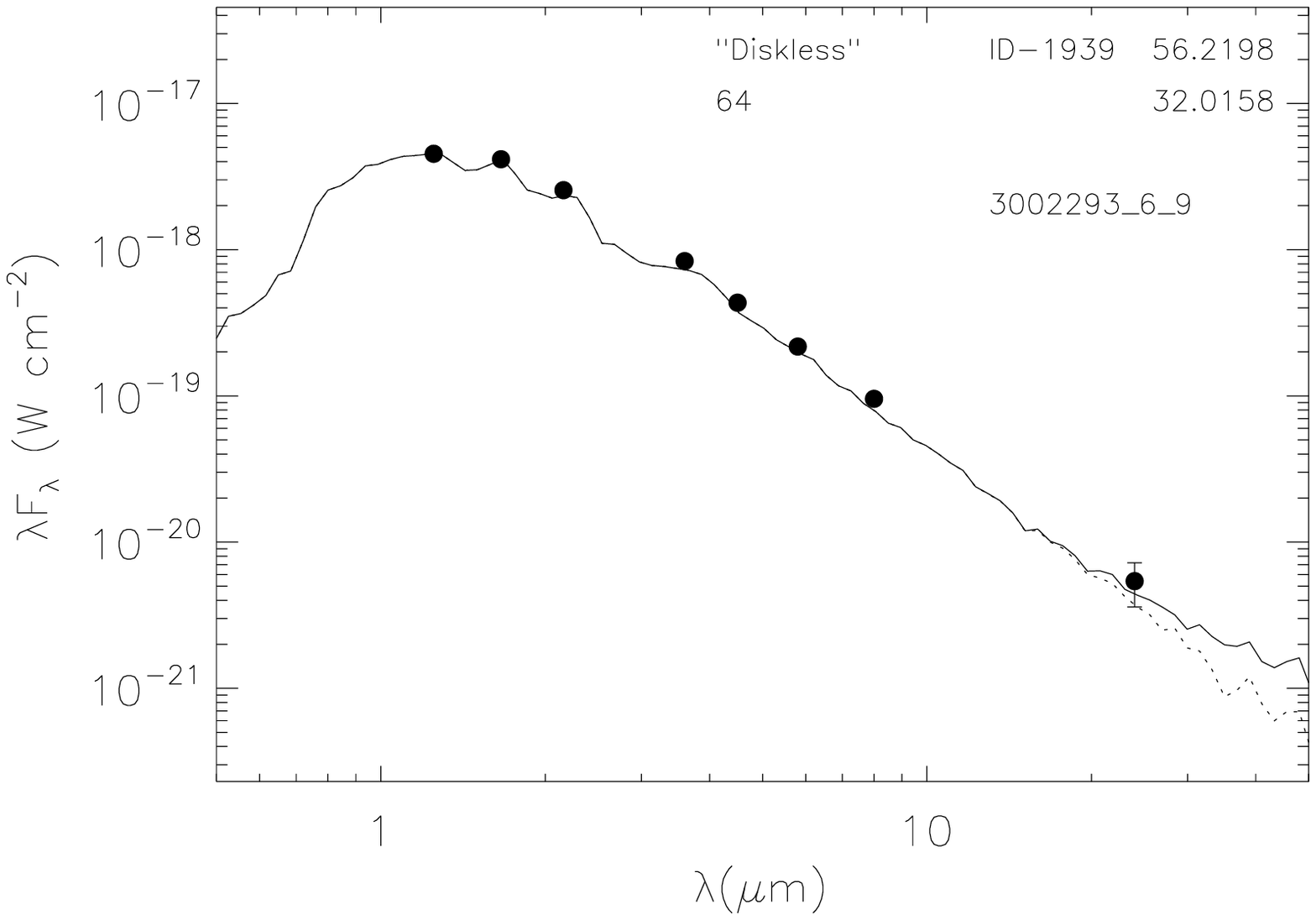}
\caption{SED of representative debris disk candidates.  
  Overplotted are the best-fit SEDs from \citet[][solid line]{Rob06} along with 
the stellar photosphere(dotted line).  Also listed are the \citet{La06} IRAC slope 
and ID number, the numerical spectral type (below the IRAC slope), the J2000 source position (in degrees), 
and the model number for the best-fit \citet{Rob06} model SED.
Photometric uncertainties (1 $\sigma$) 
are typically smaller than the symbol sizes.  }
\label{noremnantSED}
\end{figure}

Figure \ref{noremnantSED} shows SED model fits to the six sources.  Table \ref{noremnant} 
summarizes  the model parameters for each source.
The best-fit \citet{Rob06} models accurately reproduce the observed fluxes from these sources.  
Assuming a gas-to-dust ratio of 100, the required disk masses for the best-fit \citeauthor{Rob06} models 
are $\sim$ 2--8 $\times$ 10$^{-8}$ M$_{\odot}$.  The extremely low disk masses are consistent with 
the disks being very optically thin.  The required masses in dust are $\approx$ 
10$^{-10}$--10$^{-9}$ M$_{\odot}$.  These dust masses are lower than typical dust masses of primordial disks 
but similar to debris disk masses.  Unless these sources had enormous reservoirs of extremely cold dust, 
which would not significantly contribute to the observed IRAC and MIPS 24 $\mu m$ emission, the \textit{total} mass 
of dust in small grains is also likely much less than typical dust masses for primordial disks.

We can place further constraints on the temperature and luminosity of the dust.  
Most debris disk candidates have clear excess emission only at 24 $\mu m$ with dust
temperatures colder than 200 K \citep{Carpenter09}.  
However, IDs 6 and 8 have 8 $\mu m$ excess emission, which indicates that their 
dust, if confined to a narrow ring like in a debris disk, is warmer.  Indeed, a 
single temperature blackbody fit to the SEDs of these sources yields dust temperatures 
of 400 K and 270 K.  These temperatures are similar to those for known terrestrial 
planet-forming debris disks such as h and $\chi$ Per-S5, EF Cha, and HD 113766A \citep{Cu07b, 
Rh07a, Li07}.  The fractional infrared luminosities (L$_{D}$/L$_{\star}$) of debris disk candidates from 
one and two-temperature blackbody fits range from 5$\times$10$^{-3}$ (ID-6) to 10$^{-4}$ (ID-3).  
These fractional luminosities are two to three orders of magnitudes less than those for 
primordial disk-bearing stars but very similar to the stars with warm, terrestrial planet-forming 
disks as well as those with colder debris disks \citep[e.g., $\beta$ Pic][]{Rebull2008}.

\subsubsection{Dust Dynamics and Removal Timescales}
To provide better constraints on the evolutionary state of debris disk candidates, we investigate
dust dynamics.  If the dust must be second generation, then the disks are debris 
disks \citep{Bp93}.  We determine grain removal timescales in 
optically thin conditions in two cases: where there is a significant amount of undetected 
circumstellar gas (1--10 M$_{\oplus}$) and where there is little or no gas.  
If the timescale for removal is much less than the age of the star in either case, the 
grains require a replenishment source through collisions 
between larger bodies.  Disks around such stars would then
 fulfill the standard definition of a debris disk \citep[e.g.][]{Bp93}.

\begin{itemize}
\item\textbf{Case 1: Optically-Thin Disk with Residual Circumstellar Gas}\\
\citet{Tak01} extensively investigated the dynamics of circumstellar dust 
in optically-thin disks with residual gas.  Under these conditions, the dust grains 
migrate outward at a rate primarily determined by $\beta$, the ratio of the 
force from radiation pressure to the gravitational force:
\begin{equation}
\beta = \frac{3L_{\star}<Q_{pr}(a)>}{16\pi GM_{\star}cs\rho_{s}}.
\label{tpr}
\end{equation}

When $\beta$ $\ge$ 0.5 the dust is removed from the system entirely on 
orbital timescales, which are $\approx$ 100--1000 yr \citep[see also][]{StrCh06}. 
 {\textit{Removal occurs independent of whether there is residual gas left in the disk}} 
\citep[][Takeuchi 2008, pvt. comm.]{Tak01}.
Equation \ref{tpr} can be rearranged to find the size of dust grain below which 
radiation pressure rapidly removes dust from the system \citep[see also][]{Bp93, Bu79}:
\begin{equation}
s_{max, \mu m} < 1.14 \mu m (\frac{L_{\star}}{L_{\odot}})(\frac{M_{\odot}}{M_{\star}})(\frac{1 g cm^{-3}}{\rho_{s}}) <Q_{pr}>.
\end{equation}
Typically, the sizes of mid-IR emitting dust grains range from $\approx$ 1 $\mu m$ to 10 $\mu m$ 
\citep[e.g.][]{Au99, Fu06}.

For early-type stars listed in Table \ref{noremnant}, s$_{max}$ ranges from 2.2 to 
17.8 $\mu m$, assuming $<Q_{pr}>$ $\approx$ 1 and $\rho_{s}$ 
$\approx$ 1 g cm$^{-3}$.  Using a more realistic dust grain density of $\sim$ 0.1 g cm$^{-3}$ \citep[e.g.][]{Au99} 
essentially removes all dust grain sizes that can plausibly produce IRAC and MIPS disk emission.  
 Grains responsible for producing this emission are blown out; mid-IR emission ceases unless the new grains are produced 
from collisions between larger bodies.  

Dust grains much larger than s$_{max}$ ($\beta$ $<$ 0.5) 
can produce 8--24 $\mu m$ broadband excess emission in these early type stars and must also be second generation.  
For a 1--10 M$_{\oplus}$ gas disk around an early-type star, the combined influence of radiation pressure, Poynting-Robertson 
drag, and gas drag push $\approx$ 10--100 $\mu m$-sized dust from 10 AU to $\sim$ 300 AU 
in $\approx$ 600--60,000 yr \citep{Tak01}.  For the early-type stars in Table \ref{noremnant}, 120--200 K dust 
is located at $\approx$ 5--35 AU.  If pushed to $\sim$ 300 AU, the dust temperature 
 drops by up to a factor of 5, making the dust too cold 
to emit strongly at 24 $\mu m$.  If new dust at 5--35 AU is not produced, then 
the disks' 8--24 $\mu m$ emission will become undetectable.   Thus, the observed mid-IR emitting dust 
must be second generation.

The effect of radiation pressure on dust surrounding late-type stars is less clear.
  The maximum grain size for ID-39 is $\sim$ 1.5 $\mu m$, a value that is slightly smaller than 
s$_{max}$ for the earlier-type stars.  However, s$_{max}$ for ID-1939 (and, by inference, all 
M stars that are debris disk candidates) 
is $\sim$ 0.98 $\mu m$, which begins to approach the sizes of large grains in the 
interstellar medium.   While radiation pressure may be able to remove 
grains with these sizes, it is probably ineffective 
at completely removing grains with sizes typical of circumstellar disks (up to 10 $\mu m$) 
unless the grains are extremely porous.  Therefore, the dust responsible for 
 24 $\mu m$ disk emission is not necessarily removed from the disk by radiation pressure.
However, grains with sizes typical of circumstellar disks and 
and densities of $\rho_{s}$ $\approx$ 1 g cm$^{-3}$ have $\beta$ up to 0.3 and could be 
transported from their original locations to $\approx$ 300 AU scales \citep{Tak01}.

\item\textbf{Case 2: No Gas or Trace Amounts of Gas}\\
In the absence of residual circumstellar gas, the dust dynamics are more simple.  
If $\beta$ for a given dust size is greater than 0.5, then dust is 
rapidly removed by radiation pressure as in Case 1.  If $\beta$ 
is $<$ 0.5 then Poynting-Robertson drag is the dominant removal mechanism.
For $\rho_{s}$ $\sim$ 1 g cm$^{-3}$ and $<Q_{abs}>$ $\sim$ 1,
the P-R drag timescale can be parameterized \citep[e.g.][]{Bp93, Au99}:
\begin{equation}
t_{P-R} \sim 7\times10^{4} (\frac{L_{\odot}}{L_{\star}})(\frac{s}{1 \mu m})(\frac{r}{10 AU})^{2} yr,
\end{equation}
where s is the grain size.

To derive the drag timescale for grains, we consider typical 
parameters for r$_{AU}$ (the grain's distance from the star), $<Q_{abs}>$ (the 
absorption coefficient), and $\rho_{s}$ (grain volume density).  Most sources have weak or 
negligible 8 $\mu m$ excesses and stronger 24 $\mu m$ excesses; thus, 
their dust temperatures are probably $\lesssim$ 125--250 K \citep{Cu08a}.  While 
 120 K dust can produce a strong 24 $\mu m$ excess but weak/no 8 $\mu m$ excess, colder dust with 
24 $\mu m$ emission further down the Wien tail can also produce these trends.  
Given the lack of constraints on dust's temperature and location, we calculate 
t$_{P-R}$ for a range of distances -- 1, 10, and 30 AU -- which cover the range of 
distances for dust at 100--200 K for most stars listed in Table \ref{noremnant} assuming 
simple radiative equilibrium.

As shown in Table \ref{noremnant}, P-R drag removes small grains surrounding early-type 
IC 348 on much timescales shorter than the $\sim$ 2--3 Myr age of the cluster.  For 1 $\mu m$ grains, 
the removal timescales range from being several hundred years to $\sim$ 100,000 years 
for IDs 6,8, 20, and 30.  In all cases, the removal timescale is less than $\approx$ 1/20th of the 
median age for IC 348 stars.  Thus, without a replenishment source, dust grains 
surrounding these stars are removed rapidly, on timescales much less than the age of the star.

The effect of P-R drag on dust surrounding late-type stars is less clear.  
If the dust orbits closely to the stars (1 AU), then P-R drag can 
remove the dust on rapid timescales.  However, for both ID-39 and ID-1939, it is 
not clear that the dust requires a replenishment source if it is located at 
much larger distances (30 AU).  For ID-1939, dust at intermediate distances (10 AU) 
could also be remnant protoplanetary dust which are plausible.  The \citet{Rob06} 
models that best fit the IDs 39 and 1939 SED assumes inner hole sizes of $\approx$ 100 AU and 
31 AU, respectively.  Thus, the dust need not be second generation.

\end{itemize}

In summary, the dust grains surrounding the AFG stars listed in Table \ref{noremnant} 
must be 2nd generation.  This result is independent of whether the disk has 
no gas or has residual gas.  Because the dust must be 2nd 
generation, disks surrounding these stars must be debris disks.
Other characteristics of these disks -- lack of gas accretion and minimum required dust mass
 -- support this conclusion.
The only situation in which the dust can be primordial is if the disk 
retains an enormous reservoir of gas, $>>$ 100-1000 M$_{\oplus}$.  In such a 
case, the gas-dust coupling is strong enough that the 
removal timescale from radiation pressure may be comparable 1--3 Myr 
(Takeuchi, pvt. comm.).  However, the extreme gas to dust ratios 
implied by these conditions, $>>$ 10$^{8}$--10$^{9}$, would be highly unusual 
if not unphysical.  Spectroscopy of sources listed in Table \ref{noremnant} fail to identify any with active 
gas accretion, which would be expected if the disks had an enormous reservoir of gas left.  
By far the most simple explanation is that these sources are very gas poor or gas free.  
The dust responsible for their MIPS excesses then
most likely requires a replenishment source: collisions between larger bodies.  
Thus, these disks fit the standard definition of a debris disk. 

In contrast, it is not clear if dust surrounding late type stars (e.g. ID-39 and ID-1939) must be 
second generation.  Both ID-39 and ID-1939 lack evidence for accretion and have a low disk mass.  
However, for ID-39 it is not clear that its dust must second generation if the dust is 
 located far from the star and is large ($\beta$ $<$ 0.5).  The dust surrounding ID-1939 
need not be second generation if it is located beyond $\approx$ 1 AU of the star and or 
is entrained in a substantial reservoir of gas ($\gtrsim$ 1 M$_{\oplus}$).  Both of these situations are at least 
plausible in principle.  Other removal mechanisms such as
corpuscular wind drag \citep{Plav05,Plav09} may be important for late-type, low-mass stars 
are neglected here.  Analyzing dust removal timescales after including these effects,
 may affect our conclusion that the dust surrounding late-type stars need not be from a debris disk.

Based on the above analysis, we classify all AFG stars with characteristics typical of those 
listed in Table \ref{noremnant} (mid IR colors, lack of accretion signatures) as debris disk 
candidates.  Debris disk candidates surrounding K and M stars with similar characteristics could 
be either debris disks or evolved primordial disks/transitional disks with very low dust masses.  
To be conservative, we classify them as homologously depleted disks\footnote{While 
the disk models best fitting the SEDs have large inner holes, the low disk vs. photosphere 
contrast afforded by K and M stars means that the disk 
mass may simply be extremely low \citep{Er09}.}.

\section{Global Analysis of the IC 348 Disk Population}
We now combine our new optical spectroscopy and MIPS photometry of 
IC 348 stars with archival data to investigate the properties of 
the entire IC 348 disk population.  Our analysis has several goals.  
 To investigate the frequency of gas accretion as a function of spectral type,
we combine our analysis of optical spectra with archival spectroscopic data. 
We consider whether the frequency of signatures of circumstellar gas 
depends on the spectral type of stars, a trend consistent with a 
stellar-mass dependent dispersal of gas disks.

Second, we investigate how the rate of dust evolution 
varies with stellar mass.   For a simple, empirical probe 
of disk evolution, we first analyze the K$_{s}$-[8] and K$_{s}$-[24] colors 
as a function of spectral type and IRAC slope.
We consider the evidence for a spectral-type dependent 
dust luminosity at 8 $\mu m$ and 24 $\mu m$.
To further see whether disks surrounding stars of different stellar masses 
evolve at different rates, we conclude by constraining the relative 
frequency of disks in different evolutionary states\footnote{We do not include 
sources that lack near-IR photometry because SED modeling of such sources (e.g., IDs 746, 2096)
was unsuccessful in constraining their evolutionary states}.  

\subsection{Gas Disk Evolution as a Function of Stellar Mass}
To investigate gas disk evolution in IC 348, we analyze data presented 
in this work, archival data summarized in \citet{La06}, and 
 data from \citet{Dahm08}.  We use empirical criteria 
to identify accretors (primarily the H$_{\alpha}$ equivalent width) 
from \citet{Wb03} for stars later than G0.  Stars with very early spectral 
types (BAF) have a deep H$_{\alpha}$ absorption lines produced from the stellar photosphere.  
For these stars, we use our results and those from \citet{Dahm08} to 
identify stars with H$_{\alpha}$ emission line reversals consistent with accretion.
Data for late-type stars summarized in \citet{La06} typically draw from 
\citet{Lu03} and \citet{Lu99}.  Some sources lacking H$_{\alpha}$ equivalent widths 
from optical spectroscopy have near-IR spectra that either confirm their status 
as accretors (\citealt{Lu05, Dahm08}) or rule out clear evidence of accretion 
\citep[][]{Dahm08}.  We include these sources in our statistics.  
We remove from the sample all other stars lacking published H$_{\alpha}$ equivalent widths.

\begin{figure}
\plottwo{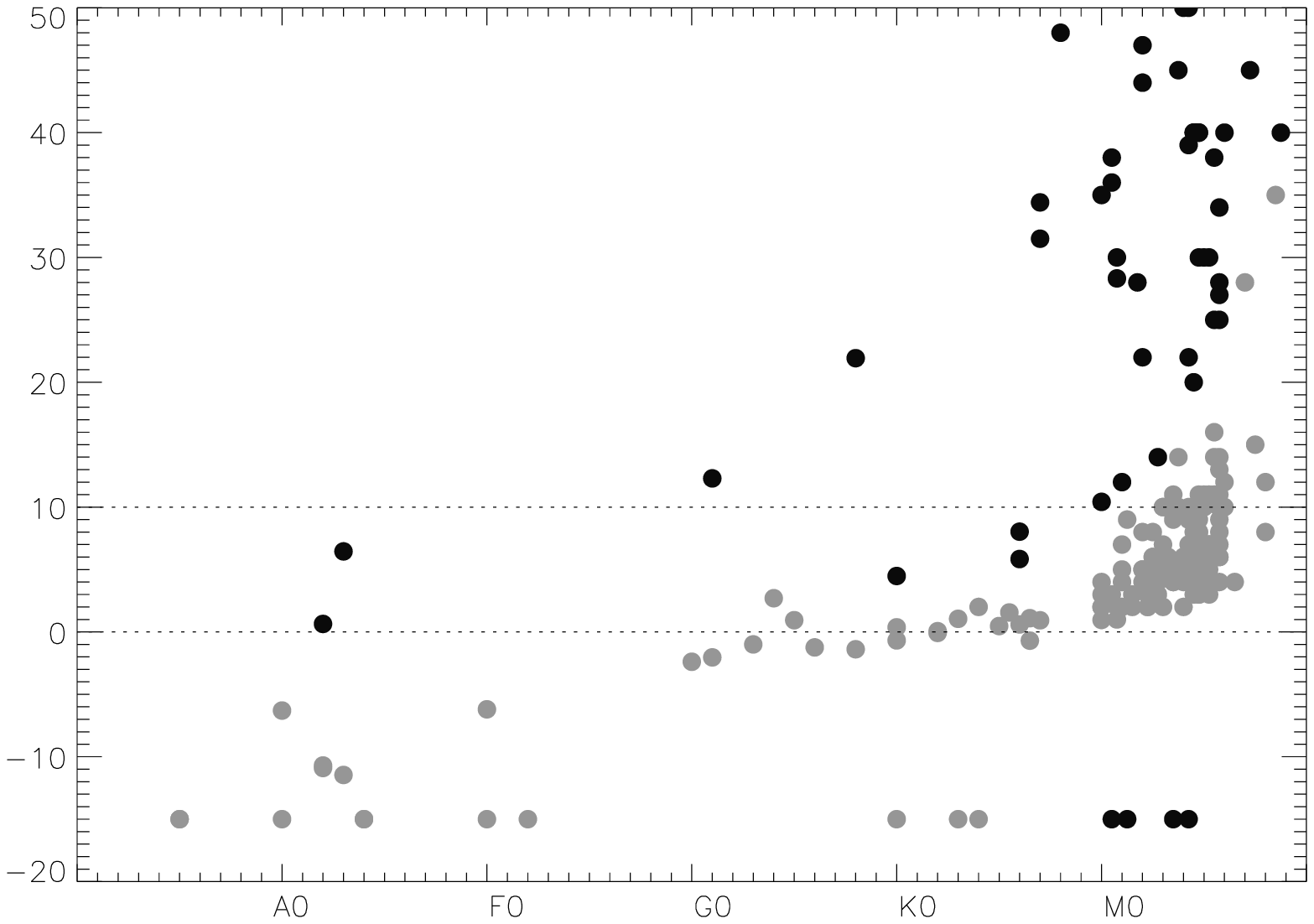}{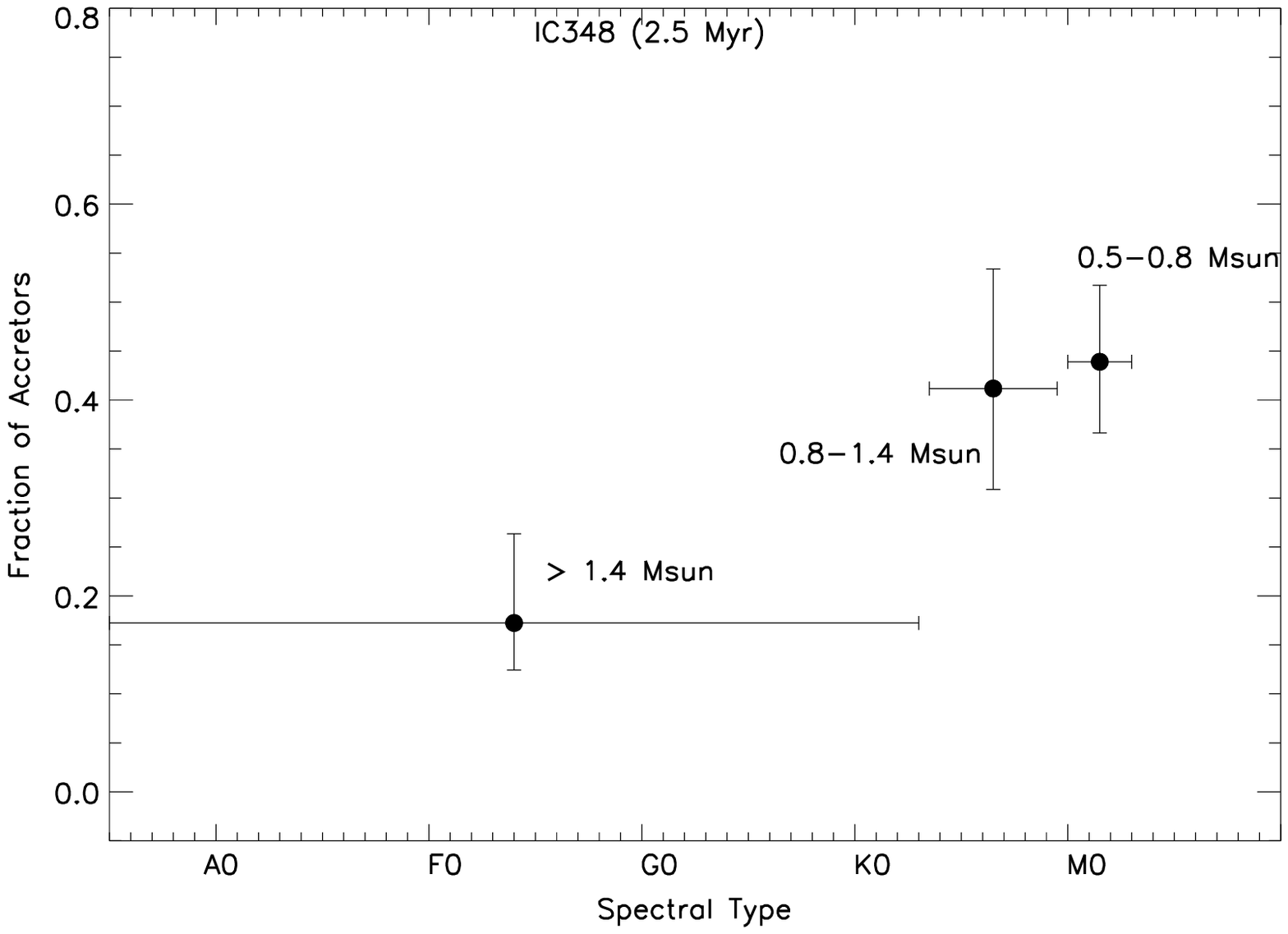}
\caption{(left) H$_{\alpha}$ equivalent width versus spectral type for IC 348 stars.  
Accretors are shown as black dots; nonaccretors are grey dots.  Sources without 
measured EW(H$_{\alpha}$) values have EW(H$_{\alpha}$) set to -15 \AA.  For reference, we 
draw a line corresponding to the division between Classical T Tauri stars and 
Weak-line T Tauri stars.  (right) Frequency of accretors as a function of 
spectral type for three spectral type bins.  The horizontal error bars correspond 
to the bin size in spectral type -- B5--K3, K3--M0, and M0--M2.5 -- and the 
vertical error bar identifies the 1 $\sigma$ uncertainty in the frequency of accretors 
for each spectral type bin based on binomial statistics.}
\label{eqha}
\end{figure}

Figure \ref{eqha} (left panel) plots the H$_{\alpha}$ equivalent width versus spectral type.  
Stars identified as accretors are shown as black dots while those lacking 
evidence for accretion are shown as grey dots. 
Stars without H$_{\alpha}$ in emission that lack evidence for accretion are shown along a
line corresponding to EW(H$_{\alpha}$)= -15 \AA(in absorption).  
For reference, we draw lines corresponding to EW(H$_{\alpha}$)=0 (the division between emission 
and absorption) and 10 \AA, which 
is the division typically used to distinguish between classical T Tauri stars 
and weak-line T Tauri stars.  According to our criteria, 
several stars with spectral types earlier than K5 
have EW(H$_{\alpha}$) $\le$ 10 \AA\ but are identified as accretors, while 
others with spectral types later than M3 have EW(H$_{\alpha}$) $\ge$ 10 \AA\ 
but are not accreting.

Figure \ref{eqha} (right panel) plots the frequency of accretors as a function of spectral type.
We divide the sample into three spectral type bins: sources earlier than K3, between K3 and M0, and between 
M0 and M2.5.  For a 2--3 Myr-old cluster, these divisions correspond to stars with masses $\gtrsim$ 
1.4 M$_{\odot}$, 0.8--1.4 M$_{\odot}$, and 0.5--0.8 M$_{\odot}$ \citep{Ba98}.
Sources with probable masses $\gtrsim$ 1.4 M$_{\odot}$ have the lowest frequency of accretion 
(5/29: 17.2$^{+9}$$_{-4.8}$\%).  Nearly half of the stars with probable masses of 0.8--1.4 M$_{\odot}$ and 0.5--0.8 M$_{\odot}$ 
show evidence for gas accretion (7/17, 41.2$^{+12.2}$$_{-10.3}$\%, and 18/41, 43.9$^{+7.8}$$_{-7.3}$\%, respectively).

In deriving these frequencies, we acknowledge that 
using H$_{\alpha}$ as a diagnostic of accretion, even following the prescription of \citet{Wb03}, 
is prone to some uncertainties.  The contrast between the H$_{\alpha}$ emission line and the 
photosphere is stronger for later type, lower-luminosity stars.  The large number of 
 stars later than $\sim$ M5 lacking optical spectra 
yields an uncertain frequency of accretors for the lowest-mass stars.  
Nevertheless, the observed trend of accretion 
frequency is consistent with the trend of infrared excess as a function of spectral type 
from \citet{La06} at a $\sim$ 1--2 $\sigma$ significance: the frequency is 
lowest for early type, high-mass stars and highest for 0.6--1 M$_{\odot}$ 
stars.  The results are consistent with a stellar-mass dependent dispersal of gaseous 
circumstellar disks \citep[see also][]{Kennedy2009}.

\subsection{Circumstellar Dust Evolution as a Function of Stellar Mass}
\begin{figure}
\includegraphics[scale=0.45]{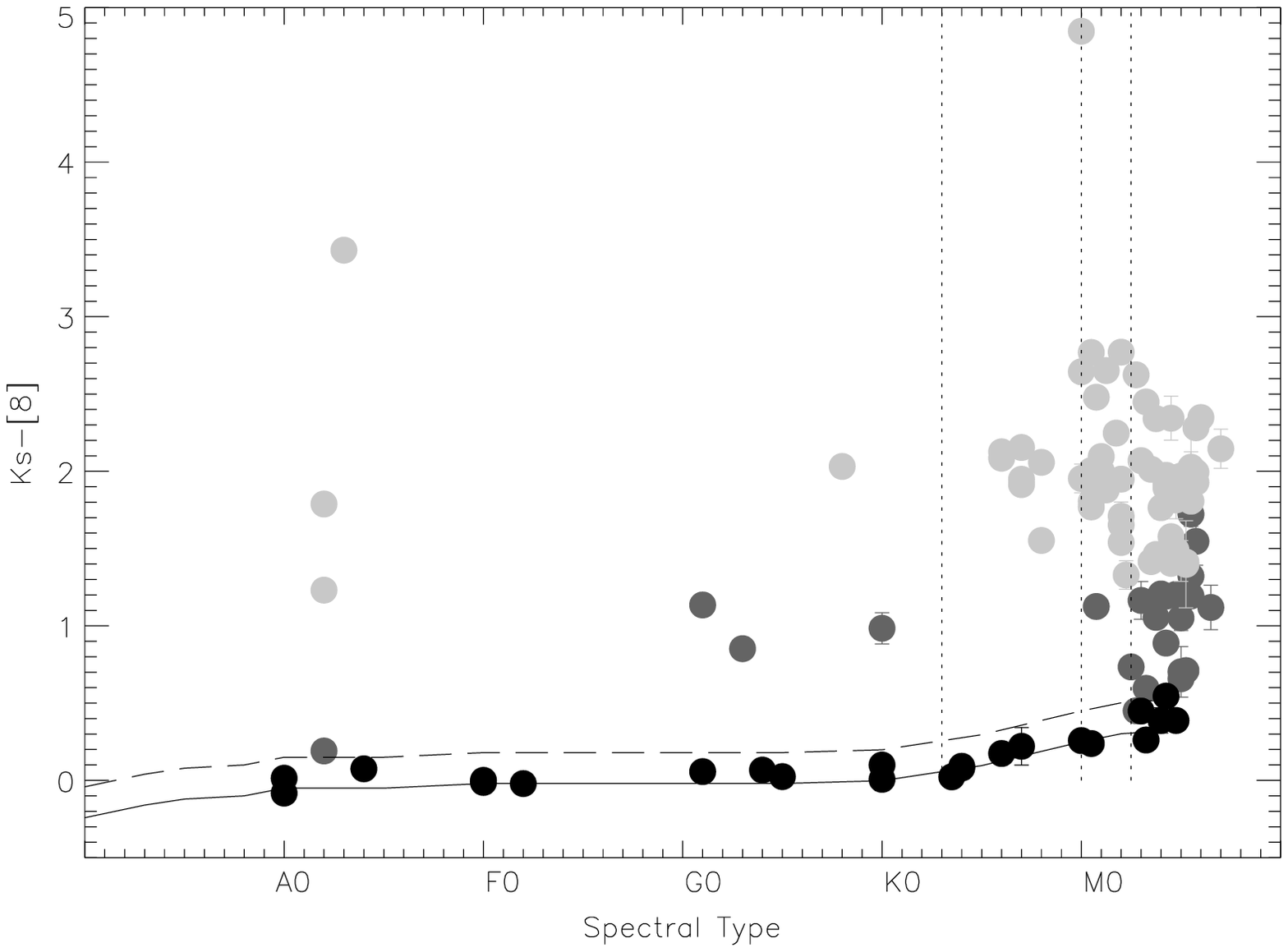}
\includegraphics[scale=0.45]{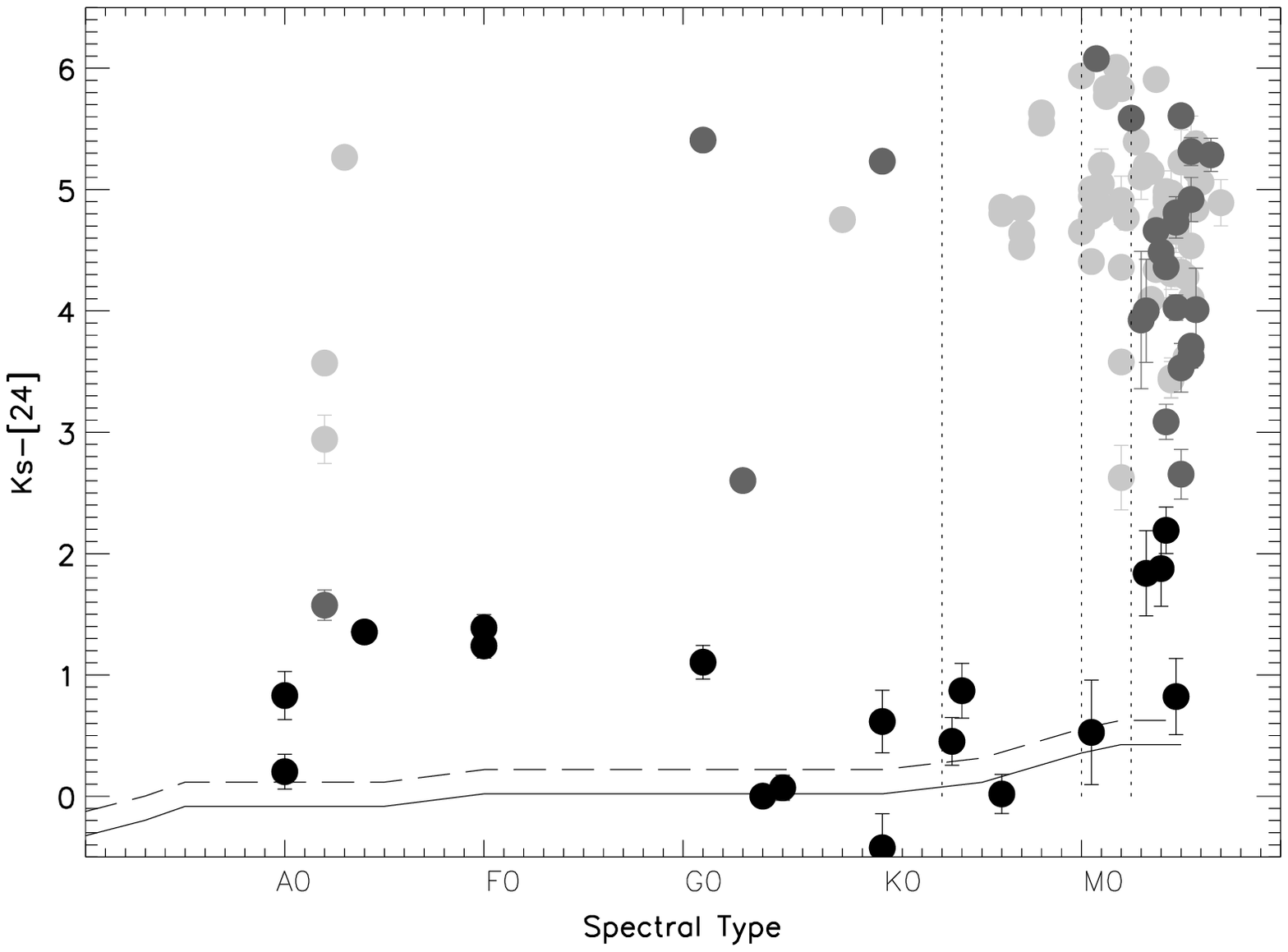}
\caption{K$_{s}$-[8] (left panel) and K$_{s}$-[24] (right panel) colors vs. spectral type for sources with MIPS 24 $\mu m$ detections. 
Sources are divided by their IRAC slope: 'thick' sources are light grey dots, 'anemic' sources are dark grey dots, and 'diskless' 
sources are black dots.  The vertical dotted line divides the sample into stars with probable masses $>$ 1.4 $M_{\odot}$, 0.8--1.4 M$_{\odot}$, 
and $<$ 0.8 M$_{\odot}$.  The horizontal solid line shows the locus of photospheric colors according to the Kurucz-Lejeune stellar atmosphere 
models.  The long-dashed horizontal line identifies sources with a $>$ 20\% IR excess.
}
\label{k8}
\label{k24}
\end{figure}

To investigate the dust luminosity as a function of stellar mass, 
we compare the intrinsic 8 $\mu m$ and 24 $\mu m$ luminosity for each star 
relative to its intrinsic K-band luminosity.
The dereddened K$_{s}$-[8] colors as a function of spectral 
type and IRAC slope are shown in the left panel of Figure \ref{k8}.  
The figure includes photometric errors for each source along with the 
locus for photospheric colors from the Kurucz-Lejeuene stellar atmosphere models.
The diskless, anemic, and thick sources are clearly separated into 
a blue to red sequence of colors across the entire 
range of spectral types.  Diskless sources (black dots) have 
K$_{s}$-[8] colors ranging from $\sim$ 0 to 0.2 from A0 to M6/M7.  
Anemic sources (grey dots) are slightly redder (K$_{s}$-[8] 
$\sim$ 0.2-1.25) and have weak excesses.  Finally, thick sources (light grey dots)
 are systematically redder, with K$_{s}$-[8] $\sim$ 1-3.  

The dereddened K$_{s}$-[8] color qualitatively shows the same separation into thick, anemic, and 
diskless classes as the observed $\alpha_{[3.6]-[8]}$ from \citet{La06}.
Together with \citet{La06}, our results show that the median 8 $\mu m$ excess is larger for later-type 
stars.  This result suggests that inner disks around early type, high-mass stars evolve faster 
than around solar and subsolar-mass stars.

The distribution of dereddened K$_{s}$-[24] colors (right panel) is more complicated. 
There is considerable overlap in 24 $\mu m$ excesses between sources 
with varying levels of 8 $\mu m$ excess, especially for 
stars with spectral types between M1 and M7.  The 'thick' and 'anemic' stars in this 
spectral type range have a wide range (2--6 mag) of 24 $\mu m$ excesses.

The distribution of K$_{s}$-[24] colors suggests that disks around later-type stars 
are more luminous relative to the star even though the contrast between 
the star and a disk of a given dust mass is likely lower for M stars \citep{Er09}.
The fractions of MIPS-detected stars with strong 24 $\mu m$ excesses (K$_{s}$-[24] $>$ 4) 
for stars earlier than K3 ($>$ 1.4 M$_{\odot}$), between K3 and M0 (0.8--1.4 M$_{\odot}$), 
and between M0 and M2.5 (0.5--0.8 M$_{\odot}$) are 5/18 (27.8\%), 7/10 (70\%), and 20/24 (83.3\%).

Despite of the large number of stars lacking detections,  
this general trend clearly holds for stars with masses $\gtrsim$ 0.5 M$_{\odot}$. 
The 24 $\mu m$ upper limits for the 12 stars earlier than K3 without MIPS detections constrain 
the stars' excesses to be $\lesssim$ 3--4 magnitudes.  Assuming that all stars between K3 and M0
currently lacking MIPS detections (10) have weak excesses or no excesses, more than 35\% of these 
stars must have strong excesses (7/20), a percentage that is still 
much higher than that for earlier type, higher-mass stars (4/31; 12.9\%).

To compare the median level of excesses for high/intermediate to low-mass IC 348 stars, including 
sources with 24 $\mu m$ upper limits, we use the ASURV Survival 
Analysis package Rev 1.2 \citep{If90, La92}.  Because our goal is to 
compare the K$_{s}$-[24] colors for two stellar mass bins, we treat the 
data as univariate and compare the two populations using the Gehan's General 
Wilcoxon Test (Permutated Variance and Hypergeometric Variance), the Peto-Prentice 
Generalized Wilcoxon Test, and the Kaplan-Meier Estimator.   
The three Wilcoxon tests yield the probability that the high/intermediate  and low-mass 
 populations are drawn from the same population.
The Kaplan-Meier Estimator yields the likely median value  
of the K$_{s}$-[24] colors for both populations while accounting for non detections.
We divide the sample into stars with spectral types earlier than K5 and 
those with spectral types between K5 and M2.5.  For a 2.5 Myr-old cluster, these spectral type 
ranges correspond to stars with M$_{\star}$ $>$ 1 M$_{\odot}$ and M$_{\star}$ = 0.5--1 M$_{\odot}$ 
\citep{Ba98}.

The ASURV statistical tests indicate that a) the excesses for high/intermedate mass and low-mass stars 
have statistically significant differences and b) the excesses are stronger for low-mass stars.  
The Wilcoxon tests find that the two populations have a probability of less than 0.14\% (0.001-0.0014)
of being drawn from the same parent population.   The Kaplan-Meier estimator yields 
a small median K$_{s}$-[24] color for high/intermediate-mass sources (K$_{s}$ -[24] $\sim$ 1.55 $\pm$ 0.31); the 
median color for low-mass sources is $\sim$ 3.19 $\pm$ 0.40.  Very similar results are obtained if 
the division between samples is made anywhere from G0 to K5.  Thus, the late type, low-mass IC 348 population 
has larger 24 $\mu m$ excesses from disks.  Under the general picture that the luminosity of disks at a given 
wavelength decline with time, these results imply that disks around low-mass stars are less evolved 
than those around high/intermediate-mass stars.

For this analysis to be completely reliable, the censoring of the data must be 'random'. 
If nondetections are primarily due to high nebular background 
emission at the source positions then the censoring is random.  If the sources lacking detections 
are intrinsically fainter, the censoring is not random  About 56\% and 51\% of the stars with spectral types earlier than 
 K5 and between K5 and M2.5 are detected at 24 $\mu m$.  Because these frequencies are comparable,
 the censoring of data \textbf{is} very close to being random for stars earlier 
than M2.5.   The frequency of 24 $\mu m$ detections drops substantially for stars 
later than M2.5 ($\sim$ 30\%).  Comparing the disk luminosities for $>$ 0.5 M$_{\odot}$ stars with those 
for even lower-mass stars is then impeded by sample incompleteness for the latter population.

\subsection{Evolutionary State of Disks vs. Stellar Mass in IC 348}
We now consider the frequencies of disks in different evolutionary states 
as a function of stellar mass.  For all stars, we quantify 
the evidence for disk emission at 24 $\mu m$.  Figure \ref{colmagall} suggests 
that stars with K$_{s}$- [24] colors $\gtrsim$ 1$\sigma(K_{s},[24])$ 
away from the distribution of photospheric sources have significant excesses.  
Sources with smaller dereddened K$_{s}$-[24] colors are then identified as diskless photospheres.
To guide our identification of primordial disks, transitional disks, and debris disks, 
 we use the results from \S 3.3.3 and \S 3.4.  For 
stars later than K0, we use the semi-empirical division based 
on Figure \ref{colcolevo} to identify primordial disks and transitional disks\footnote{Within 
the transitional disk category, we include sources first identified in \S 3.3.3 as debris disk 
candidates from Figure \ref{colcolevo} because the dust around these sources need not be 2nd generation (see \S 3.4.3).  
Separating these debris disk candidates from transitional disks does not qualitatively change our results.}.   
For stars earlier than K0, we use the results of \S 3.3.3 and \S 3.4 to select 
primordial disks, transitional disks, and debris disks. 

\begin{figure}
\centering
\plotone{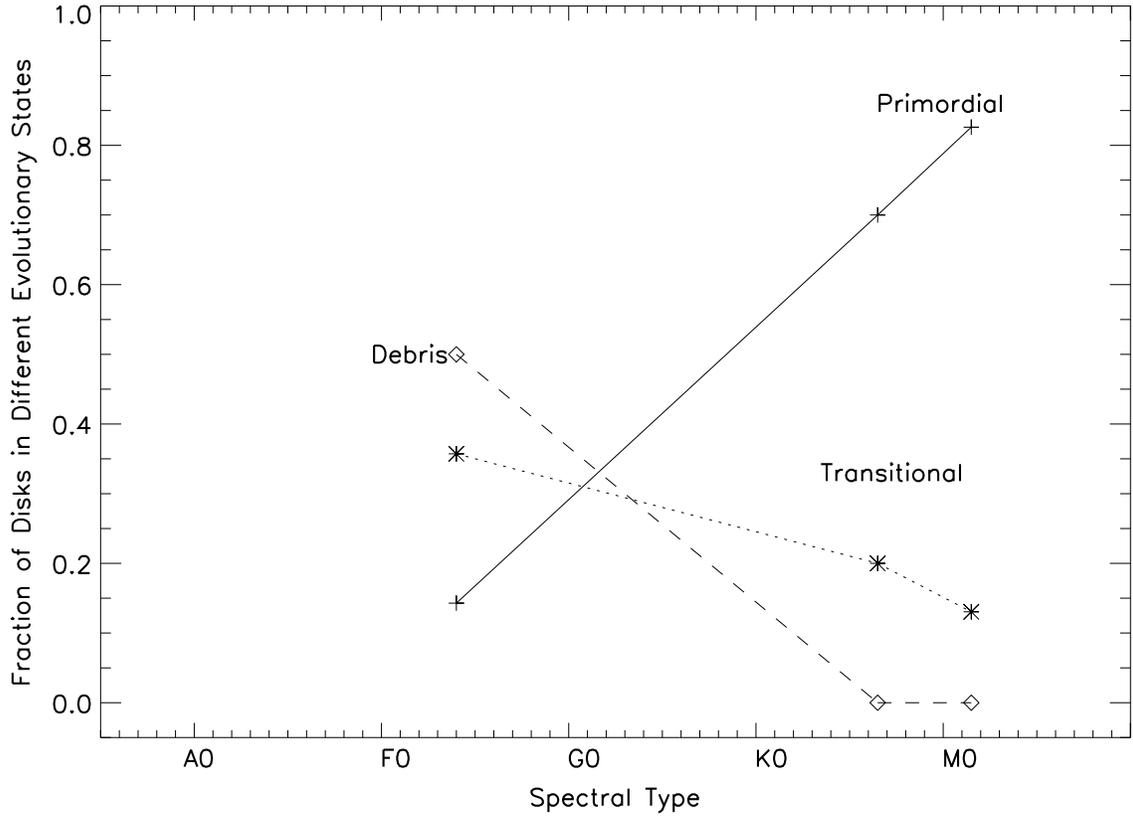}
\caption{Disk evolutionary state from mid-IR colors as a function of stellar mass.  The frequencies are given for 
three stellar mass bins (from left to right): MIPS-detected stars with probable masses $>$ 1.4 M$_{\odot}$, 
 0.8--1.4 M$_{\odot}$, and 0.5--0.8 M$_{\odot}$.}
\label{evostateall}
\end{figure}

Figure \ref{evostateall} shows the frequency of primordial disks, transitional disks, 
and debris disks as a function of spectral type/stellar mass.  We divide the sample into the same three stellar mass 
bins used to compute the gas accretion frequency: $>$ 1.4 M$_{\odot}$ (earlier than K3), 0.8--1.4 M$_{\odot}$ 
(K3--M0), and 0.5--0.8 M$_{\odot}$ (M0-M2.5).  We do not compute the relative disk fractions for later type, lower-mass 
stars where completeness at 24 $\mu m$ drops to $\approx$ 30\%.  

Based on our analysis, the disk evolutionary states in IC 348 depend on stellar spectral type.  
Half of the MIPS-detected disks around intermediate/high-mass stars are likely debris disks (dashed line/diamonds).  
Primordial disks (solid line/crosses) comprise about 15\% of the total disk population.  
The MIPS-detected disk populations for stars with probable masses 
between 0.5 M$_{\odot}$ and 1.4 M$_{\odot}$ show very different distributions.
Primordial disks comprise greater than 70\% of the MIPS-detected disk population for stars 
in this mass range.  The fact that stars in this mass range have (relative to the stellar photosphere) 
more luminous disks than their high/intermediate-mass counterparts is consistent with 
our determination that these stars typically have primordial disks.

The frequency of transitional disks (dotted line) for the MIPS-detected disk population 
ranges from $\sim$ 15\% for 0.5--0.8 M$_{\odot}$ stars, to 20\% for 0.8--1.4 M$_{\odot}$ stars, 
to 35\% for high/intermediate-mass stars.  This frequency is slightly higher than that computed for 
Taurus-Auriga ($<$ 10\%).   Moreover, the transition disk frequency 
 \textit{exceeds} the frequency of primordial disks for high/intermediate-mass stars, a result 
similar to that found for solar/subsolar-mass stars in 5 Myr-old NGC 2362 \citep{Clp09}.  
Depending on the nature of IC 348 sources currently lacking MIPS detections, mid-IR colors 
presented in Figure \ref{k8} indicate that the transition disk 
population for the lowest-mass stars (M$_{\star}$ $<$ 0.5 M$_{\odot}$) may also 
comprise a significant fraction of the total disk population ($\gtrsim$ 30\%).

\section{Summary and Discussion}
\subsection{Summary of Results}
Using new optical spectra and new, deep MIPS 24 $\mu m$ and 70 $\mu m$ photometry we investigated the 
disk population of the 2.5 Myr-old IC 348 Nebula.  Combining these data with previous work from \citet{La06}, 
we analyzed optical spectroscopic data for all stars, performed SED modeling, and examined the mid-IR 
colors of MIPS-detected members to constrain the disks' evolutionary states and probe how disk properties 
vary with stellar mass.  Our study yields the following major results:
\begin{itemize}
\item IC 348 sources with MIPS-70 $\mu m$ detections include flat-spectrum protostars and 
pre-main sequence stars with optically-thick, luminous far IR disk emission.  
Some optically-thick disks show evidence for depleted inner disks; others do not.

\item IC 348 stars with new and previous MIPS detections have disks in evolutionary states 
ranging from primordial disks to debris disks. In agreement with recent work \citep{La06, He07a, Clp09}, we find 
evidence for two separate pathways from primordial disks to debris disks.  In homologously depleted disks, 
 disks deplete their reservoir of small dust grains at all disk locations simultaneously.  In a second 
sequence ('disks with inner holes'), disks clear their supply of small dust grains from the inside out.

\item At a $\sim$ 1--2 $\sigma$ significance, signatures of circumstellar gas accretion are more frequent for solar 
and subsolar-mass stars than for high/intermediate-mass stars.  This result 
is consistent with a gas disk dispersal timescale that is shortest for high/intermediate-mass stars.

\item The mid-IR disk luminosities of MIPS-detected disks are stellar-mass dependent.  Relative 
to the stellar photosphere, 24 $\mu m$ emission from disks is lower for high/intermediate-mass stars than 
for solar/subsolar-mass stars.  If disks generally decline in luminosity as a function of time, this result 
implies that disks around high/intermediate-mass stars evolve faster.

\item The evolutionary states of MIPS-detected disks are also stellar mass dependent.  Most MIPS-detected disks surrounding high/intermediate-mass stars 
stars appear to be debris disks; primordial disks comprise only $\sim$ 15\% of the disk population for M$_{\star}$ $>$ 1.4 M$_{\odot}$.  
In contrast, most MIPS-detected disks around solar and subsolar-mass stars are primordial disks.  For stars of all masses, 
transitional disks (homologously depleted or disks with inner holes) comprise $\sim$ 15--35\% of the MIPS-detected disk population.
\end{itemize}

\subsection{The Evolutionary State of \textit{Anemic} Disks}
Our SED modeling and investigation of the mid-IR colors of IC 348 stars 
clarifies the nature of \textit{anemic} disks identified 
by \citet{La06}.  Among late type stars, most anemic disks 
are homologously depleted disks or are disks with inner holes.  
Several anemic disks are probably primordial disks.
Among earlier-type stars with probable masses $\gtrsim$ 1.4 M$_{\odot}$, 
anemic disks comprise a broader range of evolutionary states.
  At least two anemic disks (surrounding IDs 6 and 8) are probably 
debris disks.  In contrast, ID-31 has strong mid-IR emission and 
gas accretion signatures more consistent with a transitional disk, specifically one with 
an inner hole.  

Thus, as pointed out by \citet{Clp09}, a single IRAC slope serves as a useful first-order 
probe of disk properties, but a full analysis of IRAC \textit{and} MIPS data
combined with gas accretion signatures is required for an accurate taxonomy of disks.  
Our results indicate that a full analysis is especially necessary if the stellar population 
includes both high/intermediate-mass stars and lower-mass stars.  Given the diversity 
in disk properties for 2.5 Myr-old IC 348, we suggest that a single flux slope may be 
best suited as a probe of disk evolution in the youngest regions (e.g. Taurus; NGC 1333), 
where less diversity in disk states (e.g. few if any debris disks) is expected.

\subsection{Constraints on the Primordial-to-Debris Disk Transition}
Recent Spitzer studies have placed strong constraints on the timescales for the evolution of 
primordial disks into debris disks.  By 5 Myr, most high/intermediate-mass 
stars and solar-mass stars either lack evidence for 
a disk or have disk properties suggestive of a debris disk \citep{Ca06, He08, Clp09}.  At 
this age, many (detectable) subsolar-mass stars have weaker levels of mid-IR 
disk emission than typical primordial disks \citep{Clp09,Dahm09}.  

This work shows that by 2.5 Myr most disks around high/intermediate-mass stars are either actively leaving 
the primordial disk phase or have already reached the debris disk phase.  Together with previous 
results \citep[e.g.][]{Ca06, He07a, Clp09}, our results clearly support a stellar-mass dependent timescale 
for the disappearance of primordial disks and the emergence of debris disks.   Within the context of 
planet formation, gas giant planets around high/intermediate-mass stars have much less 
time to form than around low-mass stars.  If gas giant planets are more frequent 
around high/intermediate-mass stars \citep{Jj07}, their formation must be very rapid 
and efficient.  Only recently have realistic models \citep{Kb09} that form gas giant planets via core accretion been 
successful in forming the cores of such planets in the short timescales implied from this paper (2.5 Myr).
Some trends in exoplanet properties, such as the semimajor axis distribution, may emerge from the 
competing effects of core formation timescales and primordial disk dispersal timescales \citep{Currie2009}.

The fraction of IC 348 disks in a transitional phase serves as 
a useful contrast to results obtained for younger clusters like Taurus and older clusters 
like NGC 2362.  Based primarily on IRAS data, the computed frequency of transitional disks 
in Taurus is small, $<<$ 10\% \citep[e.g.][]{Sk90, Sp95, Ww96}.
  These authors then argued that the lifetime of transitional disks is $\approx$ 1--10\% the age of Taurus: $\approx$ 0.01--0.1 Myr.  
If this inference were correct, the total disk lifetime may be several Myr, but most of the disk is dissipated rapidly.

However, analysis of Spitzer data for 5 Myr-old NGC 2362 finds a much higher frequency of transitional 
disks \citep{Clp09}.  Both types of transitional disks (inner holes, homologously depleted) greatly outnumber primordial disks.  
This result argues for far longer typical transition timescale ($\approx$ 1 Myr).

In the intermediate age IC 348, the frequency of transition disks is intermediate between the frequency 
for 1 Myr-old Taurus and 5 Myr-old NGC 2362 for solar/subsolar-mass stars.  
Thus, the number of transition disks relative to primordial disks around solar/subsolar-mass stars 
appears to be an increasing function of stellar age.  This result is expected if the typical transition 
timescale is an appreciable fraction of the typical primordial disk lifetime.  Thus, the transition disks 
in Taurus and IC 348 could remain in such a state for an extended period.  However, it is possible 
that transition disk lifetimes, like primordial disk lifetimes, also have an intrinsic dispersion (i.e., a 
gaussian distribution of lifetimes), where sources in Taurus represent those that make the 
primordial-to-debris disk transition fastest and sources in 5--10 Myr-old clusters make the transition the slowest.  
Quantifying the range of times a disk spends in a transitional phase and whether the typical 
timescale depends on stellar mass requires observations of many more 1--10 Myr-old clusters.

While debris disks typically only have excess emission longwards of $\approx$ 20 $\mu m$, the debris 
disk population in IC 348 contains at least two sources (ID-6 and ID-8) that have warmer dust 
more indicative of terrestrial planet formation than icy planet formation.  
Thus, warm debris disks consistent with the observable signatures of terrestrial planet 
formation may emerge as early as $\approx$ 2.5 Myr around high/intermediate-mass stars.  
Observations of more 2--5 Myr-old clusters are needed to determine if the frequency of 
warm debris disks is higher than the $\sim$ 4\% derived for 10--15 Myr-old clusters \citep{Cu07a, Cu08}.
The existence of both warm debris disks and colder debris disks (lacking IRAC excess emission) suggests 
that even at 2.5 Myr debris disk populations may exhibit a range of dust temperatures consistent with 
a range of locations over which the debris-producing stages of planet formation are ongoing.

\section*{Acknowledgements}
We thank Taku Takeuchi for extensive discussions on the dynamics of circumstellar 
dust in optically-thin disks with and without circumstellar gas and for confirming
 the validity of the arguments presented in \S 3.4.3.  We also thank Perry Berlind, Gautum Narayan, and 
Nathalie Martinbeau  for scheduling, taking, and reducing spectra of IC 348 stars on short notice.  
Jesus Hernandez provided valuable discussions on Cycle 5 MIPS data for IC 348.
Finally, we thank Charles Lada for informative discussions about previous Spitzer observations of IC 348.
This work is supported by Spitzer GO grant 1320379, NASA Astrophysics Theory grant NAG5-13278, 
NASA grant NNG06GH25G, and Smithsonian Institution Restricted Endowment funds.

\clearpage
\begin{deluxetable}{lllllllllllll}
\tiny
\setlength{\tabcolsep}{0.02in}

\tablecolumns{13}
\tabletypesize{\scriptsize}
\tablecaption{Spectroscopic data for Selected IC 348 sources}
\tablehead{{ID Number}&{Spectral}&{IRAC}&{SNR}&{EW} &{Emission?}&{FWHM} 
& {EW H$_{\alpha}$ }&{R$_{c}$(est.)} & {R$_{\star}$} & {M$_{\star}$} &{Flux(H$_{\alpha}$)} 
& {$\dot{M}/\dot{M}_{ID-21}$}\\
{} & {Type} &{Slope}& {}   &{H$_{\alpha}$(\AA)} & {}& {H$_{\alpha}$(\AA)} 
&{Em.(\AA)}&{}&{(R$_{\odot}$)} &{(M$_{\odot}$)} &{(W m$^{-2}$)}}
\startdata
2 & A2 & thick & 143 & -10.7 & maybe &-- &$<$ 0.015 & 7.2 & 4.5 & 3.1 &$<$ 3.8$\times$10$^{-17}$ & $<$0.07\\
3 & A0 & diskless & 170 & -6.3 & no &--&$<$ 0.02 & 7.3 & 4.0 & 3.0& $<$ 4.6$\times$10$^{-17}$ & $<$0.08\\
6 & G3 & anemic & 105 & -1.0& probably not&--& $<$ 0.1 & 9.0 & 3.9 & 2.3 & $<$ 4.8$\times$10$^{-17}$ & $<$ 0.10\\
8 & A2 & anemic & 156 & -10.9 & no &--&$<$0.02& 8.51 & 2.4 & 2.3 & $<$1.5$\times$10$^{-17}$&$<$0.02\\
19 & A2 & thick & 132 & -10.0 & yes & 2.9&0.64&9.15&1.7 & 2.1&2.6$\times$10$^{-16}$&0.3\\
21 & K0 & anemic & 80 & 5.0 & yes & 7.8 &5.0&10.48&2.5&1.8&6$\times$10$^{-16}$&1\\
30 & F0 & diskless & 123 & -6.2 & no & --&$<$0.02&10.06&1.6&1.6&$<$3.5$\times$10$^{-18}$&$<$0.004\\
\enddata
\label{spectra}
\tablecomments{Negative EW(H$_{\alpha}$) indicate absorption and positive 
values indicate emission.  IC348 ID-19 has a central emission core in
 the deep absorption line.  The H$_{\alpha}$ line for ID-6 is weak 
for G3 stars, though there is no clear central emission feature and the 
star is very x-ray luminous.
}
\end{deluxetable}

\begin{deluxetable}{lllllllllllll}
\tiny
\setlength{\tabcolsep}{0.02in}
\tabletypesize{\tiny}
\tabletypesize{\scriptsize}

\tablecolumns{13}
\tabletypesize{\scriptsize}
\tablecaption{IC 348 Stars with New MIPS 24 $\mu m$ Detections}
\tablehead{{ID}&{RA}&{DEC}&{Offset(")}&{Spectral Type}&{IRAC Slope}&{[24]}&{$\sigma$[24]}&{K$_{s}$-[24] (dereddened)}}
 \startdata
   4 & 56.1300 & 32.1061 & 1.80 & F0 & 3 & 6.402 & 0.097 & 1.24\\
  39 & 56.2573 & 32.2411 & 3.68 & K4 & 3 & 8.818 & 0.225 & 0.87\\
  47 & 55.9813 & 32.1590 & 1.43 & K0 & 3 & 10.488 & 0.280 & -0.42\\
  49 & 55.9900 & 32.0271 & 3.72 & M0.5 & 3 & 8.076 & 0.430 & 0.53\\
  53 & 56.0684 & 32.1653 & 0.29 & K0 & 3 & 9.545 & 0.257 & 0.62\\
  56 & 56.0208 & 32.1649 & 1.62 & K3.5 & 3 & 9.838 & 0.195 & 0.45\\
  83 & 56.1559 & 32.1503 & 0.18 & M1 & 1 & 5.733 & 0.286 & 5.05\\
  85 & 56.1172 & 32.2668 & 0.45 & M3.25 & 3 & 9.078 & 0.350 & 1.84\\
  88 & 56.1365 & 32.1544 & 2.97 & M3.25 & 2 & 6.819 & 0.425 & 4.00\\
  91 & 56.1634 & 32.1624 & 1.90 & M2 & 1 & 8.264 & 0.265 & 2.63\\
  95 & 56.0913 & 32.2032 & 1.16 & M4 & 3 & 9.118 & 0.311 & 1.88\\
  124 & 55.9776 &32.0084 & ---    & M4.25 & 3 & 9.144 & 0.190 & 2.19\\
  167 & 56.1716 & 32.1695 & 0.47 & M3 & 2 & 7.246 & 0.565 & 3.93\\
  168 & 56.1306 & 32.1797 & 1.15 & M4.25 & 1 & 7.145 & 0.263 & 4.37\\
  193 & 56.1584 & 32.1936 & 0.85 & M4 & 1 & 7.141 & 0.182 & 4.76\\
  261 & 55.9526 & 32.2308 & 1.96 & M5 & 2 & 10.037 & 0.203 & 2.65\\
  300 & 56.1624 & 32.0555 & 1.49 & M5 & 1 & 7.735 & 0.265 & 5.23\\
  336 & 56.1349 & 32.0576 & 0.95 & M5.5 & 1 & 8.058 & 0.335 & 5.27\\
  341 & 56.0541 & 32.2210 & 1.08 & M5.25 & 1 & 8.853 & 0.304 & 4.28\\
  365 & 56.0426 & 32.1263 & 1.26 & M5.75 & 2 & 9.383 & 0.340 & 4.01\\
  373 & 56.1166 & 32.0888 & ---    & M5.5 & 2 & 8.418 & 0.107 & 5.31\\
  402 & 56.1898 & 32.3055 & 0.41 & M5.5 & 2 & 8.611 & 0.174 & 4.92\\
  407 & 56.2672 & 32.0846 & 0.25 & M7 & 1 & 9.578 & 0.164 & 4.89\\
  746 & 56.2082 & 32.1041 & 1.40 & M5 & 1 & 7.683 & 0.260 &?\\
  1124 & 56.2364 & 32.2844 & 0.86 & M5 & 2 & 8.406 & 0.198 & 3.53\\
  1939 & 56.2198 & 32.0158 & 1.11 & M4.75 & 3 & 10.584 & 0.313 & 0.82\\
  2096 & 56.0539 & 32.2234 & 0.39 & M6 & 1 & 8.713 & 0.233 &?\\
\enddata
\label{m24det}
\tablecomments{The IRAC slope follows the convention of \citet{La06}: 1 = 'thick', 2 = 'anemic', 
and 3 = 'diskless'.  The equivalent designations in \citet{Clp09} are the following: 
1 = 'strong IRAC', 2 = 'weak IRAC', and 3 = 'photosphere'.
}
\end{deluxetable}

\begin{deluxetable}{lllllllllllll}
\tiny
\setlength{\tabcolsep}{0.02in}
\tabletypesize{\tiny}
\tabletypesize{\scriptsize}

\tablecolumns{4}
\tabletypesize{\scriptsize}
\tablecaption{IC 348 Stars With MIPS 70 $\mu m$ Detections}
\tablehead{{IC348-ID}&{Spectral Type}&{Source Type} & {F$_{70}$ (mJy)}& 
{$\sigma$(F$_{70}$) (mJy)} & {Offset(")}}
 \startdata
13    & M0.5&1 & 1408 & 103.7 & 2.85\\
31    & G1 & 2& 556.2 & 140.3 & 2.86\\
51    & -- & 1 & 3972.0 & 127.0 & 1.74\\
67    & M0.75 & 2& 125.5 & 18.4 & 2.08\\
10343 & M3.75 & 1& 115.6 & 15.9 & 1.96\\
10352 & M1 & 1 & 322.1 & 25.1 &2.37\\
\enddata
\label{m70det}
\end{deluxetable}

\begin{deluxetable}{llllllllllllllllllllllll}
\tiny
\setlength{\tabcolsep}{0.025in}
\tabletypesize{\tiny}
\tabletypesize{\scriptsize}
\tablecolumns{24}
\tabletypesize{\scriptsize}
\tablecaption{Catalog of IC 348 Stars}
\tablehead{{ID}&{RA}&{DEC}&{J}&{H}&{K$_{s}$}&{[3.6]}&{$\sigma$[3.6]}&{[4.5]}&{$\sigma$[4.5]}
 & {[5.8]} &{$\sigma$[5.8]} &{[8]} & {$\sigma$[8]} & {[24]} &{$\sigma$[24]} & {[70]} & {$\sigma$[70]} & {ST} & {A$_{V}$}& {IRAC Slope}}
 \startdata
1&56.1425& 32.1630&  6.790& 6.640& 6.510& 6.740& 0.010& 6.540& 0.020& 6.580& 0.020& 6.500& 0.030& 1.874& -9& -2.267& -9& 15& 3.10& 3\\
2&56.1474& 32.1679&  7.950& 7.530& 7.250& 7.090& 0.020& 6.810& 0.020& 6.460& 0.040& 5.820& 0.040& 3.375& 0.038& -2.176& -9& 22& 3.20& 1\\
3&56.2110& 32.3185&  8.370& 7.900& 7.660& 7.530& 0.010& 7.610& 0.030& 7.470& 0.040& 7.500& 0.040& 7.086& 0.141& -1.009& -9& 20& 3.90& 3\\
4&56.1300& 32.1061&  8.340& 8.020& 7.860& 7.790& 0.010& 7.740& 0.020& 7.660& 0.020& 7.730& 0.030& 6.402& 0.097& -0.898& -9& 30& 2.30& 3\\
5&56.1084& 32.0751&  10.070& 8.870& 8.140& 6.970& 0.010& 6.520& 0.010& 6.320& 0.020& 5.630& 0.020& 2.655& 0.012& -0.699& -9& 48& 7.70& 1\\
\enddata
\tablecomments{Full catalog of IC 348 sources from \citet{La06} including MIPS photometry for newly-detected stars from this work.  
The column labeled "ST" indicates the numerical spectral type (e.g., 15=B5, 22=A2).}
\label{mcat}
\end{deluxetable}

\begin{deluxetable}{lllllllllllll}
\tiny
\setlength{\tabcolsep}{0.02in}
\tabletypesize{\tiny}
\tabletypesize{\scriptsize}

\tablecolumns{13}
\tabletypesize{\scriptsize}
\tablecaption{Evolutionary State of Disks with New 24 $\mu m$ Detections}
\tablehead{{ID}&{Spectral Type}&{IRAC Slope}&{K$_{s}$-[8] (dereddened)}&{K$_{s}$-[24] (dereddened)}&{Disk State}}
 \startdata
     4& F0&3& -0.01&  1.24&4\\
    39& K4&3&  0.08&  0.87&4\\
    47& K0&3&  0.10& -0.42&5\\
    49& M0.5&3&  0.24&  0.53&5\\
    53& K0&3&  0.01&  0.62&4\\
    56& K3.5&3&  0.02&  0.45&4\\
    83& M1&1&  2.00&  5.05&1\\
    85& M3.25&3&  0.26&  1.84&2\\
    88& M3.25&2&  0.60&  4.00&3\\
    91& M2&1&  1.71&  2.63&2\\
    95& M4&3&  0.39&  1.88&2\\
   124& M4.25&3&  0.55&  2.19&2\\
   167& M3&2&  1.16&  3.93&2\\
   168& M4.25&1&  1.91&  4.37&1\\
   193& M4&1&  1.76&  4.76&1\\
   261& M5&2&  0.66&  2.65&2\\
   300& M5&1&  1.97&  5.23&1\\
   336& M5.5&1&  1.99&  5.27&1\\
   341& M5.25&1&  1.40&  4.28&1\\
   365& M5.75&2&  1.55&  4.01&1\\
   373& M5.5&2&--&  5.31&3\\
   402& M5.5&2&  1.32&  4.92&1\\
   407& M7&1&  2.15&  4.89&1\\
   746& M5&1&--& --&-\\
  1124& M5&2&  0.70&  3.53&2\\
  1939& M4.75&3&  0.39&  0.82&4\\
  2096& M6&1&--& --&-\\
\enddata
\label{tabevo24}
\tablecomments{The IRAC slope follows the convention of \citet{La06}: 1= 'thick', 2= 'anemic', 
and 3 = 'diskless'.  The disk states are 1 = primordial, 2 = homologously depleted, 3 = disks with inner holes, 
4 = debris disk candidates, 5 = photospheres (no disk).
}
\end{deluxetable}

\begin{deluxetable}{lllllllclcll}
 \tiny
\setlength{\tabcolsep}{0.01in}
\tabletypesize{\tiny}
\tablecolumns{10}
\tablecaption{Dust Removal Timescales and Dust Masses for Representative Debris Disk Candidates}
\tablehead{{ID}&{IRAC$^{a}$}&{Spectral} & {log({L$_{\star}$}/{L$_{\odot}$})}&{M/M$_{\odot}$}& 
{t$_{pr}$ (Myr)}&{s$_{max}$}&{Disk$^{b}$}&{ M$_{dust}$}\\
{}&{Slope}&{Type}&{}&{}&{(1, 10, 30 AU)} &{($\mu m$)}&{Model}&{($\times$10$^{-10}$M$_{\odot}$)}\\}
\startdata
\\
6&Anemic&G3&1.209&2.6&4.3$\times$10$^{-5}$, 4.3$\times$10$^{-3}$, 0.039&7.1&3006897-5-9&8.77\\
8&Anemic&A2&1.509&2.8&2.2$\times$10$^{-5}$, 2.2$\times$10$^{-3}$, 0.020&13.1&3004289-5-9&2.48\\
20&Diskless&G1&0.719&2.6&1.3$\times$10$^{-4}$, 0.013, 0.120&2.3&3001439-5-9&2.09\\
30&Diskless&F0&0.806&2.6&1.1$\times$10$^{-4}$, 0.011, 0.098&2.8&3003786-5-9&2.95\\
\\
39&Diskless&K4&0.229& 1.3& 4.1$\times$10$^{-4}$, 0.041, 0.371& 1.5&3008796-8-9&20.0\\
1939&Diskless&M4.75&-0.792& 0.2& 4.4$\times$10$^{-3}$, 0.44, 3.9& 0.98 & 3002293-6-9& 14.0\
\enddata
\label{noremnant}
\tablecomments{Constraints on the dust removal timescales and disk masses for debris disk candidates.
  The Poynting-Robertson drag timescale is calculated assuming 1 $\mu m$-sized grains.  Disk models 
listed represent the file number from the \citet{Rob06} grid; dust masses for those models are listed as 
M$_{dust}$.
a) From \citet{La06}.
b) From \citet{Rob06}.} 
\end{deluxetable}

\clearpage



\clearpage
\appendix
\section{Atlas of SEDs for IC 348 Stars with New MIPS 24 $\mu m$ Detections}
\begin{figure}
\epsscale{0.9}
\centering
\plottwo{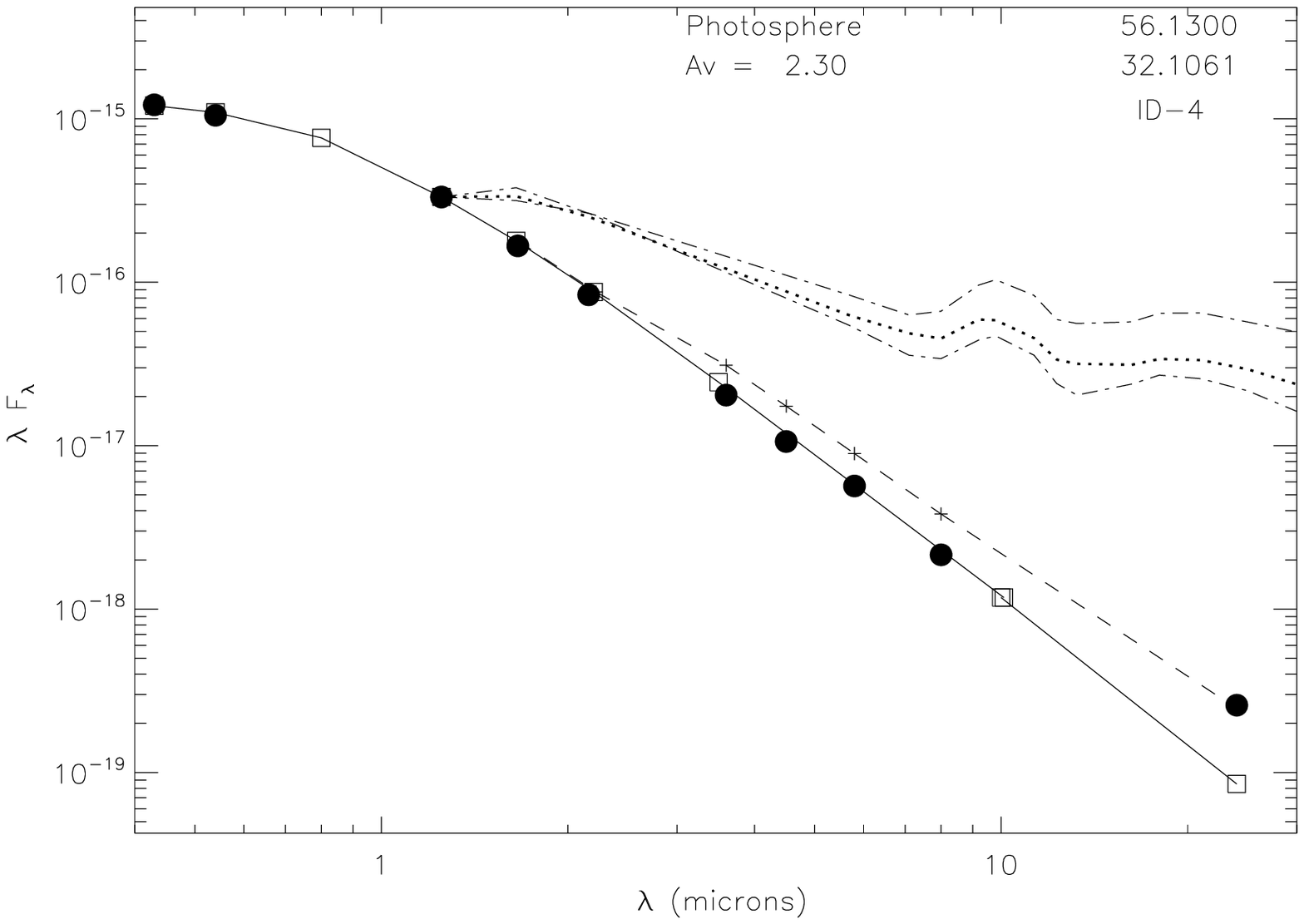}{ID-39.ps}
\plottwo{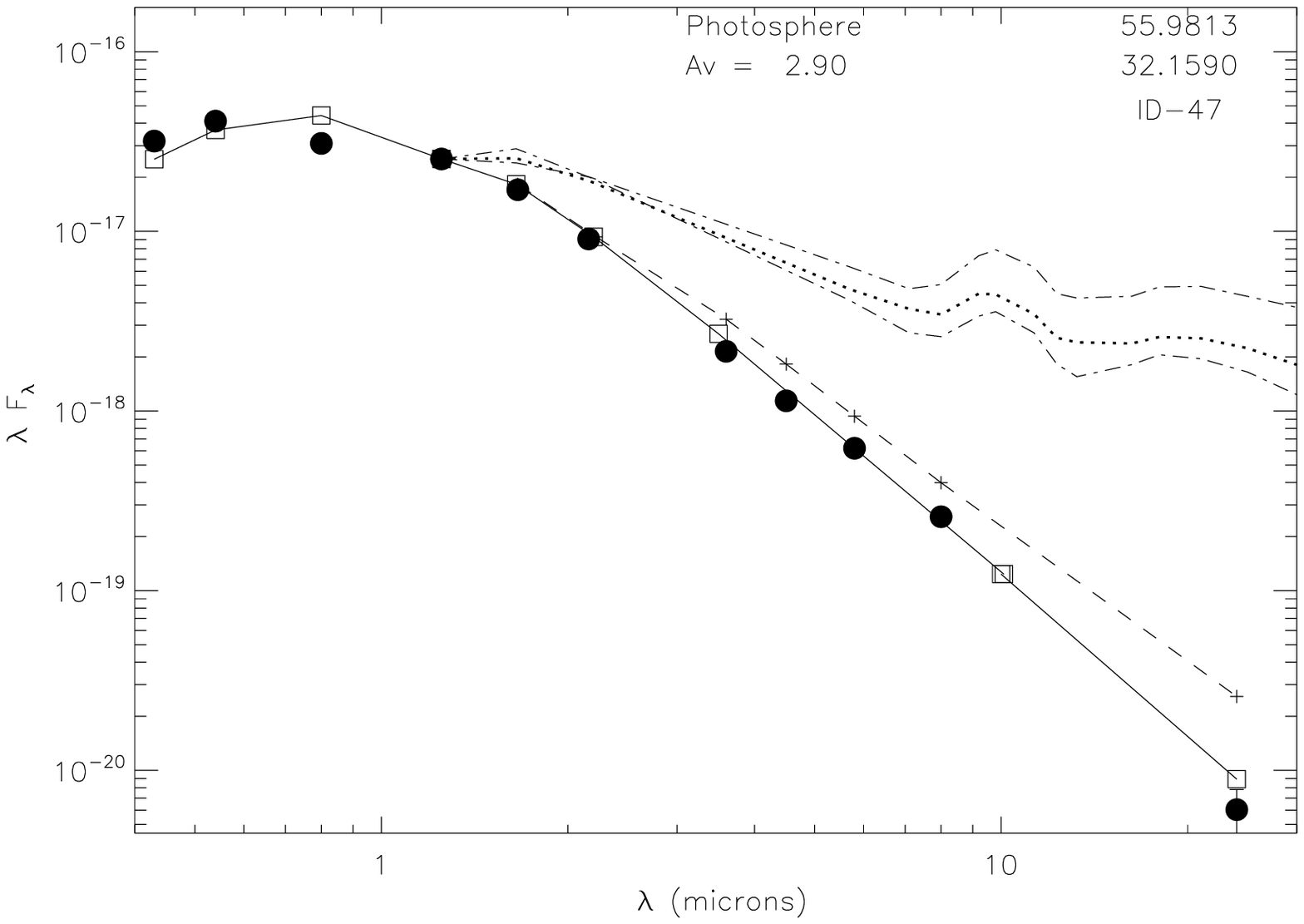}{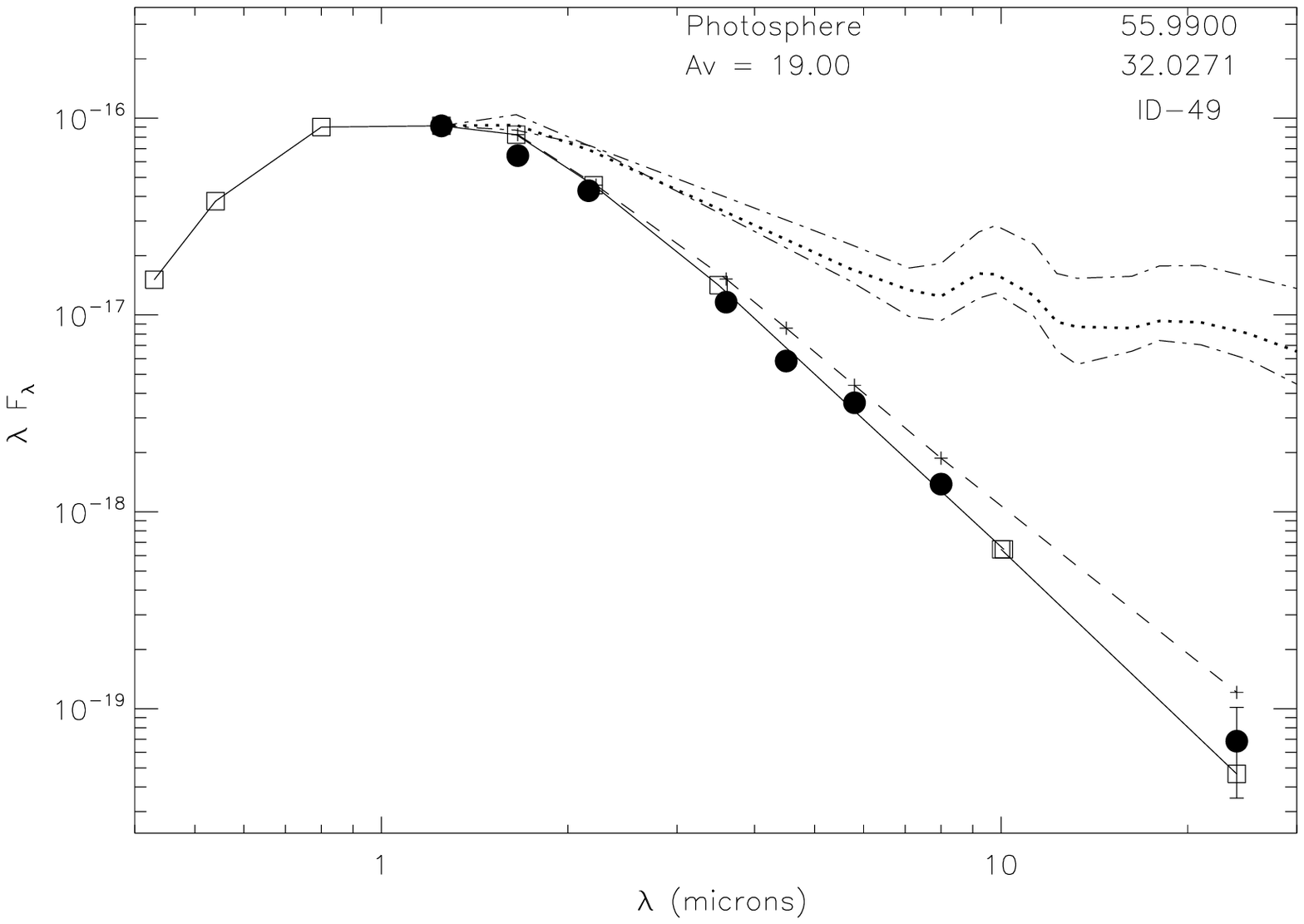}
\caption{Atlas of spectral energy distributions of IC 348 stars with new MIPS 24 $\mu m$ 
detections.  Overplotted are the median Taurus SED (dotted line) with upper and lower quartiles (dot-dashed line).
A terrestrial zone debris disk model is shown as a dashed line with mid-IR flux slightly greater
than the photosphere (solid line, connected by open squares).
}
\end{figure}
\begin{figure}
\epsscale{0.9}
\centering
\plottwo{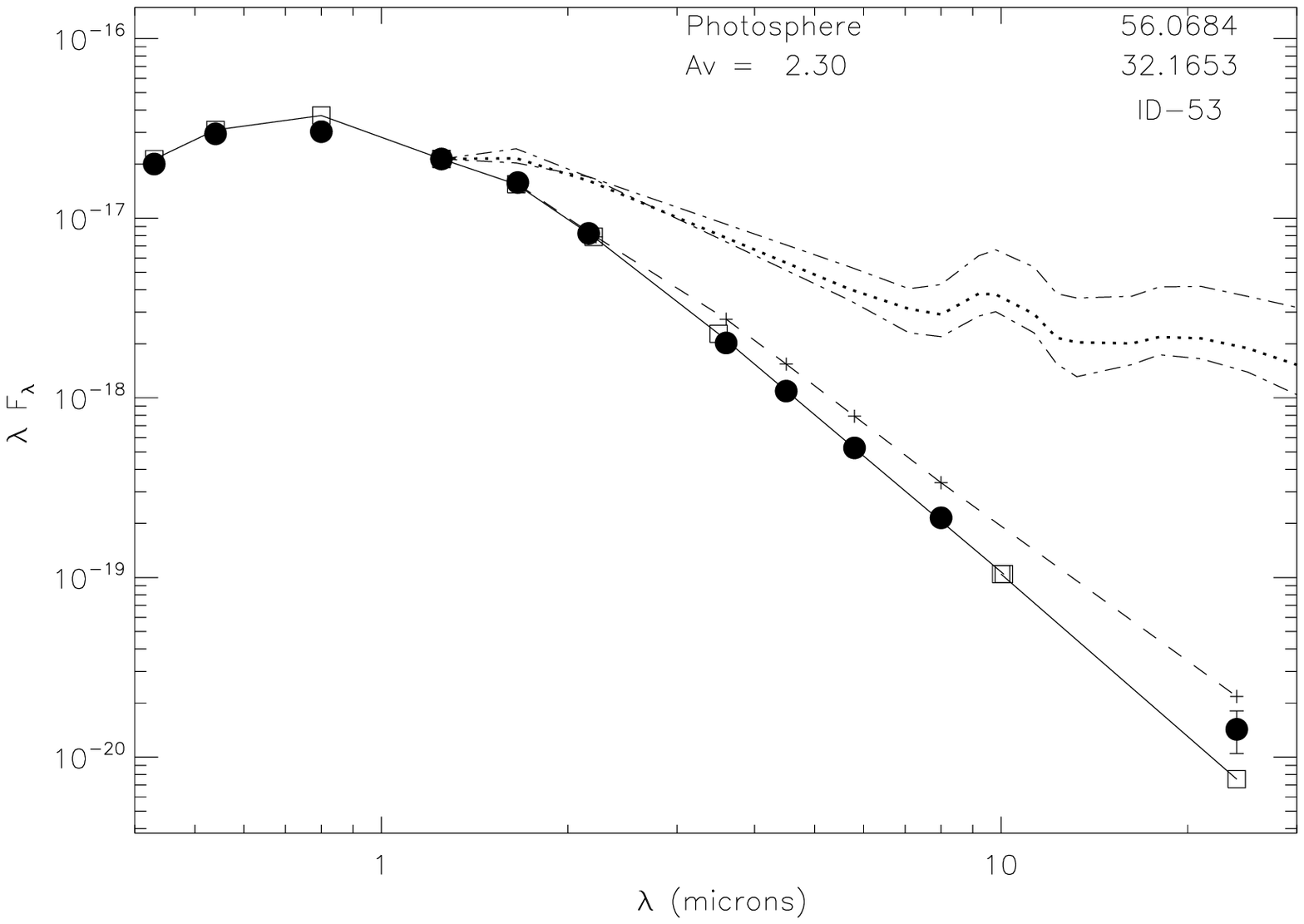}{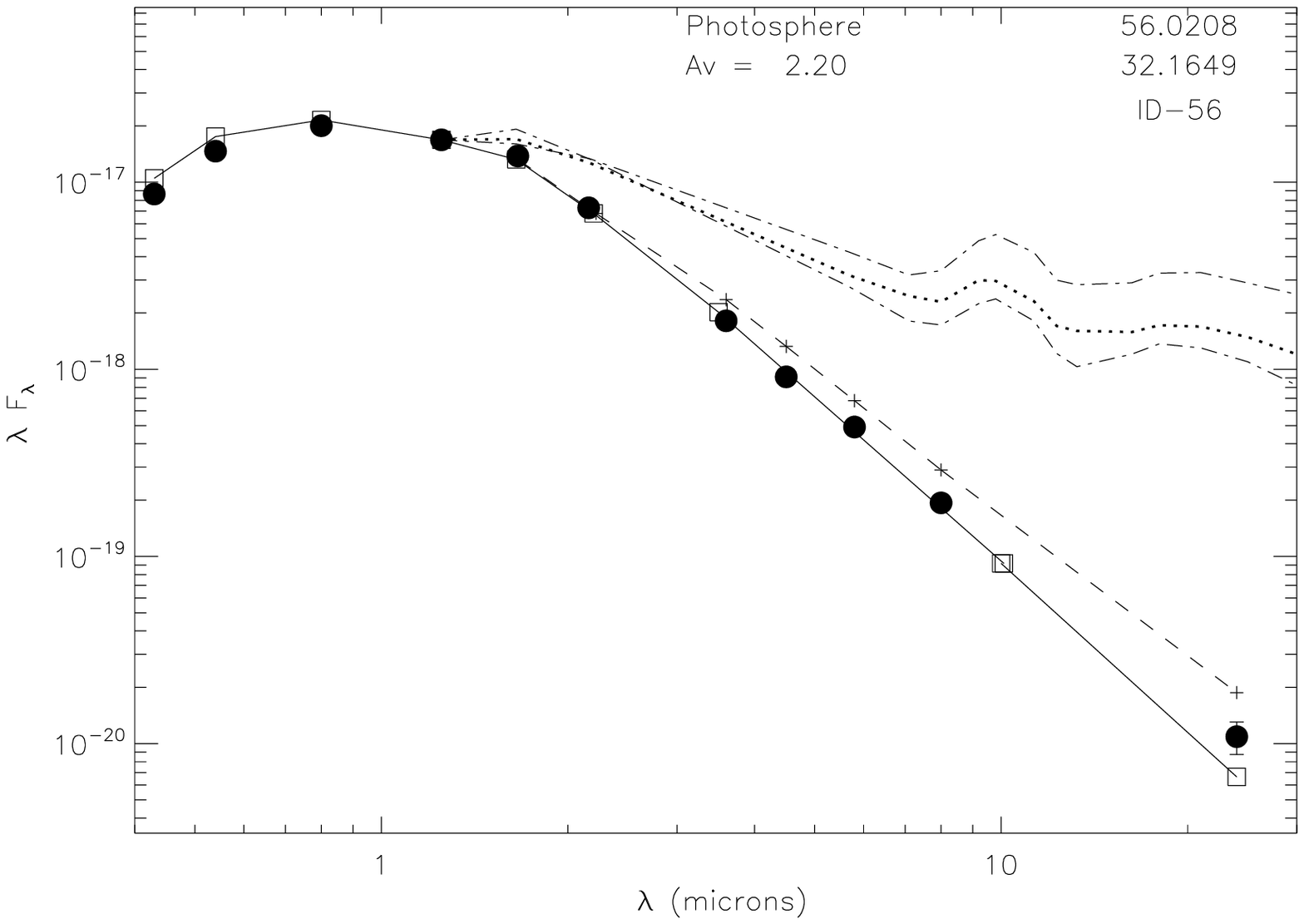}
\plottwo{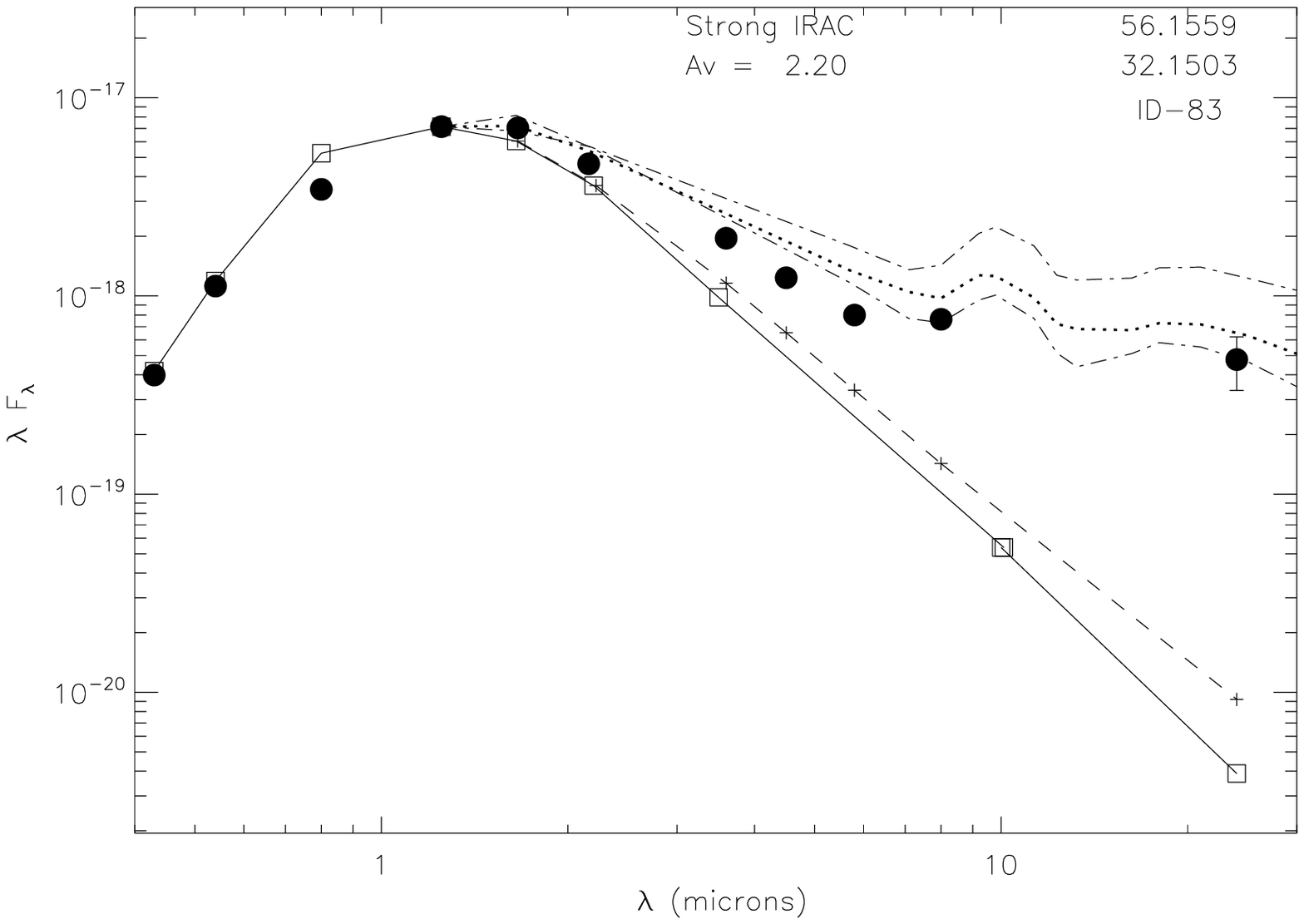}{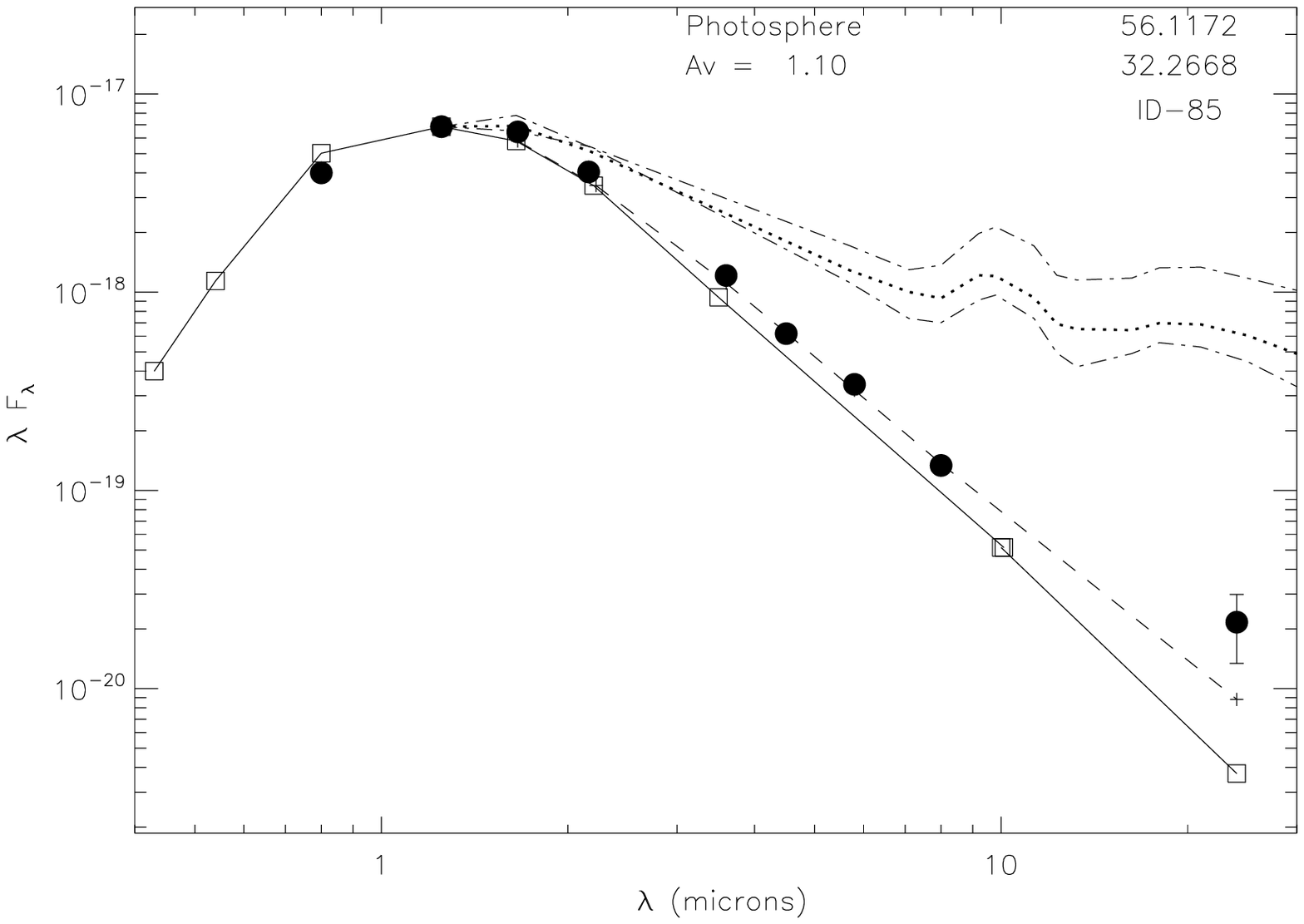}
\end{figure}

\begin{figure}
\epsscale{0.9}
\centering
\plottwo{ID-88.ps}{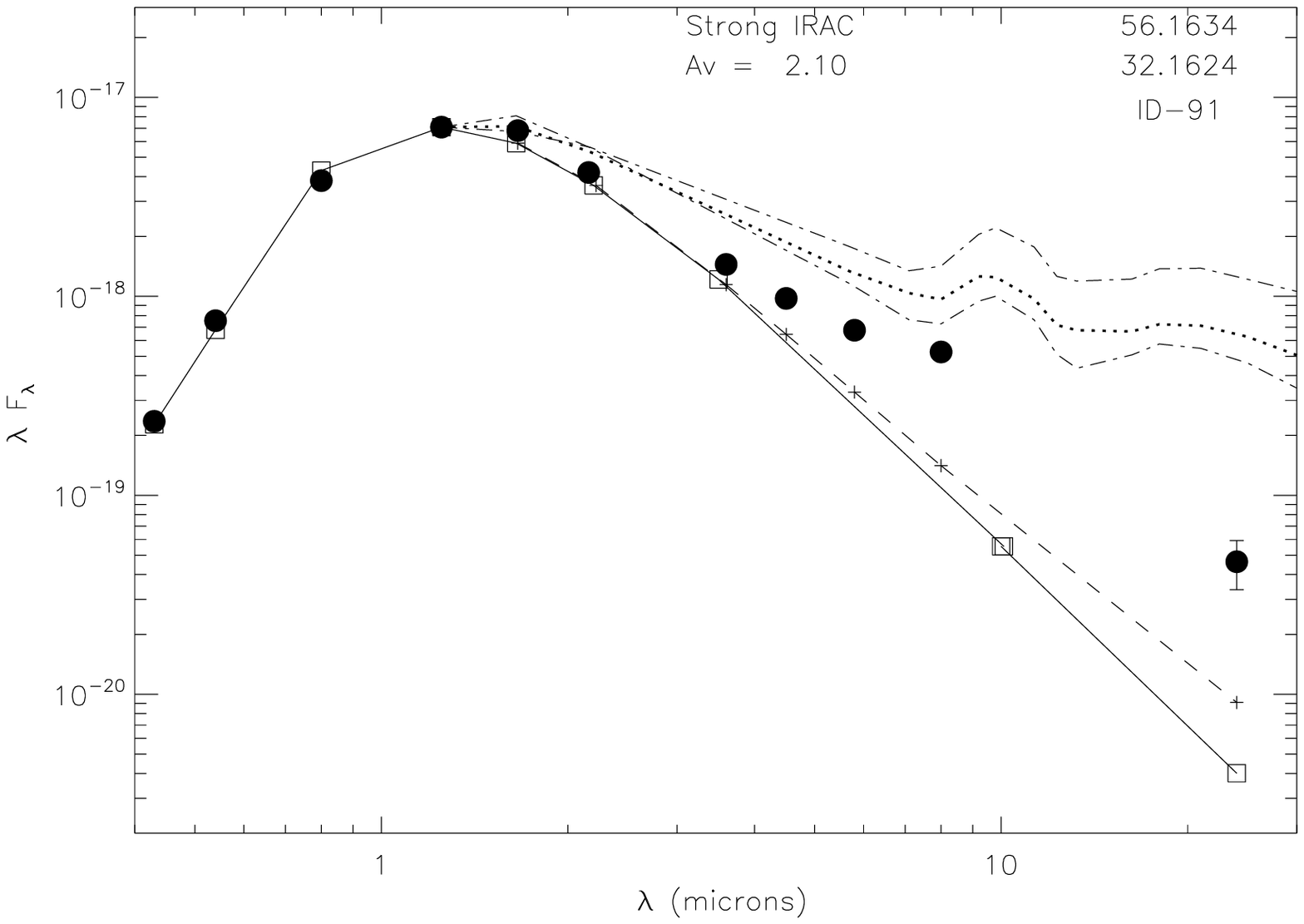}
\plottwo{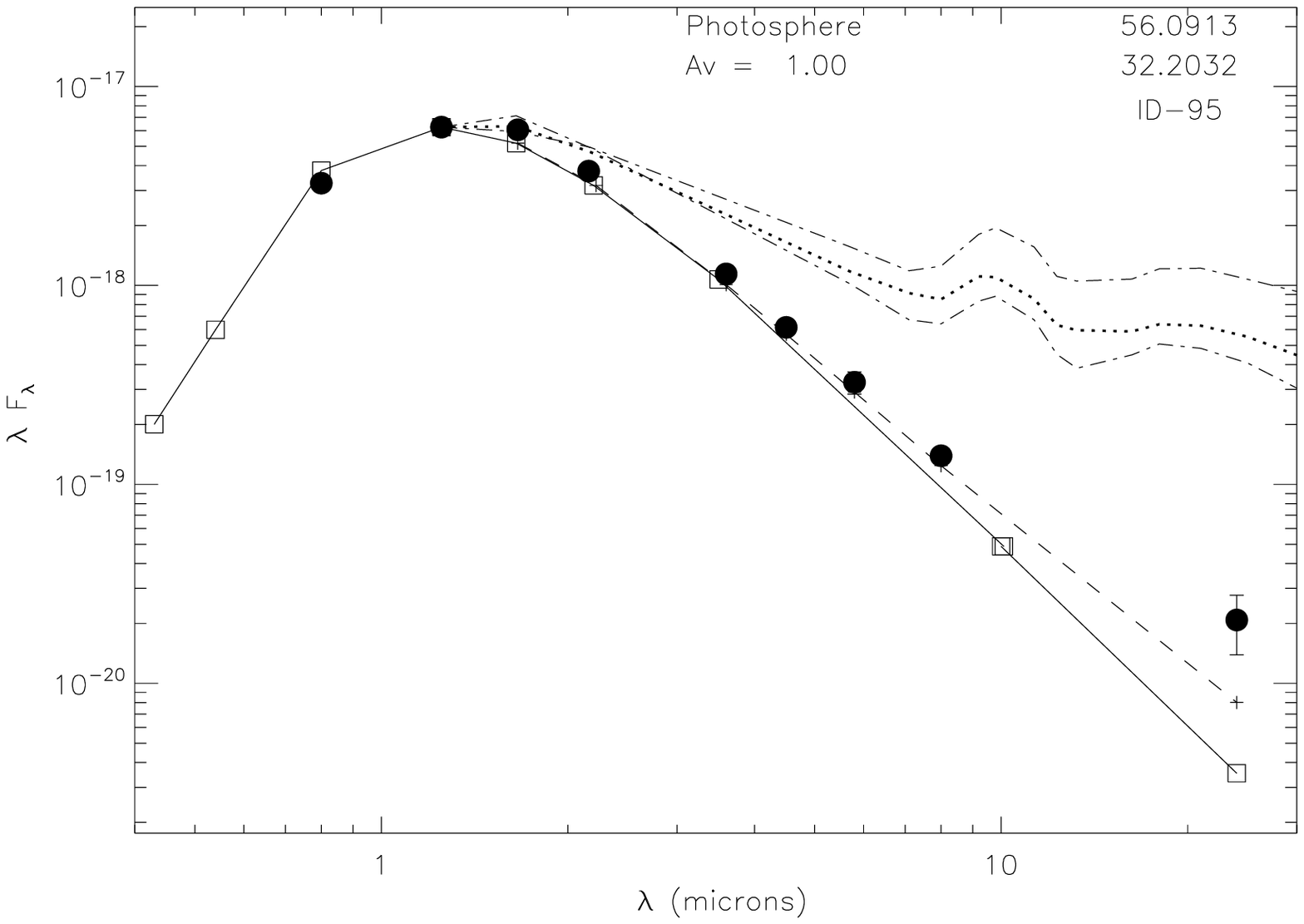}{ID-124.ps}
\end{figure}

\begin{figure}
\epsscale{0.9}
\centering
\plottwo{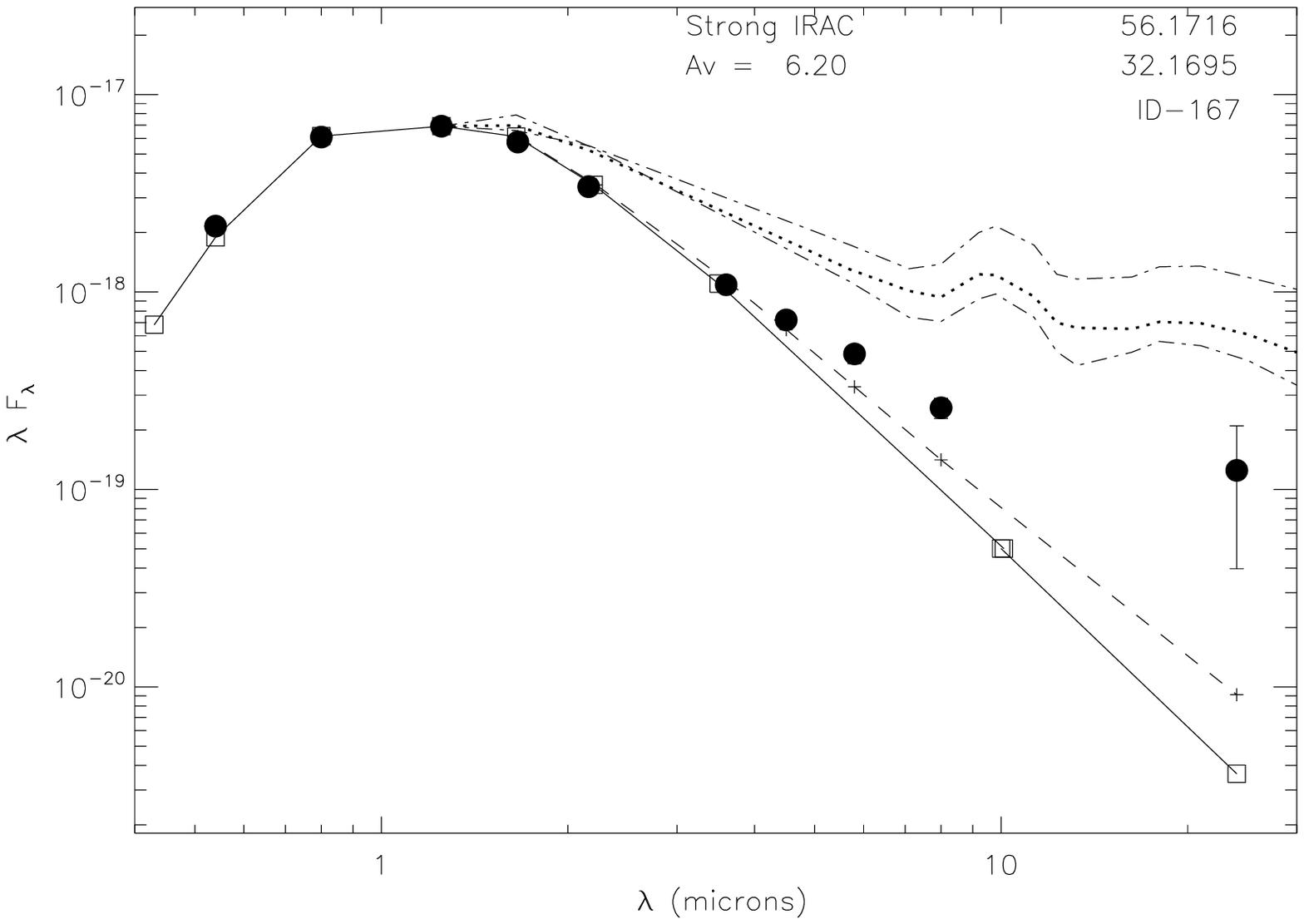}{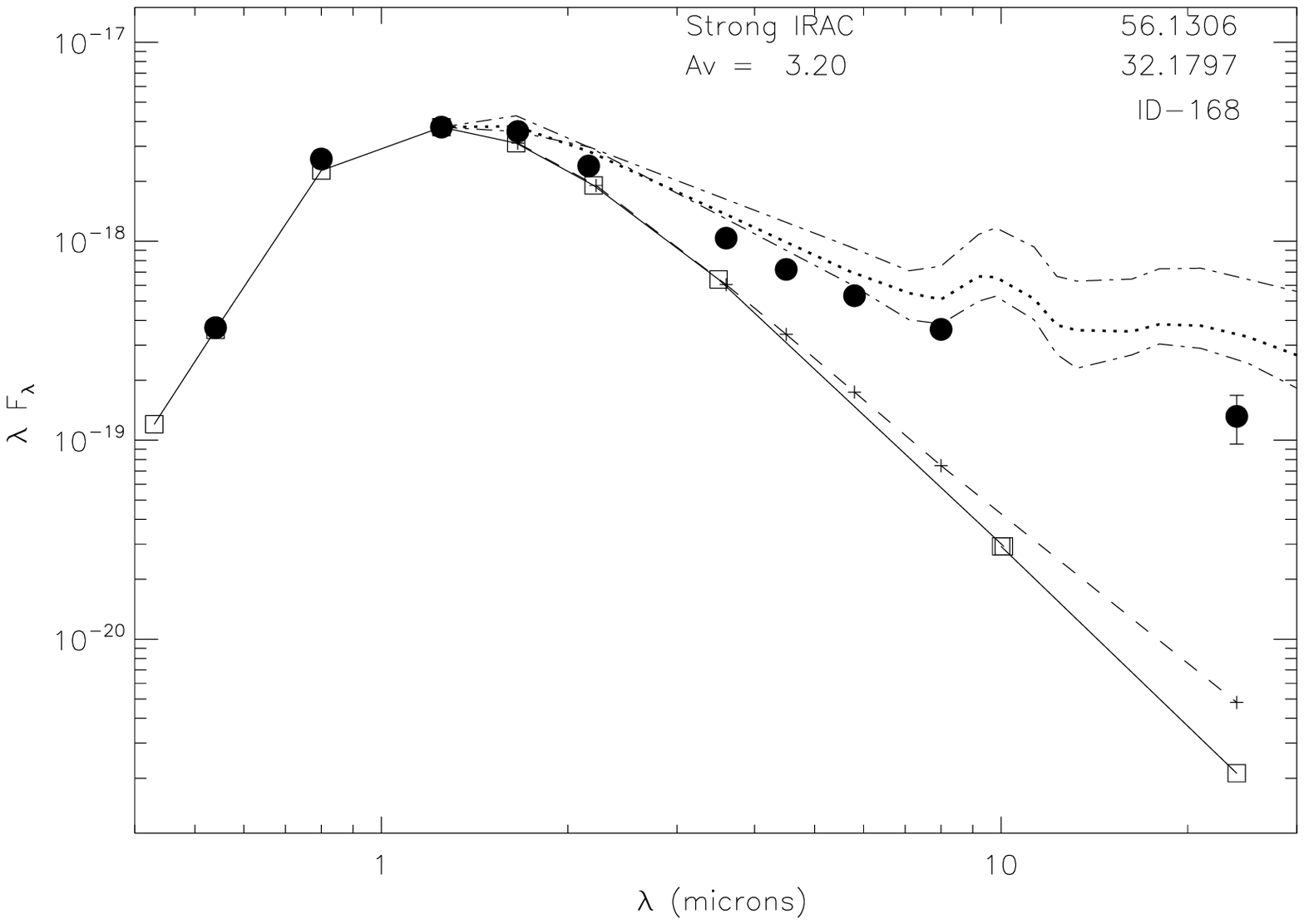}
\plottwo{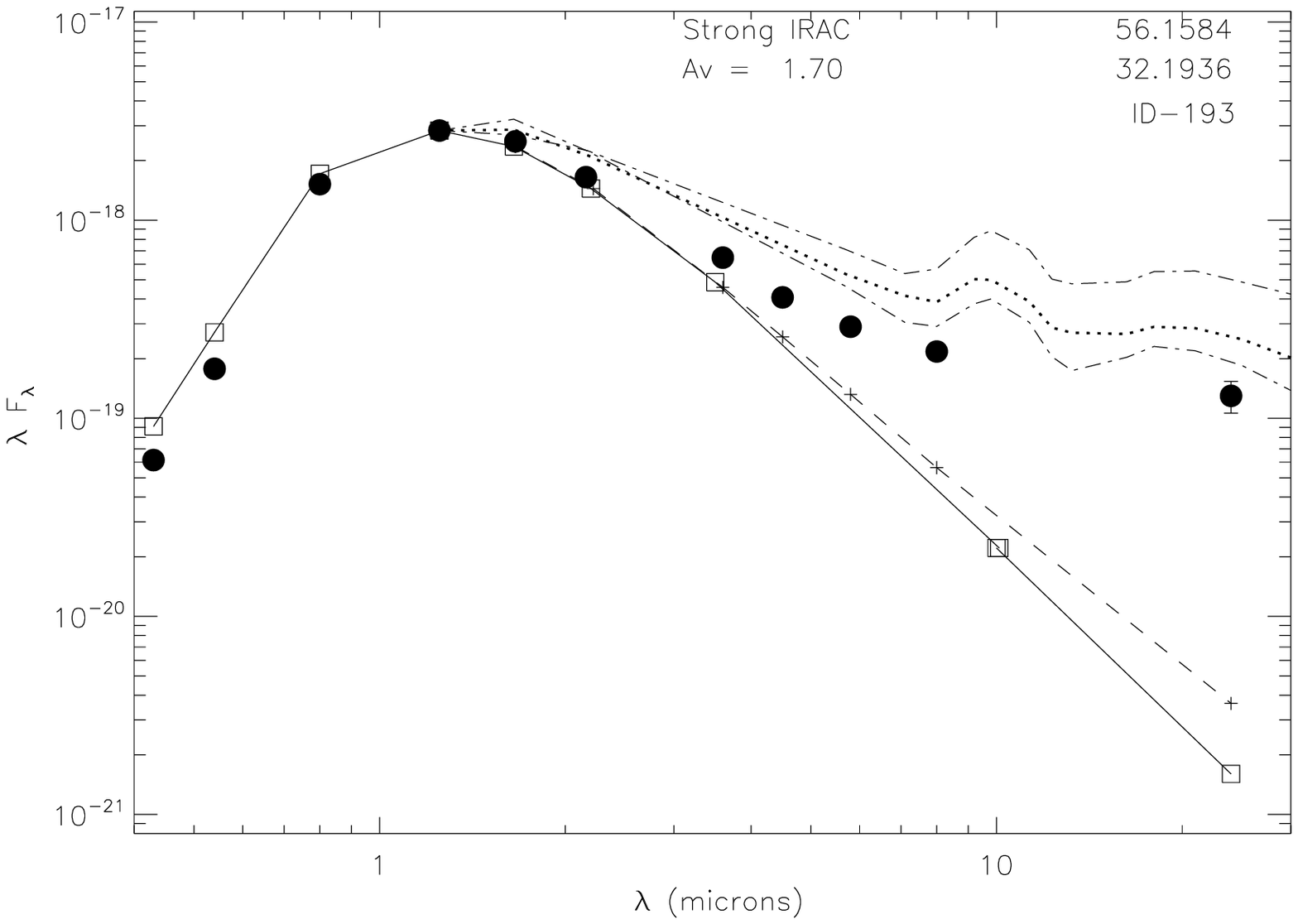}{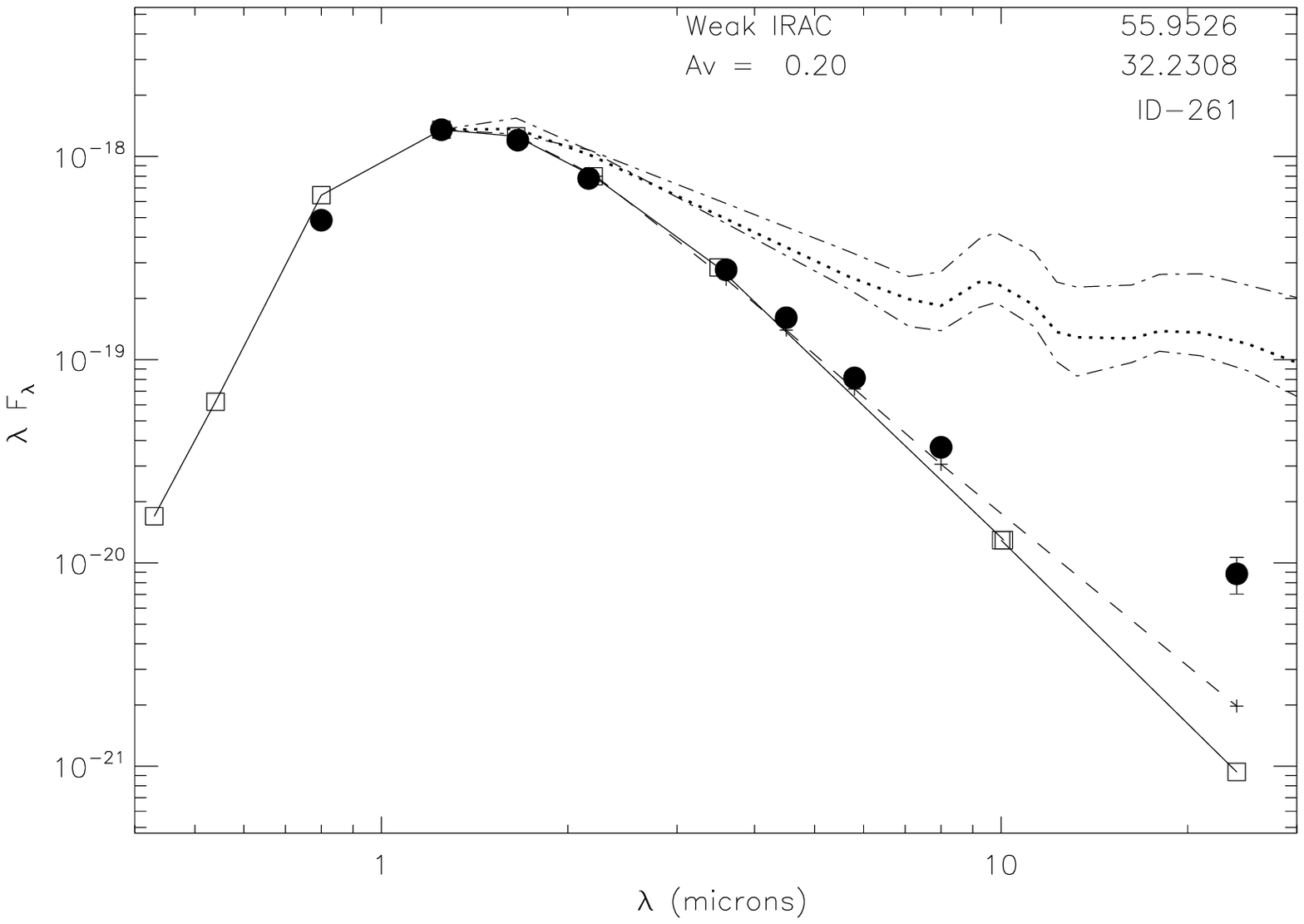}
\end{figure}

\begin{figure}
\epsscale{0.9}
\centering
\plottwo{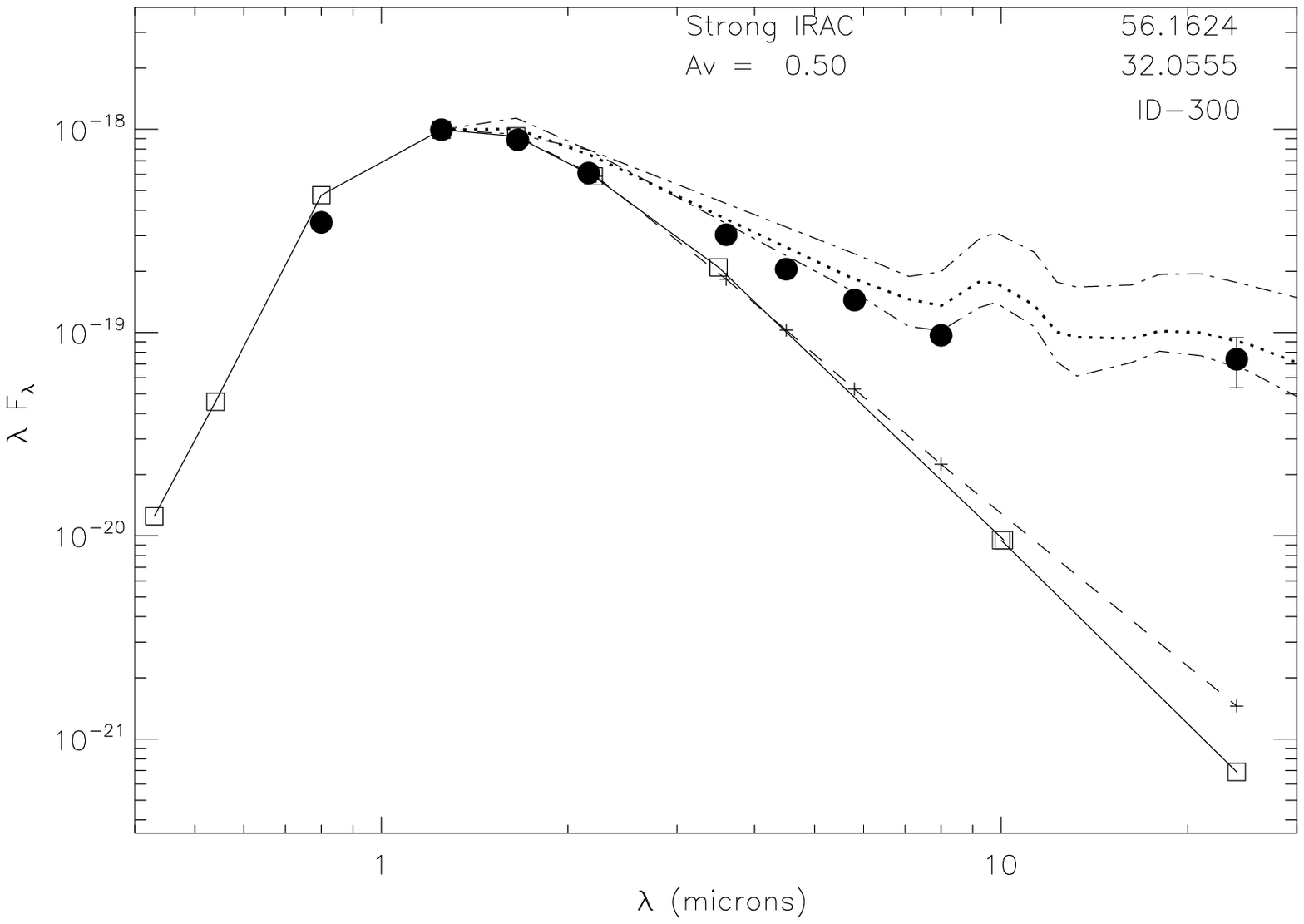}{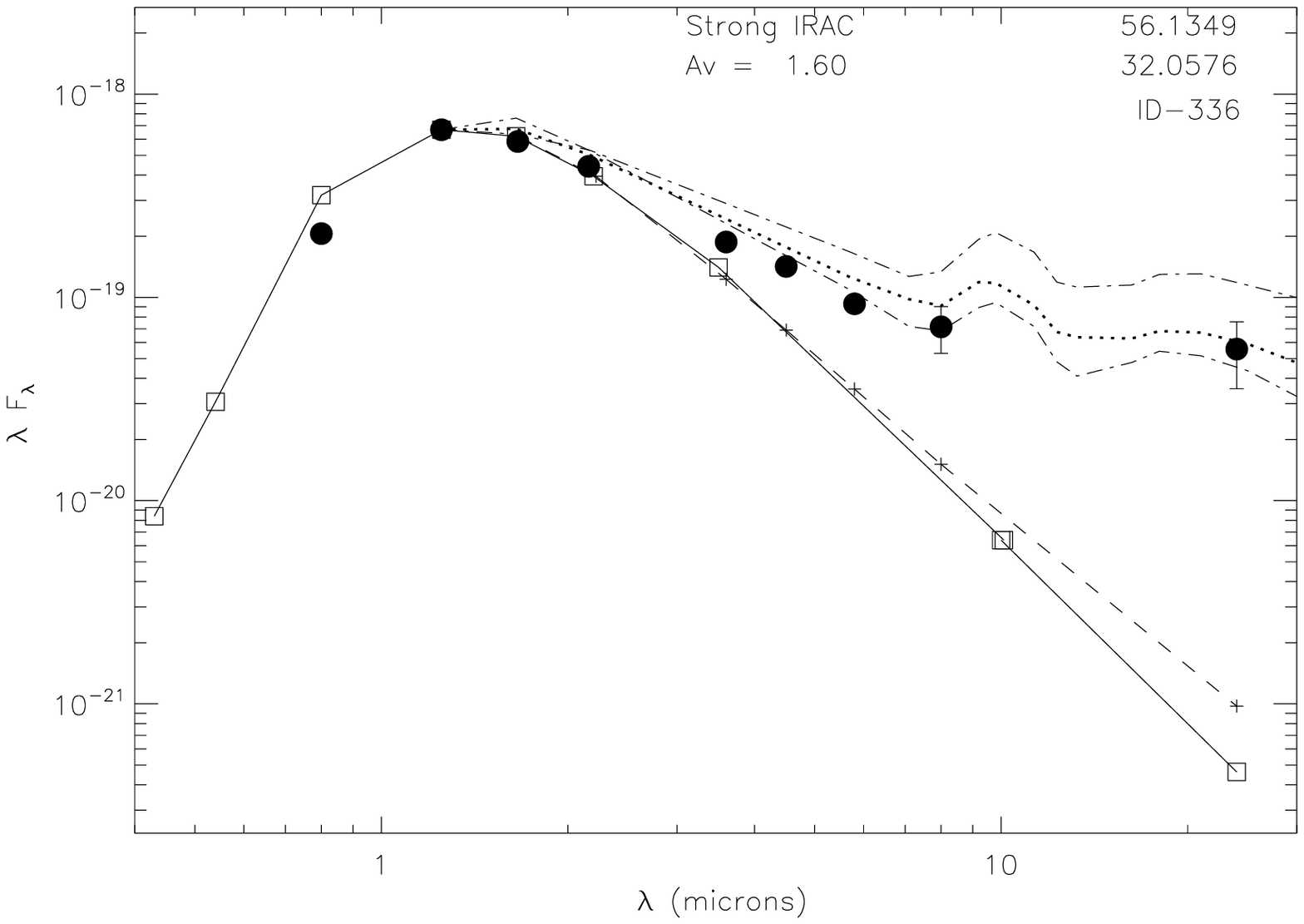}
\plottwo{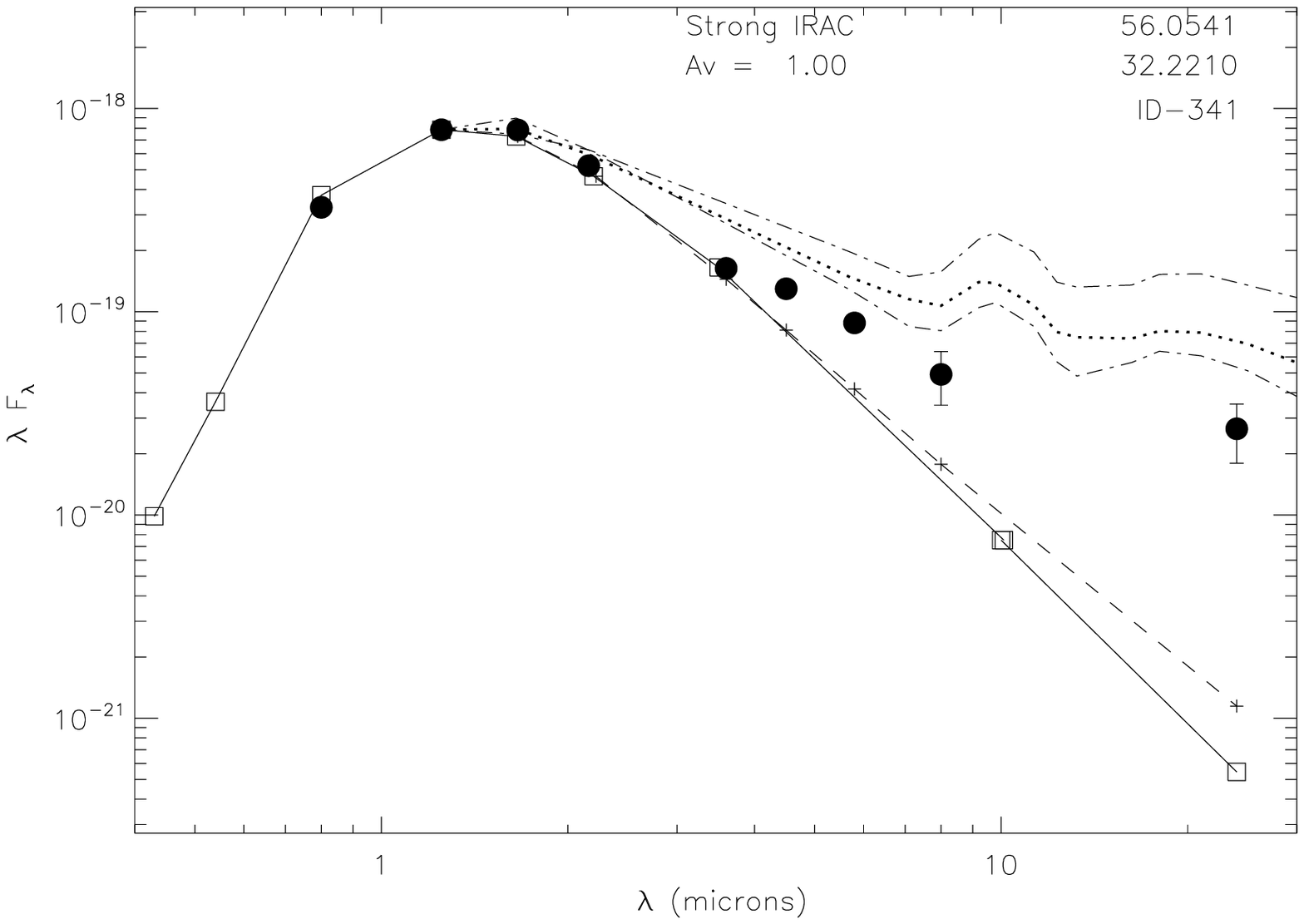}{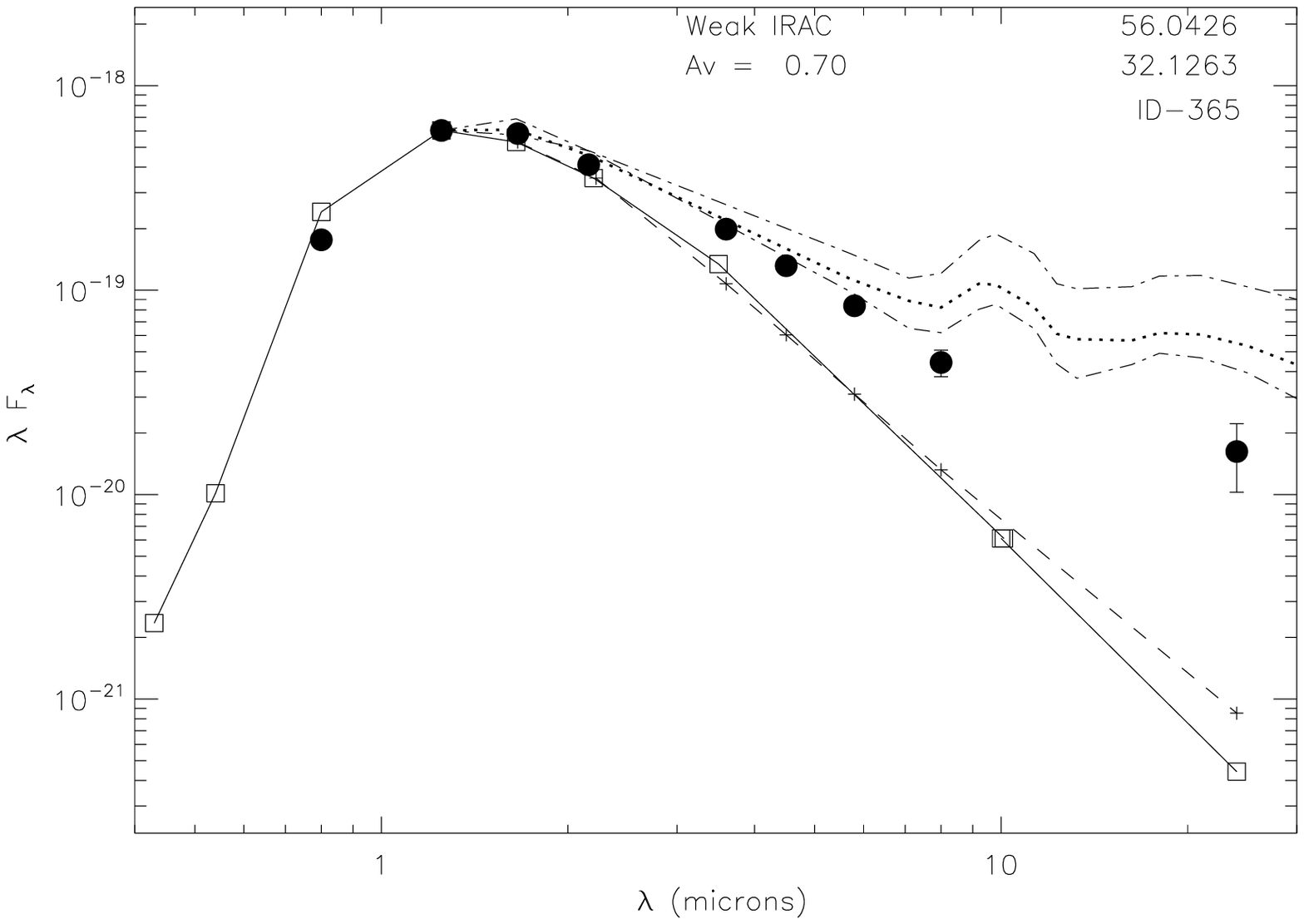}
\end{figure}

\begin{figure}
\epsscale{0.9}
\centering
\plottwo{ID-373.ps}{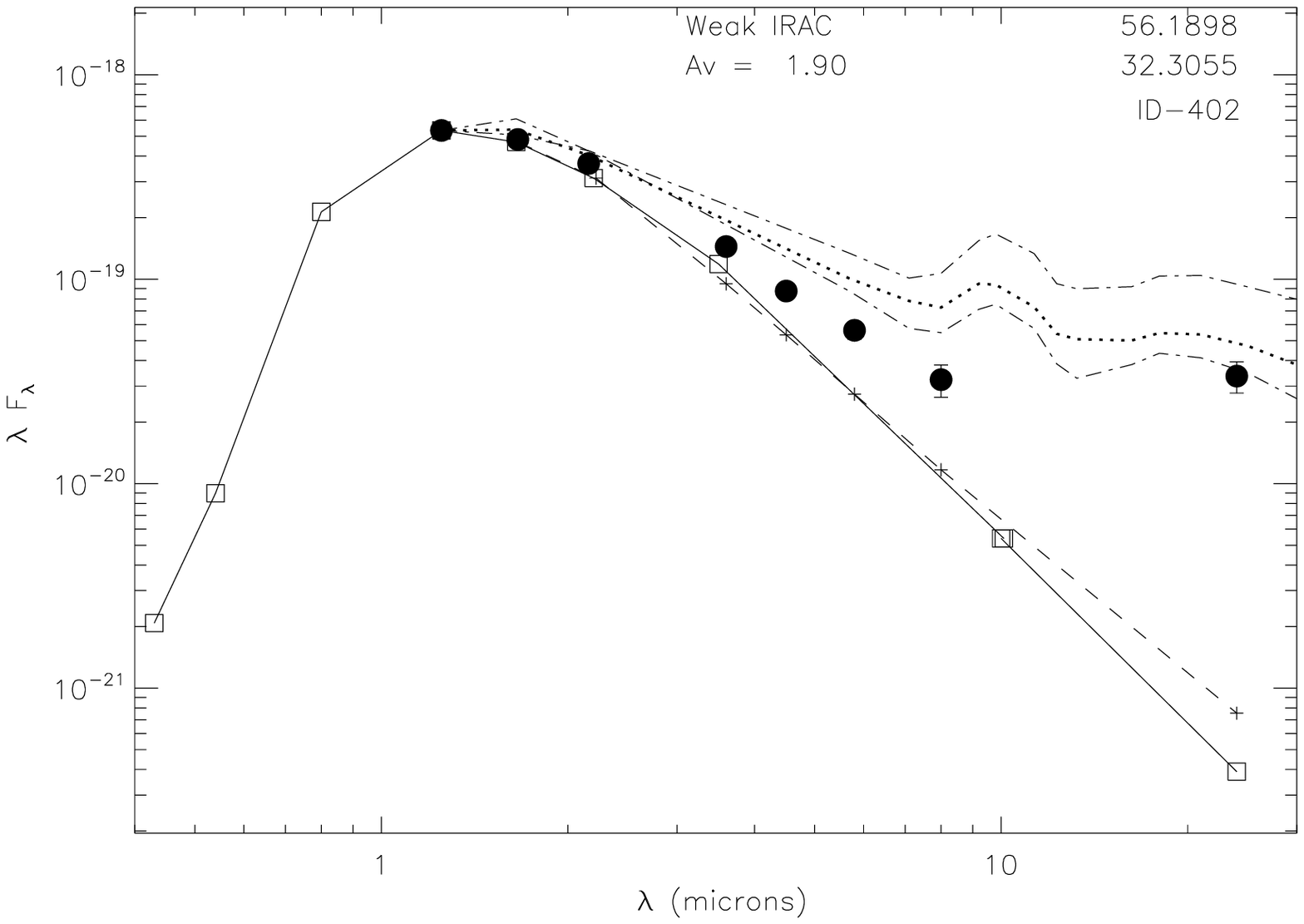}
\plottwo{ID-407.ps}{ID-1124.ps}
\centering
\epsscale{0.45}
\plotone{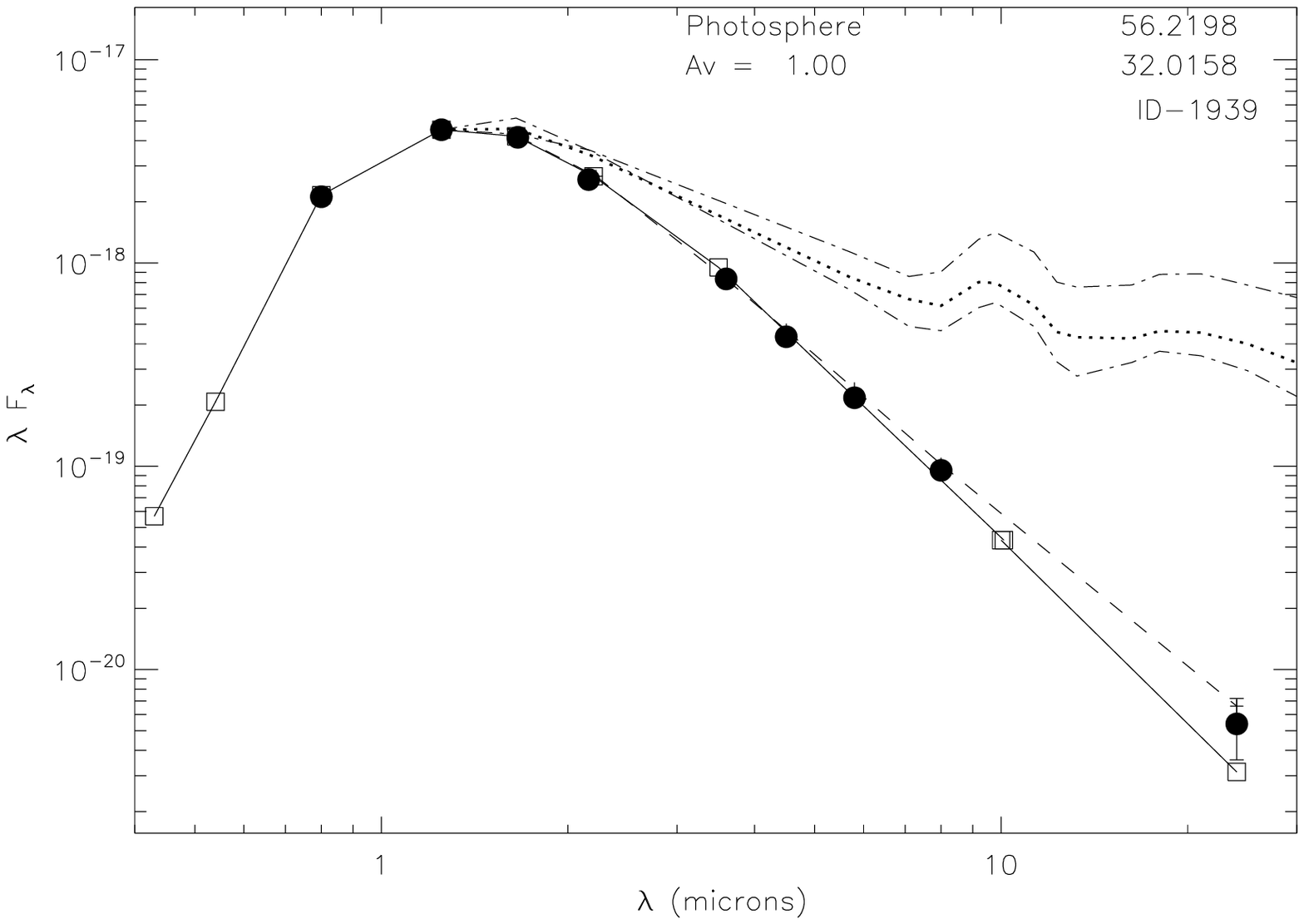}
\end{figure}

\end{document}